\documentclass[12pt]{article}

\advance\hoffset by -.35in
\begin{document}
\renewcommand{\theequation}{\arabic{section}.\arabic{equation}}
\makeatletter
\@addtoreset{equation}{section}
\makeatother
\bibliographystyle{unsrt}
\begin{flushright}
LPENSL-TH-05/2001\\
UNIL-IPT-01-13
\end{flushright}
\hfill {\tt hep-th/0110131}
\vskip0.3truecm
\begin{center}
\vskip 2truecm
{\Large\bf
Superfield Noether Procedure}
\vskip 1truecm

Marc Magro~\footnote{Marc.Magro@ens-lyon.fr} \\
Laboratoire de Physique,
\'Ecole normale sup\'erieure de Lyon, \\
46, All\'ee d'Italie, 69364 Lyon - Cedex 07, France\\

\medskip

Ivo Sachs~\footnote{ivo@theorie.physik.uni-muenchen.de} \\
Theoretische Physik, Ludwig-Maximilians Universit\"at, \\
Theresienstrasse 37, D-80333, M\"unchen, Germany \\

\medskip

Sylvain Wolf~\footnote{Sylvain.Wolf@ipt.unil.ch} \\
Institut de Physique Th\'eorique, BSP,
Universit\'e de Lausanne\\
1015 Lausanne, Switzerland

\end{center}
\vskip 2truecm
\begin{abstract}
\noindent
We develop a superspace Noether procedure for supersymmetric
field theories in $4$-dimensions for which an off-shell
formulation in ordinary superspace exists.
In this way we obtain an elegant and compact
derivation of the various supercurrents in these theories. We then apply
this formalism to compute the central charges for a variety of effective
actions. As a by-product we also obtain a simple derivation of the
anomalous superconformal Ward-identity in ${\cal{N}}=2$ Yang-Mills theory.
The connection with linearized supergravity is also discussed.
\end{abstract}

\newpage
\section{Introduction}
\setcounter{section}{1}

Noether currents play an important role in any theory
with a continuous global
symmetry. Moreover, when there is more than one invariance, the various
Noether currents themselves form a multiplet for the extended symmetry group.
This property is used extensively in supersymmetric theories. Indeed,
soon after the first $4$-dimensional supersymmetric field theory was proposed
by Wess and Zumino~\cite{WZ1}, the corresponding multiplet of Noether
currents, containing the energy momentum tensor, the supersymmetry current and
the R-current was constructed~\cite{FZ1}. This multiplet structure plays an
important role in exploring non-perturbative properties of the quantum theory.
So, for example, the rigid multiplet structure of the Noether currents
made it possible to obtain the $\beta$-function to all orders for
minimally supersymmetric Yang-Mills theory~\cite{Gr1,NSVZ1,GW1}. Furthermore,
the multiplet structure of these currents was crucial in explaining
higher loop finiteness of
theories with extended supersymmetry~\cite{SW1,HSW1,PS}.
Similarly, the supercurrent of ${\cal{N}}=2$ Yang-Mills theory played an
important role in deriving the Seiberg-Witten low energy effective action
for that theory~\cite{Seiberg-Witten,FMRSS,MRS}. On another front, the
Noether currents of supersymmetric matter theories can be
used to construct linear off-shell supergravities~\cite{OS1}.

As with any symmetry, the multiplet structure is best discussed by using a
manifestly covariant formalism. For supersymmetric theories, the manifestly
covariant formulation is in terms of
superfields~\cite{SS1}. However, as superfields are not
adapted to the canonical formalism, the approach usually
followed consists of first working in components to determine the
different conserved currents and then constructing the
corresponding supercurrent. This
can be circumvented by using a variational approach to obtain the Noether
currents~\cite{OS2,HST1,Shizuya,Osborn}.
However, one typically encounters  constrained superfields when
formulating realistic supersymmetric theories. Solving these
constraints is possible for all ${\cal{N}}=1$ theories but has been worked
out in ordinary superspace only for a limited subset of theories
with extended supersymmetry. Thus, the variational approach is complicated
by having to deal with constraint preserving variations.
Alternatively, if a superspace description of the coupling of the theory
in question to supergravity is available, the supercurrent can be
obtained by variation with respect to the supergravity
fields (see e.g. \cite{Buchbinder} for a review and references).

In spite of these drawbacks, a
manifestly covariant derivation of the multiplet of Noether currents is
certainly desirable. The purpose of this paper is to develop a general
formalism (Superfield Noether Procedure) to determine the
supercurrent  associated with
the super-Poincar\'e/superconformal invariance of a generic
theory that can be formulated in terms of ${\cal{N}}$-extended,
unconstrained superfields. For ${\cal{N}}=1$,  a procedure to extract a
supercurrent was proposed in~\cite{OS2,Clark-Piguet-Sibold}.
Here we elaborate on a superfield Noether procedure first proposed
in~\cite{Shizuya,Osborn}. The starting point is an abstract
supersymmetric theory whose action can be expressed in
terms of
${\cal{N}}$-extended superfields. We then obtain the supercurrent by considering
the variation of the action under a local transformation
on the superfield level. In this way we obtain directly
the supercurrent
for an arbitrary action of the type
described above. The various component Noether currents are then obtained
by an appropriate projection of the supercurrent. While this
approach may not appear
to be very economic to recover the known multiplets of Noether
currents in the simplest models,
it is rather powerful in generalizing these results to more complicated
Lagrangians, such as low energy effective Lagrangians in supersymmetric
quantum field theory and string theory. In particular, we obtain the
supercurrents for a variety
of ${\cal{N}}=1$ and ${\cal{N}}=2$ multiplets with arbitrary local action.
Furthermore, our procedure enables us to add improvement terms at the
superfield level, that is, in a
manifestly supersymmetric manner.  This allows us to give a uniform
description of all multiplets differing by improvement terms including the
so-called
canonical multiplet which, among its components contains the central
charges of the supersymmetry algebra. As a result we obtain an elegant
and economic derivation of central charges of the supersymmetry algebra in models with
arbitrary Lagrangians, not just holomorphic ones. As another
simple application, we will give a simple derivation of the
anomalous superconformal Ward-Identity in ${\cal{N}}=2$ Yang-Mills
theory~\cite{Howe96}. Finally, the superfield Noether procedure provides
a tool for a simple construction of linearized supergravities directly at
the superfield level. In particular, we recover various known, linearized
${\cal{N}}=1$ supergravities~\cite{SW1}, \cite{FN1}-\nocite{Aku1,B1,GG1}\cite{GOS1}
as well as some ${\cal{N}}=2$
supergravities~\cite{SG2}-\nocite{SG3,SG4,SG5,SG6}\cite{SG6}
in a simple and uniform manner.

The rest of this paper is organized as follows.
In section~\ref{Noether},  we formulate the super-Noether procedure
for an abstract supersymmetric action formulated in terms of unconstrained
superfields. In order to be self contained  we begin
with a review
of the superspace diffeomorphism transformations and of the
superconformal and super-Poincar\'e subgroups. In section~\ref{secn1},
we illustrate the use of this formalism by applying it to concrete models
with ${\cal{N}}=1$ supersymmetry. In particular we
obtain their supercurrents and compute the corresponding central charges.
In section~\ref{secn2},  we then
refine the general formalism to deal with the constrained superfields in
${\cal{N}}=2$  and then apply it to the ${\cal{N}}=2$ vector
and tensor multiplets. As a result we obtain a simple derivation of the
anomalous superconformal Ward Identity for the vector
multiplet and compute the effective central charge including
fermions in the theory
as well as the contributions from the non-holomorphic
part of the
effective action for that model.
For the tensor multiplet, we derive the supercurrent
and discuss the central charge. In section~\ref{eq1000},
we discuss the construction of
${\cal{N}}=1$ and ${\cal{N}}=2$ linearized superfield supergravities
using the superfield Noether procedure. In section $6$, we present the
algebraic relations between the component Noether
currents. Finally, we present the conclusions in the last section.

\label{general}

\section{Superfield Noether Procedure}
\label{Noether}
The purpose of this section is to develop the general
formalism to obtain the supercurrent for an arbitrary
super-Poincar\'e/superconformal invariant theory formulated in
terms of unconstrained, ${\cal{N}}$-extended,
superfields. To extract the Noether currents associated
with the global super-Poincar\'e/superconformal symmetries, we need to
consider the corresponding local transformations. Therefore, in the next
subsection we first discuss some aspects of the larger group of
superdiffeomorphisms.

\subsection{Superdiffeomorphisms}
\label{980}

The field theories considered in this paper are formulated
on ${\mathcal N}$-extended superspace with coordinates
$z=(x^\mu,\theta_{\alpha{\mathbf i}},\bar\theta_{\dot{\alpha}}^{\mathbf i})$
$\in {{\rm I} \kern -.19em {\rm R}}^{4|4{\mathcal N}}$,
${\mathbf i}$ being the $SU({\mathcal N})$ index, or, on the complex
chiral- and anti chiral superspaces parametrized
by~\footnote{We use essentially Wess and Bagger conventions~\cite{WessBagger},
see appendix~\ref{eq231}.}
$z_+=(x_+^\mu,\theta_{\alpha{\mathbf i}})$ and
$z_-=(x_-^\mu,\bar\theta_{\dot{\alpha}}^{\mathbf i})$.
Concretely we consider  those
superdiffeomorphisms which preserve chirality. Such
transformations can be described in terms of a superfield
$h^{\alpha{\dot{\alpha}}}$, subject to the constraints~\footnote{For detailed discussions of
the multiplet of superconformal transformations
we refer to~\cite{Sohnius76}-\nocite{Lang81,Conlong-West,Howe}\cite{Park},\cite{Shizuya}-\nocite{Osborn}\cite{Buchbinder},\cite{Howe96}. 
We will give generalizations where necessary.}
\begin{eqnarray}
&{\bar D}^{({\dot{\beta}}}_{\mathbf i} h^{\alpha{\dot{\alpha}})}=0\ ,\qquad
\qquad
D^{(\beta{\mathbf i}}{\bar h}^{\alpha){\dot{\alpha}}}=0\ .
\label{cons1b}
\end{eqnarray}
The corresponding transformations of the chiral coordinates are given by
\begin{equation}\begin{array}{ll}
\delta x^\mu_+=h^\mu(z)+2i\lambda_{\mathbf i}(z_+)\sigma^\mu\bar
\theta^{\mathbf i}\ ,\qquad\qquad
&
\delta\theta^\alpha_{\mathbf i}=\lambda^\alpha_{\mathbf i}(z_+)
\ ,\\[3mm]
\delta x^\mu_-={\bar h}^\mu(z)-2i\theta_{\mathbf i}\sigma^\mu
{\bar\lambda}^{\mathbf i}(z_-)\ ,\qquad\qquad
&
\delta\bar\theta^{{\dot{\alpha}}{\mathbf i}}=
{\bar\lambda}^{{\dot{\alpha}}{\mathbf i}}(z_-)\ ,
\end{array}\label{defdeltax}\end{equation}
with
\begin{equation}
\lambda^\alpha_{\mathbf i}(z_+)=-\frac{i}{8}
{\bar D}_{{\dot{\alpha}}{\mathbf i}} h^{\alpha{\dot{\alpha}}}\ ,\qquad\qquad
{\bar\lambda}^{{\dot{\alpha}}{\mathbf i}}(z_-)=\frac{i}{8}D_\alpha^{\mathbf i}
{\bar h}^{\alpha{\dot{\alpha}}}\ .
\label{lambda}\end{equation}
The representation of~(\ref{defdeltax}) in terms of the
corresponding differential operators
acting on chiral and antichiral superfields are then
\begin{equation}
{\mathcal L}_+=h^\mu\partial_\mu+\lambda^\alpha_{\mathbf i}
D_\alpha^{\mathbf i}\ ,\qquad\qquad
{\mathcal L}_-={\bar h}^\mu\partial_\mu
+{\bar\lambda}^{\mathbf i}_{\dot{\alpha}}{\bar D}_{\mathbf i}^{\dot{\alpha}}\ .
\label{defLL}\end{equation}
These commute with the chirality constraint, $[
{\bar D}_{{\dot{\alpha}}{\mathbf i}},{\mathcal L}_+]=0$ and
$[D_\alpha^{\mathbf i},{\mathcal L}_-]=0$, as a consequence of~(\ref{cons1b}).

An important subgroup of the superdiffeomorphisms containing the
super-Poincar\'e transformations is the superconformal
group \cite{Sohnius76,Lang81} obtained by imposing
the constraint
\begin{eqnarray}
&h^{\alpha{\dot{\alpha}}}= {\bar h}^{\alpha{\dot{\alpha}}}\ .\label{cons1}
\end{eqnarray}
From~(\ref{cons1b}) and (\ref{cons1}), we can easily extract the
Killing equation $\partial_\mu h_\nu+\partial_\nu h_\mu=
\frac{1}{2}\eta_{\mu\nu}\partial^\rho h_\rho$,
the general solution of which is given in
appendix \ref{1000}. Finally, super-Poincar\'e transformations are those for
which the  chiral and traceless, hermitian superfields, 
\begin{equation}
\sigma =\frac{1}{6}\left( D^{\alpha{\mathbf i}}\lambda_{\alpha{\mathbf i}}
-\frac{1}{2}\partial_{\alpha{\dot{\alpha}}}h^{\alpha{\dot{\alpha}}}\right)
\label{defsigma}\end{equation}
and
\begin{equation}
\Lambda^{\mathbf i}{}_{\mathbf j} =
-\frac{i}{4}\left( D_\alpha^{\mathbf i}\lambda^\alpha_{\mathbf j}+
{\bar D}_{{\dot{\alpha}}{\mathbf j}}{\bar\lambda}^{{\dot{\alpha}}{\mathbf i}}
-\frac{1}{{\mathcal N}}\delta^{\mathbf i}_{\mathbf j}\left(
D_\alpha^{\mathbf k}\lambda^\alpha_{\mathbf k}
+{\bar D}_{{\dot{\alpha}}{\mathbf k}}{\bar\lambda}^{{\dot{\alpha}}{\mathbf k}}
\right)\right)
\end{equation}
both vanish.

\subsection{Noether Procedure}

Let us denote by $O^{A{\mathbf J}}$ a generic, unconstrained
superfield, where $A$ is a collective vector and spinor index
and ${\mathbf J}$ transforms in a representation of the $SU({\mathcal N})$
internal group.
On $O^{A{\mathbf J}}$, the superconformal group acts as
\begin{equation}
\delta O^{A{\mathbf J}}(z) =
\begin{array}[t]{l}
 -{\mathcal L}  O^{A{\mathbf J}}
 +(\Omega_\alpha{}^\beta{\mathcal S}_\beta{}^\alpha+{\bar\Omega}_{\dot{\alpha}}
{}^{\dot{\beta}}{\bar{\mathcal S}}_{\dot{\beta}}{}^{\dot{\alpha}})_B{}^A\
O^{B{\mathbf J}}
\\[3mm]
+i\Lambda^{\mathbf i}{}_{\mathbf j} ({\mathcal R}^{\mathbf j}
{}_{\mathbf i})^{\mathbf J}{}_{\mathbf K}\ O^{A{\mathbf K}}
-2(q\sigma+{\bar q}{\bar\sigma})\ O^{A{\mathbf J}} \ ,
\end{array}\label{rep}\end{equation}
where 
\begin{equation}
{\mathcal L}=\frac{1}{2}(h^\mu+{\bar h}^\mu)\partial_\mu
+\lambda^\alpha_{\mathbf i} D_\alpha^{\mathbf i}
+{\bar\lambda}^{\mathbf i}_{\dot{\alpha}}{\bar D}^{\dot{\alpha}}_{\mathbf i}
\ .
\label{defLLgen}\end{equation} 
The symbols ${\mathcal S}_\beta{}^\alpha$ and ${\mathcal R}^{\mathbf j}
{}_{\mathbf i}$ stand for the Lorentz and
$SU({\mathcal N})$ generators
of the representation to which $O^{A{\mathbf J}}$ belongs
and $\Omega_\alpha{}^\beta$ is defined by
eq.(\ref{defsol}) in appendix \ref{1000}.
The dimension $d$ and $U(1)_R$-weight of $O^{A{\mathbf J}}$
are related to $q$ and ${\bar q}$ through
\begin{equation}
d=\frac{4-{\mathcal N}}{3}(q+{\bar q})\ ,\qquad\qquad\qquad
R=-\frac{2{\mathcal N}}{3}(q-{\bar q})\ .
\label{relRdqbq}\end{equation}

For a global super-Poincar\'e transformation,
we have $\delta S[O^{A{\mathbf J}}]= 0$, where $S[O^{A{\mathbf J}}]$ is
the action.
As in ordinary Noether procedure, we then consider a local
transformation, by letting the parameters have an arbitrary $x$-dependence.
We implement this by removing the
reality condition~(\ref{cons1}) but still maintaining~(\ref{cons1b}),
{\em i.e.} the chirality
preserving constraint.
Of course, there is always the ambiguity of adding terms to $\delta
O^{A{\mathbf J}}$
which are proportional to derivatives of the parameters of super-Poincar\'e
transformations.
By construction, the terms in
$\delta S[O^{A{\mathbf J}}]$ induced by them
 are of the form ``derivatives of the parameters'' times
``equations of motions'' and thus
induce in the currents terms which vanish on-shell.
We will make use of this freedom
when dealing with constrained superfields.
Using~(\ref{rep}),  the variation of the action can then
 always be written as
\begin{equation} \label{16}
\delta S[O^{A{\mathbf J}}] \begin{array}[t]{l}\displaystyle
= \frac{i}{16}\int d^{4+4{\mathcal N}}z\ \left\{\left(
h^{\alpha{\dot{\alpha}}} -{\bar h}^{\alpha{\dot{\alpha}}}\right)
T_{\alpha{\dot{\alpha}}} +i\left( h^{\alpha{\dot{\alpha}}} +
{\bar h}^{\alpha{\dot{\alpha}}}\right) K_{\alpha{\dot{\alpha}}}\right\}
\\[3mm]\displaystyle
\phantom{=}
-\frac{1}{2}\int  d^{4+2{\mathcal N}}z_+\ \sigma J
-\frac{1}{2}\int  d^{4+2{\mathcal N}}z_-\ {\bar\sigma}{\bar J}\ ,
\end{array}\label{h+-h}\end{equation}
with $T_{\alpha{\dot{\alpha}}}$ and $K_{\alpha{\dot{\alpha}}}$ real and $J$
chiral. The vanishing of $\delta S$ for global
super-Poincar\'e transformations furthermore
implies~\footnote{For ${\mathcal N}=1$, this was shown in~\cite{Osborn}.
For ${\mathcal N}=2$, this can be proved following the same argument as
in~\cite{Osborn}.
Note however that the ${\mathcal N}=2$ case is more elaborated than in $
{\mathcal N}=1$
as one has to use not only super-Poincar\'e invariance but also
$SU(2)_R$ invariance.
The proof of~(\ref{KX}) for arbitrary ${\mathcal N}$
should follow the same pattern.}
\begin{equation} \label{KX}
K_{\alpha{\dot{\alpha}}}=
-\frac{i}{4} [D_\alpha^{\mathbf i},{\bar D}_{{\dot{\alpha}}{\mathbf i}}](X-
{\bar X})
-\frac{{\mathcal N}}{2}\partial_{\alpha{\dot{\alpha}}}(X+{\bar X})
\ ,
\label{K}\end{equation}
for some $X$. However, the superfields
$T_{\alpha{\dot{\alpha}}},K_{\alpha{\dot{\alpha}}}$ and
$J$ are not unique. Indeed,
as a consequence of~(\ref{cons1b}) we have the two identities
\begin{equation}\begin{array}{rcl}
\displaystyle
-\frac{1}{2}[ D_\alpha^{\mathbf i},{\bar D}_{{\dot{\alpha}}{\mathbf i}}](h+
{\bar h})^{\alpha{\dot{\alpha}}}-48i(\sigma-{\bar\sigma})
-i(4-{\mathcal N})\partial_{\alpha{\dot{\alpha}}}(h-
{\bar h})^{\alpha{\dot{\alpha}}}=0
\ ,\\[3mm]
\displaystyle
-\frac{1}{2}[ D_\alpha^{\mathbf i},{\bar D}_{{\dot{\alpha}}{\mathbf i}}](h-
{\bar h})^{\alpha{\dot{\alpha}}}-48i(\sigma+{\bar\sigma})
-i(4-{\mathcal N})\partial_{\alpha{\dot{\alpha}}}(h+
{\bar h})^{\alpha{\dot{\alpha}}}=0
\ ,
\end{array}\label{rel2}\end{equation}
which, upon integration by parts lead to the equivalence relation
\begin{equation}
\left(\begin{array}{c}
T_{\alpha{\dot{\alpha}}}\\[3mm]
K_{\alpha{\dot{\alpha}}}\\[3mm]
J
\end{array}\right)
\simeq
\left(\begin{array}{c}
T_{\alpha{\dot{\alpha}}}+i\frac{4-{\mathcal N}}{2}
\partial_{\alpha{\dot{\alpha}}}(A-{\bar A})
-\frac{{\mathcal N}}{4(4-{\mathcal N})}[D_\alpha^{\mathbf i},
{\bar D}_{{\dot{\alpha}}{\mathbf i}}](A+{\bar A})\\[3mm]
K_{\alpha{\dot{\alpha}}}+\frac{i}{4}[D_\alpha^{\mathbf i},
{\bar D}_{{\dot{\alpha}}{\mathbf i}}](A-{\bar A})+\frac{{\mathcal N}}{2}
\partial_{\alpha{\dot{\alpha}}}(A+{\bar A})\\[3mm]
J -\frac{6}{4-{\mathcal N}}{\bar D}^{2{\mathcal N}}\left( 2 A +({\mathcal N}-2)
{\bar A}\right)
\end{array}\right)\ ,
\label{eqv1}\end{equation}
where $A$ is any unconstrained superfield. This equivalence
relation corresponds to the freedom of adding
improvement terms to the different
Noether currents in the multiplets. One way to see that
is to note that the equivalence classes~(\ref{eqv1}) are obtained through
integration by parts
or equivalently, by adding boundary terms. We will explain this
correspondence in more detail in the following sections.

There are two distinguished multiplets of Noether currents.
We denote them by the {\it minimal} and the
{\it canonical} multiplets respectively. Indeed, starting from a  generic
supermultiplet $(T_{\alpha{\dot{\alpha}}},K_{\alpha{\dot{\alpha}}},J)$
and taking $A =  X$, we arrive at the minimal
multiplet~\footnote{In general,
this minimal multiplet may not be unique~\cite{Osborn},
but this will not affect our analysis below.}
$(T^{min}_{\alpha{\dot{\alpha}}},K^{min}_{\alpha{\dot{\alpha}}}=0,J^{min})$.
In particular, for a
superconformal invariant theory,
$\delta S =  0$ for a superconformal transformation
and hence~(\ref{16})
implies that $J^{min} =  {\bar J}^{min} =  0$.  In this case the
supercurrent $T^{min}_{\alpha{\dot{\alpha}}}$ contains the traceless ({\em i.e.}
improved)
currents.
This is the smallest possible multiplet of Noether currents.
If the theory does not have superconformal invariance,
the $K_{\alpha{\dot{\alpha}}} =  0$
multiplet of currents  is still minimal  in the sense that it has the
smallest number of components.

The other distinguished multiplet
is the canonical multiplet. This multiplet contains the generators of the
super-Poincar\'e transformations and is related to the minimal
multiplet by
total derivative terms. This multiplet has the further property that it
contains among its components central charges.

Thus far the discussion applies to any theory
formulated in terms of unconstrained superfields.
However, as already mentioned in the introduction, in realistic situations
we typically have to deal with constrained superfields.
In general, the localization of the super-Poincar\'e
or superconformal transformations in terms of an arbitrary parameter
$h^{\alpha{\dot{\alpha}}}$ subject to~(\ref{cons1b}) may not be
compatible with
the constraints on the superfield.
In that case,  one may have to impose stronger
constraints on $h^{\alpha{\dot{\alpha}}}$. The purpose of the
next sections is to illustrate the general procedure by applying it
to some ${\mathcal N} =  1$ models of interest and to discuss the
modifications for ${\mathcal N} =  2$ models with off-shell superfield
formulations.

\section{${\mathcal N}=1$ Theories}
\label{secn1}

In this section, we will apply the general formalism of
section~\ref{Noether} to
an arbitrary ${\cal{N}}=1$ theory formulated in terms
of (possibly constrained)
${\cal{N}}=1$ superfields.
Indeed, it turns out that
the general structure described above applies
directly without modification to any ${\mathcal N}= 1$ theory with an off-shell
superfield formulation.

\subsection{General Structure}
Before considering concrete models with ${\cal N}=1$ supersymmetry we
present some model independent, general results in this
subsection.

\subsubsection{Equation of Conservation}

We begin by solving the constraint~(\ref{cons1b})
in terms of a free spinor superfield $L^\alpha$ as
\begin{equation}
h^{\alpha{\dot{\alpha}}}=2{\bar D}^{\dot{\alpha}} L^\alpha\ ,
\qquad\qquad\qquad
{\bar h}^{\alpha{\dot{\alpha}}}= -2D^\alpha {\bar L}^{\dot{\alpha}}\ .
\label{defhL}\end{equation}
Using the
results in section~\ref{Noether}
we can now write the variation of
the action under an arbitrary local transformation parametrized by $L^\alpha$
as\footnote{In this paper, we will take the
localization of superconformal transformations even
for a theory which is only super-Poincar\'e invariant. This is because a
free superfield, playing the analogous role of $L_\alpha$ but for
super-Poincar\'e transformations, is not known.} \cite{Shizuya,Osborn}
\begin{eqnarray}
\delta S&=&\displaystyle\int d^8z\ \left( \frac{i}{8} L^\alpha R_\alpha +\mbox
{c.c.}\right)\
,\qquad{\quad\hbox{{with}}\quad} {\nonumber}\\\
R_\alpha&=&{\bar D}^{\dot{\alpha}} T_{\alpha{\dot{\alpha}}}+i
{\bar D}^{\dot{\alpha}} K_{\alpha{\dot{\alpha}}}-\frac{1}{6}D_\alpha J\ ,
\label{consequRa}
\end{eqnarray}
where $K_{\alpha{\dot{\alpha}}}$ is given by~(\ref{K}).
Variation with respect to $L^\alpha$ then leads
to the supercurrent conservation equation
\begin{equation}
{\bar D}^{\dot{\alpha}} T_{\alpha{\dot{\alpha}}}+i{\bar D}^{\dot{\alpha}}
K_{\alpha{\dot{\alpha}}}-\frac{1}{6}D_\alpha J = 0
\ .\label{consequ}\end{equation}
As explained in section~\ref{Noether},
the multiplets appearing in~(\ref{consequ}) are not unique.
In particular, we can always make the choice
$K_{\alpha{\dot{\alpha}}}=0$ provided we redefine $T_{\alpha{\dot{\alpha}}}$
and $J$ as
\begin{eqnarray}
T_{\alpha{\dot{\alpha}}}&\to&{\widetilde T}_{\alpha{\dot{\alpha}}}=
T_{\alpha{\dot{\alpha}}}+\frac{3i}{2}\partial_{\alpha{\dot{\alpha}}}(X-
{\bar X})
-\frac{1}{12}[D_\alpha,{\bar D}_{\dot{\alpha}}](X+{\bar X})
{\quad\hbox{{and}}\quad}{\nonumber}\\
J&\to&{\widetilde J}=J-2{\bar D}^2(2X-{\bar X})\ ,
\end{eqnarray}
where we have replaced $\ {}^{min}$ by $\ \tilde{}\ $ to avoid clutter.
The supercurrent conservation equation then reads
\begin{equation}
R_\alpha={\bar D}^{\dot{\alpha}} {\widetilde T}_{\alpha{\dot{\alpha}}}-\frac{1}
{6}D_\alpha {\widetilde J}=0\ .
\label{consequt}\end{equation}
In particular, if the theory has superconformal invariance,
there is a minimal multiplet with $\tilde J=0$. In general, $\tilde J$ contains
the trace of the energy momentum tensor and the supersymmetry current and is
then referred to as the multiplet of
anomalies~\cite{FN1,SW2}.

\noindent Using~(\ref{K}) we can recast the conservation equation~(\ref{consequ})
in a more familiar form
\begin{eqnarray}\label{conG2}
\bar D^{{\dot{\alpha}}}T_{\alpha{\dot{\alpha}}}+W_\alpha+D_\alpha \tau&=&0\ ,
\end{eqnarray}
where
\begin{equation}
W_\alpha=\frac{1}{4}\bar D^2D_\alpha\left(X-2\bar X\right)
{\quad\hbox{{,}}\quad}
\tau=
-\frac{1}{6}J
+\frac{1}{4}\bar D^2 X\ .
\label{jenesaispas}
\end{equation}
This form of the conservation equation is familiar from the
discussion of the old- and new minimal multiplets \cite{FN1,SW2,Aku1}
which will be reviewed in section \ref{MNC}.

\subsubsection{$R$-invariance}
\label{secRcurrent}

The variation of the action under a global $R$-transformation, corresponding to
$L^\alpha=-\eta \theta^\alpha{\bar\theta}^2$ is given by
\begin{equation}
\delta S=\displaystyle
\int d^4 x\
\eta\displaystyle{\Biggl(}
2\partial^{\alpha{\dot{\alpha}}}{\widetilde T}_{\alpha{\dot{\alpha}}}
-\frac{i}{6}(D^2{\widetilde J}-{\bar D}^2
{\kern .35em\bar{\kern -.35em{\widetilde J}}})
\displaystyle{\Biggr)}\displaystyle{\biggl\vert}\ ,
\label{Rvariation}\end{equation}
and as expected, this is not automatically a derivative as we
are not necessarily dealing with a $R$-invariant theory. Indeed,
the $R$-invariance condition can be written as
\begin{equation}
D^2{\widetilde J}-{\bar D}^2{\kern .35em\bar{\kern -.35em{\widetilde J}}}=24i
\partial^\mu Z_\mu\ ,
\label{condR}\end{equation}
for some $Z_\mu$. If this condition is satisfied, we can define a conserved
$R$-current by
\begin{equation}
j^{(5)}_\mu=
-4\left({\widetilde T}_\mu-Z_\mu\right)\displaystyle{\biggl\vert}
\ .
\label{Rcurrent}\end{equation}
The condition~(\ref{condR}) is solved by
\begin{equation}
{\widetilde J}={\bar D}^2 U\qquad \mbox{and}\quad
{\kern .35em\bar{\kern -.35em{\widetilde J}}}=D^2 U\ ,
\label{sc1}\end{equation}
for some real superfield $U$.
Correspondingly, we have  $Z_{\alpha{\dot{\alpha}}}=\frac{1}{3}[D_\alpha,\bar
D_{\dot{\alpha}}]U$.
Now, applying~(\ref{eqv1})
with $A=X+2U$ we get
\begin{equation}
\left(\begin{array}{l}
T_{\alpha{\dot{\alpha}}}
\\[3mm]
W_\alpha
\\[3mm]
\tau
\end{array}\right)
\to
\left(\begin{array}{rl}
T^{new}_{\alpha{\dot{\alpha}}} =& {\widetilde T}_{\alpha{\dot{\alpha}}}-\frac
{1}{3}[D_\alpha,\bar D_{\dot{\alpha}}]U
\\[3mm]
W^{new}_\alpha =& \frac{1}{2}{\bar D}^2D_\alpha\ U
\\[3mm]
&0\label{new}
\end{array}\right)
\ ,
\label{shift1}\end{equation}
where we introduced the label $new$
for the multiplet~(\ref{new}). This is because (\ref{new})
is, in fact, the {\em new minimal} multiplet~\cite{Aku1}.
We will come back to this below.
The conservation equation~(\ref{conG2}) then reads
\begin{equation}
\bar D^{{\dot{\alpha}}}T^{new}_{\alpha{\dot{\alpha}}}+W^{new}_\alpha=0\ ,
\end{equation}
where $W^{new}_\alpha$ is now a curl
($D^\alpha W^{new}_\alpha+\bar D^{\dot{\alpha}}
{\bar W}^{new}_{\dot{\alpha}}=0$).
Therefore
\begin{equation}
\partial^{\alpha{\dot{\alpha}}}T^{new}_{\alpha{\dot{\alpha}}}=0
\end{equation}
and consequently the conserved $R$-current is given by
$T^{new}_{\alpha{\dot{\alpha}}}|$
which agrees with~(\ref{Rcurrent}).

\subsubsection{Multiplets of Noether currents}
\label{MNC}
It is clear that the size of the multiplet of currents depends on
the details of the theory under consideration but also on
the choice of improvement terms. If the theory is conformally invariant,
then the variation~(\ref{16}) must vanish for $h^{\alpha{\dot{\alpha}}}=\bar
h^{\alpha{\dot{\alpha}}}$
and hence it is possible to choose $K_{\alpha{\dot{\alpha}}}=0$
and $J=0$.
In this case,
the multiplet of Noether currents is contained entirely in
$T_{\alpha{\dot{\alpha}}}$
subject to the conservation equation $D^{\dot{\alpha}}
T_{\alpha{\dot{\alpha}}}=0$.
This is the improved
multiplet with $8+8$ components first constructed by Ferarra and Zumino~\cite
{FZ1} for the Wess-Zumino model. The next bigger multiplet is described
by $T_{\alpha{\dot{\alpha}}}$ and $J$. This is the {\it old minimal}
multiplet~\cite{FN1,SW2} with $12+12$ components. Another possibility is that
$\tau$ vanishes but not $W_\alpha$. If $W_\alpha$ is a curl then we obtain the
{\it new
minimal} multiplet~\cite{Aku1} with again $12+12$ components.
This multiplet
is distinguished in that it contains a conserved $R$-current as explained
in subsection~\ref{secRcurrent}.
If $W_\alpha$ is not a curl then this multiplet has $16+16$
components. If we furthermore include $J$,
we obtain a multiplet with $20+20$ components.
Finally, for a general $X$, we obtain  a multiplet with $28+28$ components.

\subsection{Applications}
\label{eq800}

In this subsection, we illustrate the general
formalism by applying it to some concrete models with
${\cal N}=1$ supersymmetry.
As we shall see, the super-Noether procedure
provides an efficient tool to obtain the supercurrent and
the central charge for general local (effective) actions
given in terms of arbitrary K\"ahler- and prepotentials.

\subsubsection{K\"ahler Sigma Models}
\label{secKahler}

Here we compute the supercurrent for the ${\cal{N}}=1$ sigma model
defined in terms of arbitrary real K\"ahler potential $K(\phi,{\bar\phi})$
and
superpotential ${\mathcal W}(\phi)$, where
$\phi$ is a chiral scalar field. Such Lagrangians arise as the local part of
quantum effective actions for supersymmetric field theories
and string theory. The general action is given by
\begin{equation}
S=\frac{1}{16}\int d^8z \ K(\phi,{\bar\phi})
-\frac{1}{4}\int d^6z_+ \ {\mathcal W}(\phi)
-\frac{1}{4}\int d^6z_- \ {\bar\mathcal W}({\bar\phi})\ ,
\label{actionkahler}\end{equation}
with the corresponding equation of motion
\begin{equation}
{\mathcal E}\equiv\frac{1}{16}{\bar D}^2 K_\phi
-\frac{1}{4}{\mathcal W}_\phi =0\ ,
\end{equation}
where a subscript $\phi$ stands for differentiation by $\phi$.
According to~(\ref{rep}),
the transformation law for $\phi$ under global super-Poincar\'e/superconformal
transformations is
$\delta\phi=-{\mathcal L}_+\phi-2q\sigma\phi$,
where $q$ is related to the $R$-weight and dimension of
$\phi$ through~(\ref{relRdqbq}).
Expressing this transformation in terms of $L^\alpha$, we have
\begin{equation}
\delta\phi=\frac{i}{4}{\bar D}^2\left( L^\alpha D_\alpha\phi+\frac{1}{3} q
D^\alpha L_\alpha\ \phi\right)\ .
\label{transff}\end{equation}
Substitution into~(\ref{actionkahler}) then leads to
the variation of the form~(\ref{consequRa}) with
\begin{equation}
R_\alpha= 2D_\alpha\phi\ {\mathcal E}-\frac{2}{3}q D_\alpha(\phi{\mathcal E})\
,
\label{ralplhaKahler}\end{equation}
from which we get
\begin{equation}\begin{array}{lll}
{\widetilde T}_{\alpha{\dot{\alpha}}}&=&
\frac{1}{12}D_\alpha\phi{\bar D}_{\dot{\alpha}}{\bar\phi} K_{\phi{\bar\phi}}
-\frac{i}{6}\partial_{\alpha{\dot{\alpha}}}\phi K_\phi
+\frac{i}{6}\partial_{\alpha{\dot{\alpha}}}{\bar\phi} K_{\bar\phi}
\ ,\\[3mm]
{\widetilde J}&=&
-\frac{1}{4}{\bar D}^2(K-q\phi K_\phi)+3{\mathcal W}-q\phi {\mathcal W}_\phi\ .
\end{array}\label{TKJKahler}\end{equation}
Referring to the discussion in subsection~\ref{secRcurrent},
we infer that the existence of
a conserved $R$-current requires that ${\widetilde J},
{\kern .35em\bar{\kern -.35em{\widetilde J}}}$ can be written in the form~(\ref
{sc1}). This, in turn implies
\begin{equation}
3{\mathcal W}-q\phi {\mathcal W}_\phi=0{\quad\hbox{{and}}\quad}
K(\phi,{\bar\phi})=H(\phi{\bar\phi})\  ,
\label{52}
\end{equation}
for some function $H(\phi{\bar\phi})$.
Superconformal invariance requires
${\widetilde J}=0$ which implies the same condition as in~(\ref{52})
on ${\cal{W}}$, but furthermore that $K-q\phi
K_\phi=0$. The solution of these two conditions are given
by $K(\phi,
{\bar\phi}) \propto (\phi{\bar\phi})^{-1/q}$ and
${\mathcal W}(\phi) \propto \phi^{-3/q}$. Thus, there is
a superconformal model for
each value of $q$.
In particular, for the Wess-Zumino model, where $K(\phi,{\bar\phi})=\phi
{\bar\phi}$,
the superconformal invariance is achieved for $q=-1$ and ${\mathcal W}(\phi)=g \phi^3$, for
some coupling constant $g$.

\subsubsection{Supersymmetric $QED$}

We now discuss the supercurrent of ${\cal{N}}=1$ $QED$ in our formalism.
This will serve as a useful preparation
for dealing with the constrained
superfield of ${\cal{N}}=2$ Yang-Mills discussed in the next section.

The ${\cal{N}}=1$ gauge multiplet is described by a curl superfield
$W_\alpha$ satisfying the constraints
\begin{equation}
{\bar D}_{\dot{\alpha}} W_\alpha =0\ ,
\qquad\qquad
D_\alpha {\bar W}_{\dot{\alpha}} =0\ ,
\qquad\qquad
D^\alpha W_\alpha={\bar D}_{\dot{\alpha}} {\bar W}^{\dot{\alpha}}\ .
\label{consW}\end{equation}
These constraints are solved in terms of a real prepotential $V$ by
\begin{equation}
W_\alpha=-\frac{1}{4}{\bar D}^2D_\alpha V\ .
\label{defV}\end{equation}
$W_\alpha$ is invariant under the gauge transformations:
\begin{equation}
\delta_{\mathcal G} V =i(\Lambda-{\bar\Lambda})\ ,
\qquad\qquad\mbox{with}\qquad\qquad
D_\alpha{\bar\Lambda}=0\ ,
\qquad\qquad
{\bar D}_{\dot{\alpha}}\Lambda=0\ ,
\label{gaugetransf}\end{equation}
and the free action is given by
\begin{equation}
S_W=\frac{1}{4}\int d^6z_+\  W^\alpha W_\alpha +\mbox{c.c.}\ .
\label{actionlibreW}\end{equation}
To couple this ${\mathcal N}=1$ gauge multiplet to a chiral multiplet $\phi$,
we
first recall its gauge transformations
\begin{equation}
\delta_{\mathcal G} \phi = -ig\Lambda\ \phi\ ,
\qquad\qquad
\delta_{\mathcal G} {\bar\phi} = ig{\bar\phi}\ {\bar\Lambda}\ .
\label{defgaugef}\end{equation}
The corresponding invariant action is then given by
\begin{equation}
S_\phi=\frac{1}{16}\int d^8z\ \left( {\bar\phi}\ {\mathrm e}^{gV}\ \phi\right)
\ .
\label{actionfW}\end{equation}
For global superconformal transformations,
we have~(\ref{rep})
\begin{equation}
\delta V=-{\mathcal L} V -2(q\sigma+{\bar q}{\bar\sigma})V\ ,
\label{covV}\end{equation}
with $q={\bar q}$ since $V$ is real. The gauge potential
$A_\mu=\frac{1}{2}\sigma_\mu^{\alpha{\dot{\alpha}}}[D_\alpha,
{\bar D}_{\dot{\alpha}}]V\displaystyle{\biggl\vert}$  is of dimension 1 and
hence  $V$ has to be of dimension $0$, which implies that we take $q=0$.
As explained in~\cite{Osborn} the gauge covariant localization of~(\ref{covV})
reads
\begin{equation}
\delta V \begin{array}[t]{l}
= -{\mathcal L} V -\frac{i}{8}(h-{\bar h})^{\alpha{\dot{\alpha}}}[D_\alpha,
{\bar D}_{\dot{\alpha}}] V \ ,
\end{array}
\label{transfV}\end{equation}
up to gauge transformations.
In particular, it is convenient to add to~(\ref{transfV})
a pure gauge term of the form
\begin{equation}
 -\frac{i}{4}{\bar D}^2(L^\alpha D_\alpha V)
+\frac{i}{4}D^2({\bar L}^{\dot{\alpha}} {\bar D}_{\dot{\alpha}} V) \ .
\end{equation}
Combining the two contributions we end up with
\begin{equation}
\delta V \begin{array}[t]{l} =i (L^\alpha W_\alpha-{\bar L}^{\dot{\alpha}}
{\bar W}_{\dot{\alpha}})\ .
\end{array}
\label{transfV2}\end{equation}
As we have added a gauge transformation with parameter
$\Lambda=-\frac{1}{4}{\bar D}^2(L^\alpha D_\alpha V)$
to the transformation law for $V$,
we have to add the corresponding modification to the transformation
law for $\phi$.
Thus~(\ref{transff}) is modified into
\begin{eqnarray}
\delta\phi&=&\frac{i}{4}{\bar D}^2\left( L^\alpha (D_\alpha\phi+gD_\alpha V\
\phi)+\frac{1}{3} q D^\alpha L_\alpha
\phi\right){\nonumber}\\
&=&\frac{i}{4}{\bar D}^2\left( L^\alpha {\nabla}_\alpha\phi+\frac{1}{3} q
D^\alpha L_\alpha \phi\right)
\ ,
\label{transfginv}
\end{eqnarray}
where ${\nabla}_\alpha\phi=D_\alpha\phi+gD_\alpha V\phi$ is
the gauge covariant
derivative for $\phi$ \cite{Grisa,Shizuya}.
Alternatively, (\ref{transfginv}) can be written as
\begin{equation}
\delta\phi=\left(\frac{1}{2} h^{\alpha{\dot{\alpha}}}
{\nabla}_{\alpha{\dot{\alpha}}}
-\lambda^\alpha{\nabla}_\alpha -2q\sigma\right)\phi\ ,
\end{equation}
with the gauge covariant space-time derivative\footnote{
For a general gauge covariant expression $X$ transforming as
$\delta_{\mathcal G} X=ig (e\Lambda-{\bar e}{\bar\Lambda})X$,
the gauge covariant derivatives are defined as
${\nabla}_\alpha X=D_\alpha X-egD_\alpha V\ X$ and
${\bar\nabla}_{\dot{\alpha}} X={\bar D}_{\dot{\alpha}} X-{\bar e}g
{\bar D}_{\dot{\alpha}} V\ X$.
Note that we still have ${\bar\nabla}_{\dot{\alpha}}\phi=0$,
{\em i.e.} $\phi$ is gauge covariantly chiral.
The covariant derivative ${\nabla}_{\alpha{\dot{\alpha}}}$ acts on $X$ as
${\nabla}_{\alpha{\dot{\alpha}}}X=\partial_{\alpha{\dot{\alpha}}}X-\frac{i}
{2}g(e{\bar D}_{\dot{\alpha}} D_\alpha V
+{\bar e}D_\alpha {\bar D}_{\dot{\alpha}} V) X$.}
${\nabla}_{\alpha{\dot{\alpha}}}\phi=\partial_{\alpha{\dot{\alpha}}}\phi-
\frac{i}{2}g{\bar D}_{\dot{\alpha}} D_\alpha V \phi$.

In order to extract the supercurrent for the combined theory
$S=S_W+S_\phi$ we then apply the, by now,
standard procedure and obtain
a variation of the form~(\ref{consequRa}) with

\begin{equation}\begin{array}{l}
R_\alpha=-\frac{1}{2}W_\alpha\ {\mathcal E}_W
+2  {\nabla}_\alpha\phi\ {\mathcal E}_{\bar\phi}
-\frac{2}{3}q D_\alpha(\phi\ {\mathcal E}_{\bar\phi})\ ,
\\[3mm]
\mbox{where}
\qquad\qquad
\left\{\begin{array}{lll}
{\mathcal E}_W&=&
\frac{1}{8}(D^\alpha W_\alpha+{\bar D}_{\dot{\alpha}} {\bar W}^{\dot{\alpha}})+
\frac{1}{16}g\left({\bar\phi}\ {\mathrm e}^{gV}\ \phi\right)\ ,
\\[3mm]
{\mathcal E}_{\bar\phi}&=&
\frac{1}{16}{\bar D}^2\left({\mathrm e}^{gV}\ {\bar\phi}\right)\ ,
\end{array}\right.
\end{array}
\label{RpourYM}\end{equation}
are the equations of motion for the gauge field and the chiral matter field,
respectively. For the first term in $R_\alpha$,
which depends only on the gauge superfield, we can use the
constraints~(\ref{consW}) to write it as
\begin{equation}
R_\alpha^{W}= -\frac{1}{8}W_\alpha {\bar D}_{\dot{\alpha}}
{\bar W}^{\dot{\alpha}}
= -\frac{1}{8} {\bar D}^{\dot{\alpha}} \left( W_\alpha {\bar W}_{\dot{\alpha}}
\right)\ ,
\label{pureYM}\end{equation}
which shows that for the Maxwell supermultiplet we have
${\widetilde T}^W_{\alpha{\dot{\alpha}}}=-\frac{1}{8}W_\alpha
{\bar W}_{\dot{\alpha}}$ and ${\widetilde J}^W=0$.

For the matter part of the supercurrent we proceed by analogy with
the Wess-Zumino case, but taking into account that the theory is
now gauge invariant.
For this, we first define
$\Phi={\mathrm e}^{\frac{g}{2}V}\phi\ $
so that the invariant matter action takes the same form as for the
Wess-Zumino model (~(\ref{actionkahler}) with $K=\Phi{\bar\Phi}$
and ${\mathcal W}=0$):
$S_\phi=\frac{1}{16}\int d^8z\ \bar\Phi\Phi$.
This suggests that the supercurrent
is just the one for the Wess-Zumino model, (\ref{TKJKahler}),
but with derivatives replaced by covariant ones \cite{Shizuya}:
\begin{equation}\begin{array}{lll}
{\widetilde T}^\phi_{\alpha{\dot{\alpha}}} &=&
\frac{1}{12} {\nabla}_\alpha\Phi {\bar\nabla}_{\dot{\alpha}}\bar\Phi
-\frac{i}{6}\bar\Phi
\stackrel{\leftrightarrow}{{\nabla}}_{\alpha{\dot{\alpha}}} \Phi
\ ,\\[3mm]
{\widetilde J}^\phi &=&
\frac{1}{4}(q-1){\bar\nabla}^2\left(\bar\Phi\Phi\right)
\ .
\end{array}\label{TKJYM}\end{equation}
Plugging this into~(\ref{consequt})
then shows that this is the correct answer,
{\em i.e.} we recover~(\ref{RpourYM}). Thus, the gauge invariant
supercurrent for the ${\mathcal N}=1$ abelian YM theory coupled to a scalar
multiplet
is
\begin{equation}\begin{array}{lll}
{\widetilde T}_{\alpha{\dot{\alpha}}} &=&
-\frac{1}{8} W_\alpha {\bar W}_{\dot{\alpha}}+\frac{1}{12} {\nabla}_\alpha\Phi
{\bar\nabla}_{\dot{\alpha}}\bar\Phi-\frac{i}{6}\bar\Phi
\stackrel{\leftrightarrow}{{\nabla}}_{\alpha{\dot{\alpha}}} \Phi
\ ,\\[3mm]
{\widetilde J} &=&
\frac{1}{4}(q-1){\bar\nabla}^2\left(\bar\Phi\Phi\right)
\ .
\end{array}\label{TKJYMf}\end{equation}

\subsubsection{Tensor Multiplet}

The last application we consider is the tensor multiplet~\cite{Siegel1}
described by a superfield $G$ constrained by:
\begin{equation}
D^2 G=0\ ,\qquad\qquad
{\bar D}^2 G =0\ ,\qquad\qquad
G=\bar G\ .
\label{defG}\end{equation}
Again, this will serve as a preparation for the ${\cal{N}}=2$ tensor discussed
in section~\ref{secn2}.
We consider a general action $S=\int d^8 z\ {\mathcal F}(G)$ depending
on an arbitrary function ${\mathcal F}$. There are two distinguished cases.
The first one is the free field action ${\mathcal F}(G)=G^2$ which is
$R$-invariant for a suitable choice of the $R$-weight of $G$, but is not
conformal invariant for any choice of the conformal weight.
The second model has ${\mathcal F}(G)=G\log G$ \cite{WitRocek1}. This model is
superconformal invariant for a suitable choice of the conformal weight of $G$.

The constraints~(\ref{defG}) are solved by
\begin{equation}
G=D^\alpha\phi_\alpha+{\bar D}_{\dot{\alpha}}{\bar\phi}^{\dot{\alpha}}\ ,
\label{defphi}\end{equation}
where $\phi_\alpha$ is an unconstrained chiral spinor superfield.
Moreover, as expected, there is a gauge freedom of the form
$\delta_{\mathcal G} \phi_\alpha={\bar D}^2 D_\alpha K$, with $K=\bar K$.
In analogy  with the vector multiplet in the last subsection,
we find for the gauge covariant localization of the global symmetry
transformation of $\phi_\alpha$
\begin{equation}
\delta\phi_\alpha \begin{array}[t]{l}
=-{\mathcal L}\phi_\alpha-3\sigma\phi_\alpha + \Omega_\alpha{}^\beta
\phi_\beta
-\frac{i}{8}{\bar D}^2\left(\left( h_{\alpha{\dot{\alpha}}}-
{\bar h}_{\alpha{\dot{\alpha}}}\right){\bar\phi}^{\dot{\alpha}}\right)
+\frac{i}{4}{\bar D}^2 D_\alpha\left( L^\beta\phi_\beta+{\bar L}_{\dot{\beta}}
{\bar\phi}^{\dot{\beta}}\right)
\\[3mm]
=-\frac{i}{4}{\bar D}^2\left( L_\alpha G\right)
\ ,
\end{array}
\label{transfpfi}\end{equation}
corresponding to $q=\frac{3}{2}$, which is the only choice which leads
to a gauge invariant transformation for $G$.
The equation of motion
for $G$ reads ${\mathcal E}_\alpha=-{\bar D}^2 D_\alpha{\mathcal F}'$,
where the prime denotes, as usual,
the derivative of ${\mathcal F}$ w.r.t. $G$.
The computation of the supercurrent conservation equation is now
straightforward and we end up with
\begin{equation}
R_\alpha=-2G{\mathcal E}_\alpha=2G{\bar D}^2 D_\alpha{\mathcal F}'=
-2{\bar D}^{\dot{\alpha}}(G{\bar D}_{\dot{\alpha}} D_\alpha{\mathcal F}'-
{\bar D}_{\dot{\alpha}} G D_\alpha{\mathcal F}')
\ .
\label{conservG}\end{equation}
It is then just a matter of separating the real part from the imaginary one
to get
\begin{equation}\begin{array}{lll}
T_{\alpha{\dot{\alpha}}}&=&-2 D_\alpha G {\bar D}_{\dot{\alpha}} G
{\mathcal F}''+ G [D_\alpha,{\bar D}_{\dot{\alpha}}] {\mathcal F}'\ ,
\\[3mm]
K_{\alpha{\dot{\alpha}}}&=&2G\partial_{\alpha{\dot{\alpha}}}{\mathcal F}'=-2
\partial_{\alpha{\dot{\alpha}}}({\mathcal F}-G{\mathcal F}')\ ,
\\[3mm]
J&=&0\ .
\end{array}\label{supercurrentG}\end{equation} From these objects,
we finally obtain the minimal multiplet:
\begin{equation}\begin{array}{lll}
{\widetilde T}_{\alpha{\dot{\alpha}}}&=&-\frac{4}{3} D_\alpha G
{\bar D}_{\dot{\alpha}} G ({\mathcal F}''-2G{\mathcal F}''')
+\frac{4}{3} G {\mathcal F}''\  [D_\alpha,{\bar D}_{\dot{\alpha}}]G\ ,
\\[3mm]
{\widetilde J}&=&4{\bar D}^2\left( G{\mathcal F}'-{\mathcal F}\right)\ .
\end{array}\label{supercurrentG2}\end{equation}
We immediately see that ${\widetilde J}$ fulfills the $R$-invariance
condition~(\ref{condR})
with $U=4( G{\mathcal F}'-{\mathcal F})$ and, correspondingly,
$Z_\mu=\frac{4}{3}\sigma_\mu^{\alpha{\dot{\alpha}}}[D_\alpha,
{\bar D}_{\dot{\alpha}}]({\mathcal F}-G{\mathcal F}')$. Thus,
the model is $R$-invariant for any function~\footnote{ This is
expected as $G$, being real, is not charged under $R$-symmetry.}
${\mathcal F}(G)$.

For superconformal invariance, we have to impose
${\widetilde J}=0$, which leads to the differential equation
${\bar D}^2({\mathcal F}-G{\mathcal F}')=0$ whose solutions
are ${\mathcal F}(G)=G$ which is not physically interesting and $
{\mathcal F}(G)=G\log G$.
Thus, the action $S=\int d^8 z\ \left( G\log G\right)$ is indeed
superconformal invariant.

\subsection{Central Charge}
\label{nn}
In ${\cal{N}}=1$, the central charges are often ignored as
they are not compatible with translation invariance. However, they play an
important role, for example, for domain
walls~\cite{shifman-chibisov,GT1}. In this subsection we obtain the
central charge as a certain projection of the supercurrent.
This approach is much simpler than computing the central
charges from the Poisson
brackets of the component Noether currents.
Indeed, the only place where components are needed is to identify
the supersymmetry current among the components of
the supercurrent. Its variation, which contains the center, is then easily obtained in superspace.

\noindent  To extract the supersymmetry current we take
$L^\alpha=-i\bar\theta^2\varepsilon^\alpha(x_+)$ and $
{\bar L}^{\dot{\alpha}}=-2i\bar\theta^{\dot{\alpha}}\theta^\alpha
\varepsilon_\alpha(x_-)$.
Then
\begin{equation}\delta S=\displaystyle
\int d^4 x \
\varepsilon^\alpha\displaystyle{\Biggl(}
-2i (\sigma^\nu{\bar\sigma}^\mu)_\alpha{}^\beta D_\beta \partial_\mu
{\widetilde T}_\nu
+\frac{i}{3}\sigma^\mu{}_\alpha{}^{\dot{\beta}}{\bar D}_{\dot{\beta}}
\partial_\mu{\kern .35em\bar{\kern -.35em{\widetilde J}}}
\displaystyle{\Biggr)}
\displaystyle{\biggl\vert}\ .
\label{susyvariation}\end{equation}
Thus, the supersymmetry current is given by
\begin{equation} \label{s1}
j_{\mu\alpha}\begin{array}[t]{l}=\displaystyle{\Biggl(}
2i (\sigma^\nu{\bar\sigma}_\mu)_\alpha{}^\beta D_\beta {\widetilde T}_\nu
+\frac{i}{3}\sigma_{\mu\alpha{\dot{\alpha}}}{\bar D}^{\dot{\alpha}}
{\kern .35em\bar{\kern -.35em{\widetilde J}}}
+\sigma_{\mu\nu\alpha}{}^\beta\partial^\nu D_\beta(a_1 (X+{\bar X}) + a_2 (X-
{\bar X}))
\displaystyle{\Biggr)}\displaystyle{\biggl\vert}
\\[3mm]
=\sigma_\mu{}^\nu{}_\alpha{}^\beta D_\beta
\left( -8i{\widetilde T}_\nu+\partial_\nu(a_1 (X+{\bar X}) + a_2 (X-{\bar X}))
\right)\displaystyle{\biggl\vert}
\ ,\end{array}
\label{susycurrent}\end{equation}
where we have used the conservation equation in the second equality.
The two complex constants $a_1$ and $a_2$ introduced here correspond to
improvement terms for the supersymmetry current as they are
automatically conserved. It is not hard to see that the improvement terms
in~(\ref{s1}) correspond to different choices for $T_{\alpha{\dot{\alpha}}}$,
$K_{\alpha{\dot{\alpha}}}$
and $J$ in~(\ref{eqv1}). Indeed, for a generic choice,
the supersymmetry current is given by
\begin{eqnarray}\label{N1SUSY}
j_{\mu\alpha}=2i(\sigma^\nu{\bar\sigma}_\mu)_\alpha{}^\beta
D_\beta T_\nu-2i(\sigma_\mu)_{\alpha{\dot{\alpha}}}\bar D^{{\dot{\alpha}}}\bar
\tau
+\frac{i}{2}(\sigma_\mu)_{\alpha}^{\;\;{\dot{\alpha}}}D^2\bar
D_{{\dot{\alpha}}}(2\bar X- X)\ .
\end{eqnarray}
A short computation then shows that applying~(\ref{eqv1})
with $A+\bar A=a_1(X+\bar X)$ and
$A-\bar A=a_2(X-\bar X)$ on~(\ref{N1SUSY}),
reproduces the supersymmetry current~(\ref{s1}).
For $a_1=a_2=0$, the supersymmetry current~(\ref{s1}) is part of the
minimal multiplet as defined in section~\ref{Noether}. A precise
discussion of the multiplet structure will be given in section
$6$.  For our purpose, the only relevant part of the
algebra  is the following variation of
the supersymmetry current
\begin{equation}
\delta_\alpha j_{\mu\beta} =
\sigma_{\mu\nu\alpha\beta}\partial^\nu\left(\frac{8}{3}
{\kern .35em\bar{\kern -.35em{\widetilde J}}}
+\frac{1}{2}D^2(a_1 (X+{\bar X}) + a_2 (X-{\bar X}))\right)\displaystyle
{\biggl\vert}\ .\label{multiplet2a}
\end{equation}

\subsubsection{Center for K\"ahler Sigma Models}

According  to~(\ref{TKJKahler}) and to~(\ref{susycurrent}), we have
\begin{equation}
j_{\mu\alpha}=\displaystyle{\Biggl(}
\sigma_\mu{}^\nu{}_\alpha{}^\beta D_\beta(\begin{array}[t]{l}
\frac{1}{6}\sigma_\nu^{\gamma{\dot{\gamma}}}D_\gamma\phi{\bar D}_{\dot{\gamma}}
{\bar\phi} K_{\phi{\bar\phi}}
-\frac{2i}{3}\partial_\nu\phi K_\phi
+\frac{2i}{3}\partial_\nu{\bar\phi} K_{\bar\phi}
\\[3mm]
+\partial_\nu(a_1 (X+{\bar X}) + a_2 (X-{\bar X})))\displaystyle{\Biggr)}
\displaystyle{\biggl\vert}\ .
\end{array}
\label{susycurrentK}\end{equation}
To compute the central charge for this model,
we first need to extract the {\it canonical} supersymmetry current. The
canonical current can be characterized
by the absence of space-time derivatives on fermions.
That is, the second term in~(\ref{susycurrentK})
must be canceled by the improvement terms, {\em i.e.}
$a_1 (X+{\bar X}) + a_2 (X-{\bar X})=\frac{2i}{3}K$. Thus,
\begin{equation}
j_{\mu\alpha}^{can}=
\sigma_\mu{}^\nu{}_\alpha{}^\beta D_\beta
\left(\frac{1}{6}\sigma_\nu^{\gamma{\dot{\gamma}}}D_\gamma\phi
{\bar D}_{\dot{\gamma}}{\bar\phi} K_{\phi{\bar\phi}}
+\frac{4i}{3}\partial_\nu{\bar\phi} K_{\bar\phi}\right)\displaystyle
{\biggl\vert}\ ,
\label{susycurrentKcanonique}\end{equation}
where ``can'' labels the canonical Noether current.
Using~(\ref{multiplet2a}),
we then read off the result for the central charge
${\mathcal Z}_{(\alpha\beta)}=\int d^3x\ \delta_{(\alpha}
j^{can\,0}_{\ \beta)}$ as
\begin{equation}
{\mathcal Z}_{(\alpha\beta)}=\int d^3x\ \sigma^{0i}_{\alpha\beta}\partial_i
\left(-\frac{4i}{3}{\kern .35em\bar{\kern -.35em{\widetilde J}}}+
\frac{i}{3}D^2K\right)\displaystyle{\biggl\vert}
=-3\int d^3x\ \sigma^{0i}_{\alpha\beta}\partial_i\ {\bar\mathcal W}
\displaystyle{\biggl\vert}\ .
\label{centralK}\end{equation}
This is in agreement with the result found in~\cite{shifman-chibisov}
but without the ``ambiguous term'' which was due to the fact that in~\cite
{shifman-chibisov} the minimal
rather than the canonical multiplet
was used in the computation of the central charge. For the canonical
multiplet,
which is the correct multiplet to use to compare with the canonical
formalism, these extra terms are absent as it should be.

\subsubsection{Center for the Tensor Multiplet}

Following the same road as for the K\"ahler Sigma Models,
we find
$a_1 (X+{\bar X}) + a_2 (X-{\bar X})=-\frac{64}{3}(G{\mathcal F}'-
{\mathcal F})$
and the canonical supersymmetry
current reads~\footnote{After adding an equation of motion.}
\begin{equation}
j^{can}_{\mu\alpha}=-8i \sigma_\mu^{\beta{\dot{\beta}}}\ D_\alpha\left(
{\mathcal F}'' D_\beta G {\bar D}_{\dot{\beta}} G \right)
\ .
\label{cancurtens}\end{equation}
Finally, using~(\ref{multiplet2a})
shows that the central charge for this model identically vanishes.

\medskip\noindent To summarize, we have shown how the Superfield Noether
Procedure works for a variety of ${\cal N}=1$
theories including constrained superfields. The simplicity is due to the
fact that, except for
the identification of the supersymmetry current, all
manipulations have been done at the superspace
level. Of course, to determine the remaining Noether currents and
their algebra one needs to go to component fields. We postpone this
discussion to the section $6$  where the currents and their corresponding
multiplet structure are discussed in detail.

\section{${\cal N}=2$ Theories}
\label{secn2}

In this section, we will explain how the general procedure
developed in section~\ref{Noether} has to be refined  in order
to deal with the various constrained superfields in
${\cal{N}}=2$ supersymmetry. Concretely the challenge is to
find the appropriate localizations
of the global symmetry transformations
compatible with the constraints.  That this is
possible at all relies on the fact that  in
${\cal{N}}=2$, contrary to ${\cal{N}}=1$, (\ref{cons1b}) can be replaced by a stronger
constraint while still localizing the
super-Poincar\'e/superconformal transformations.
This is the subject of the next subsection. In the second part, we  then
apply the formalism to
the  ${\cal{N}}=2$ vector and tensor multiplets respectively.

\subsection{${\cal N}=2$ Superconformal Group}
\label{eq305}

In analogy with the ${\cal N}=1$ case,
the constraints~(\ref{cons1b}) can be solved in terms of an unconstrained
${\cal{N}}=2$ superfield $L^\alpha_{\mathbf i}$ as
\begin{equation}
h^{\alpha\dot\alpha} = - \frac{2}{3} {\bar
  D}^{3\dot\alpha {\mathbf i}} L^\alpha_{\mathbf i} {\quad\hbox{{and}}\quad}
{\bar h}^{\alpha\dot\alpha} =  -\frac{2}{3}
  D^{3\alpha {\mathbf i}} {\bar L}^{\dot\alpha}_{\mathbf i}\ .
\end{equation}
In the global limit,
$h^{\alpha{\dot{\alpha}}}=\bar h^{\alpha{\dot{\alpha}}}$, the identity
${\bar D}_{\dot\alpha {\mathbf i}} {\bar D}_{\dot\beta {\mathbf j}}
h^{\gamma\dot\gamma} = D_{\alpha {\mathbf i}} D_{\beta {\mathbf j}}
h^{\gamma \dot\gamma} =0$
holds~\cite{Park}, and a straightforward computation
shows that a suitable $L^\alpha_{\mathbf i}$ is  given by
$L^\alpha_{\mathbf i} = -\frac{1}{12} {\bar\theta}^3_{\dot\alpha{\mathbf i}}
h^{\alpha\dot\alpha}$.
However, contrary to the ${\cal N}=1$ case, it is possible to replace
$L^\alpha_{\mathbf i}$ by a constrained superfield without losing the global
transformations. Indeed we can write
\begin{equation}
L^\alpha_{\mathbf i} \equiv D^{\alpha {\mathbf j}} L_{{\mathbf j}{\mathbf i}}
\quad\mbox{and}\quad
{\bar L}^{\dot\alpha}_{\mathbf i} \equiv - {\bar D}^{\dot\alpha {\mathbf j}}
{\bar L}_{{\mathbf j}{\mathbf i}}
\end{equation}
where $L_{{\mathbf i}{\mathbf j}}$ is  an unconstrained symmetric
superfield\footnote
{We checked that for super-Poincar\'e
transformations it is even possible to integrate further by
$L_{{\mathbf i}{\mathbf j}} = D^\alpha_{({\mathbf i}}
A_{\alpha {\mathbf j})}$ However, as this does not
seem to play a role for the Superfield Noether Procedure,
we do not explore this possibility further in the present
work.}. The expansion of $L_{{\mathbf i}{\mathbf j}}$ for global
super-Poincar\'e and superconformal transformations is given in
appendix~\ref{1001}.
This allows us to define a new scalar superfield $H$ which, like
$h^{\alpha\dot\alpha}$, is real
for superconformal transformations. Indeed we can take
\begin{equation}
H \equiv
{\bar D}^{{\mathbf i}{\mathbf j}} L_{{\mathbf i}{\mathbf j}}, \quad
{\bar H} \equiv D^{{\mathbf i}{\mathbf j}} {\bar L}_{{\mathbf i}{\mathbf j}} \
.\label{eq600}
\end{equation}
Furthermore, $h^{\alpha\dot\alpha}$ can be expressed in terms of $H$ as
\begin{eqnarray}
h^{\alpha\dot\alpha} =
\frac{1}{2} [D^{\alpha {\mathbf i}},{\bar D}^{\dot
\alpha}_{{\mathbf i}}] H \quad\mbox{and}\quad
{\bar h}^{\alpha\dot\alpha} =\frac{1}{2}
[D^{\alpha {\mathbf i}},{\bar D}^{\dot
\alpha}_{{\mathbf i}}] {\bar H}\ .  \label{eq314}
\end{eqnarray}
In terms of $L_{{\mathbf i}{\mathbf j}}$, we have
\begin{equation}
\lambda^{\alpha}_{{\mathbf i}} = \frac{i} {24}
{\bar D}^4 D^{\alpha {\mathbf j}}L_{{\mathbf i}{\mathbf j}}
\qquad\mbox{and}\qquad
\sigma = \frac{i}{144} {\bar D}^4 D^{{\mathbf i}{\mathbf j}}
L_{{\mathbf i}{\mathbf j}}\ .
\label{eq118}
\end{equation}
According to the prescription given in section~\ref{Noether},
we localize the parameters of
the global symmetry transformations by relaxing the reality condition
$h^{\alpha{\dot{\alpha}}}= \bar
h^{\alpha{\dot{\alpha}}}$, while maintaining 
the chirality preserving condition~(\ref{cons1b}). However, 
in view of the discussion above we consider only those
$h^{\alpha{\dot{\alpha}}}$ that can be written in terms of
$L_{{\mathbf i}{\mathbf j}}$ where
$L_{{\mathbf i}{\mathbf j}}$ is an arbitrary symmetric
superfield.

\subsection{${\cal N}=2$ Vector Multiplet}
\label{1.2}

As a first application of the general formalism we consider
the Abelian vector multiplet~\cite{Brink1}.
This multiplet plays an important role
for the low energy effective description of non-Abelian ${\cal N}=2$
Yang-Mills theory~\cite{Seiberg-Witten}.
The vector multiplet is described by
a chiral superfield ${\cal A}$
of $R$-weight -2 and dimension 1,
subject to the Bianchi constraint
$D^{{\mathbf i}{\mathbf j}} {\cal A} = {\bar D}^{{\mathbf i}{\mathbf j}}
{\bar{\cal A}}$.
This constraint can be solved by
${\cal A} = {\bar D}^4 D^{{\mathbf i}{\mathbf j}} V_{{\mathbf i}{\mathbf j}}$
where the
prepotential $V_{{\mathbf i}{\mathbf j}}$ is a real superfield.

The global superconformal transformation of ${\mathcal A}$ is given
by~(\ref{rep}) with $q=\frac{3}{2}$ (and, of course, $\bar q=0$ since
${\mathcal A}$ is chiral).
Although the corresponding local transformation, {\em i.e.}
when $L_{{\mathbf i}{\mathbf j}}$ is free, does preserve the chirality
of ${\mathcal A}$, it does not preserve
the Bianchi constraint. Nevertheless, we can
proceed by complete analogy with the ${\cal N}=1$ case:
this suggests to generalize the transformation~(\ref{transfV2})
of the  ${\cal N}=1$ prepotential $V$ to
\begin{equation}\label{tai}
\delta V_{{\mathbf i}{\mathbf j}} \equiv  -\frac{i}{48}({\cal A}
L_{{\mathbf i}{\mathbf j}}-{\bar
  {\cal A}} \bar L_{{\mathbf i}{\mathbf j}}).
\end{equation}
Using the definition of ${\mathcal A}$ and the constraints it satisfies,
we then compute the transformation of ${\mathcal A}$. It leads to
\begin{equation}
48 i \delta {\cal A} =  {\bar D}^4  (D^{{\mathbf i}{\mathbf j}} {\cal A}
L_{{\mathbf i}{\mathbf j}}) +  2 {\bar D}^4 (D^{\alpha {\mathbf j}}
L_{{\mathbf i}{\mathbf j}} D^{\mathbf i}_\alpha
{\cal A}) + {\cal A} {\bar D}^4
D^{{\mathbf i}{\mathbf j}}L_{{\mathbf i}{\mathbf j}} - {\bar D}^4
({\bar{\cal A}} D^{{\mathbf i}{\mathbf j}} {\bar L}_{{\mathbf i}{\mathbf j}})\
.
\end{equation}
The Bianchi constraint enables us
to write the first term in this variation as \linebreak
${\bar  D}^4({\bar D}^{{\mathbf i}{\mathbf j}} {\bar {\cal A}}
L_{{\mathbf i}{\mathbf j}} ) =
{\bar D}^4 ( {\bar{\cal A}} {\bar D}^{{\mathbf i}{\mathbf j}}
L_{{\mathbf i}{\mathbf j}})$.
Then, using the definition of
$H$ and ${\bar H}$ we end up with
\begin{equation}
\delta {\cal A} = -\frac{i}{24} {\bar D}^4 \left( D^{\alpha {\mathbf j}}
  L_{{\mathbf i}{\mathbf j}} D_\alpha^{\mathbf i} {\cal A}  \right) - \frac{i}
{48} {\bar
  D}^4 D^{{\mathbf i}{\mathbf j}} L_{{\mathbf i}{\mathbf j}} \, \, {\cal
A} - \frac{i}{48} {\bar
  D}^4 \left[ (H - {\bar H}) {\bar{\cal A}} \right]. \label{eq201}
\end{equation}
Therefore, this transformation is a suitable one as it
reduces to the transformation~(\ref{rep})
in the superconformal limit $H = {\bar H}$ and
furthermore preserves the Bianchi constraint.

\subsubsection{Variation of the holomorphic Action}

As for the ${\cal{N}}=1$ chiral multiplet, the most general local action for
the vector multiplet consists of holomorphic and a non-holomorphic terms.
We first consider the holomorphic
part, referring non-holomorphic terms to subsection~\ref{601}.

The general holomorphic action is given by
\begin{equation}
S[{\cal A}] \equiv  \frac{1}{4\pi} \, \mbox{Im} \int\!\! d^8z_+
{\cal F}({\cal A}). \label{eq303}
\end{equation}
The classical Yang-Mills action corresponds to  ${\cal
  F}({\cal A}) \equiv  \tau {\cal A}^2$ with
$\tau \equiv \frac{\theta}{2\pi} + i \frac{4\pi}{g^2}$
 where $g$ is the coupling constant and $\theta$ the
 vacuum-angle.   For convenience, we also introduce the
dual superfield  ${\cal A}_D
\equiv {\cal F}'({\cal A})$. The equations of motion are
then $D^{{\mathbf i}{\mathbf j}} {\cal A}_D =
{\bar D}^{{\mathbf i}{\mathbf j}} {\bar {\cal A}}_D$.

It is now straightforward to compute the variation of the action~(\ref{eq303})
 under
the transformation~(\ref{eq201}). Indeed, as
\begin{equation}
\delta {\cal F} = -\frac{i}{24} {\bar D}^4 \left( D^{\alpha {\mathbf j}}
  L_{{\mathbf i}{\mathbf j}} D_\alpha^{\mathbf i} {\cal F}  \right) - 3 \sigma
{\cal
  A}{\cal A}_D - \frac{i}{48} {\bar D}^4 \left[ (H -
  {\bar H}) {\bar{\cal A}} {\cal A}_D \right],
\end{equation}
the variation of the action is
\begin{equation}
\delta S =  \frac{1}{4\pi} \,\mbox{Im} \,\, \left\{ - \frac{i}{24} \int\!\!
  d^{12}z\ D^{\alpha {\mathbf j}}
  L_{{\mathbf i}{\mathbf j}} D_\alpha^{\mathbf i} {\cal F}  -3 \int\!\! d^8z_+
  \sigma {\cal
  A}{\cal A}_D - \frac{i}{48} \int\!\!
  d^{12}z\ (H -
  {\bar H}) {\bar{\cal A}} {\cal A}_D \right\}.
\end{equation}
 Then, we successively integrate by parts the first term, write it as
an integral on the chiral superspace, use the chirality
of ${\cal F}({\cal A})$ and the relation~(\ref{eq118})
between $\sigma$ and $L_{{\mathbf i}{\mathbf j}}$ to obtain:
\begin{equation}
\delta S = \frac{1}{4\pi} \, \mbox{Im} \,\, \left\{ 6 \int\!\! d^8z_+\ \sigma (
{\cal
    F} - \frac{1}{2} {\cal A} {\cal A}_D)
- \frac{i}{48} \int\!\!  d^{12}z\ (H- {\bar H}) {\bar{\cal
    A}} {\cal A}_D  \right\}.
\end{equation}
This means that the variation of the action can be
written as
\begin{equation}
\delta S = i\int\!\! d^{12}z\ (H -{\bar
  H}) T - 144 i \int\!\! d^8z_+\ \sigma {\cal J} +
144 i \int\!\! d^8z_-\ {\bar\sigma}
{\bar{\cal J}} \label{eq200}
\end{equation}
with
\begin{equation}
T = -\frac{i}{384 \pi} ({\cal A}{\bar{\cal A}}_D -  {\bar{\cal
    A}} {\cal A}_D), \quad
{\cal J} = \frac{1}{192 \pi} ({\cal F} - \frac{1}{2} {\cal A}{\cal
  A}_D)\ .
\label{eq304}
\end{equation}\label{hc1}
This result deserves some comments:

{\it 1)} Contrary to the general situation of section~\ref{Noether},
it is not possible to obtain the variation of the action~(\ref{eq303})
in terms of $h^{\alpha{\dot{\alpha}}}$ due to the
constraints on ${\mathcal A}$.

{\it 2)} Nevertheless, the invariance of the action under super-Poincar\'e
transformations is explicit as $H = {\bar H}$ and
$\sigma = {\bar\sigma}=0$ for these transformations.

Moreover, as in ${\cal N}=1$, the conservation equations
are obtained from~(\ref{eq200}) by expressing $H$,
${\bar H}$, $\sigma$ and ${\bar\sigma}$ in terms of the
free parameters $L_{{\mathbf i}{\mathbf j}}$ and $
{\bar L}_{{\mathbf i}{\mathbf j}}$. This leads to
\begin{equation}
D^{{\mathbf i}{\mathbf j}} T = -i {\bar D}^{{\mathbf i}{\mathbf j}}
{\bar {\cal J}}. \label{eq121}
\end{equation}
\par\noindent

{\it 3)} Contrary to the situation in ${\cal N}=1$, there is no
freedom in the definition of $T$ and ${\cal J}$ {\em i.e.} they are
uniquely determined in terms of the constrained
superfield ${\cal A}$.\par\noindent

{\it 4)} It is now clear from {\it 3)} that the theory is superconformal
invariant if and only if ${\cal J}=0$.
Hence, in analogy with the ${\cal N}=1$ case, ${\cal J}$ is the superconformal
anomaly and therefore our method
provides a simple derivation of the anomalous
superconformal `Ward identity' first derived in~\cite{Howe96}.\par\noindent

\subsubsection{Central Charge}
\label{605}

For classical ${\cal{N}}=2$ Yang-Mills theory, the
central charge was first computed in~\cite{Olive-Witten} by
explicit evaluation of the anticommutator of the supersymmetry charges.
The quantum corrected effective central charge of
${\cal N}=2$ supersymmetry is important
because it determines the mass of the BPS states in the quantum theory.
Indeed, the central charge formula for low-energy
effective action of ${\cal N}=2$ Yang-Mills theory which was
assumed in~\cite{Seiberg-Witten} contains the seeds of the duality
properties of this model~\cite{FMRSS,MRS}. That this assumption
is correct was proved in components in~\cite{Wolf,Iorio}.
In general, the complete computation of central charges for effective
theories in components is a rather magnificent task. On the other hand,
we have seen in  subsection~\ref{nn}
that the Superfield Noether Procedure leads to a
simple computation even for complicated actions. This method
naturally extends to the ${\cal N}=2$ case and as we shall now show it
leads to a simple and efficient computation of the
effective central charge.
In order to have equivalence with the canonical computation~\cite
{Olive-Witten,Wolf,Iorio} it is, however, important to work with the
canonical multiplet.

\noindent As in ${\cal N}=1$, we first determine the
supersymmetry current by taking
$L_{{\mathbf i}{\mathbf j}} = - \frac{i}{9}
\varepsilon^\alpha_{({\mathbf i}}(x^+)
\theta_{\alpha {\mathbf j})} {\bar\theta}^4$ and
${\bar L}_{{\mathbf i}{\mathbf j}} = \frac{2i}{9}
\varepsilon^\alpha_{\mathbf k}(x^-)
\theta^{3{\mathbf k}}_\alpha {\bar\theta}_{{\mathbf
    i}{\mathbf j}}$. This leads to
\begin{equation}
J_\alpha^{\mu {\mathbf i}} = 192 \left[
 i \sigma^\mu_{\alpha\dot\alpha} {\bar
  D}^{\dot\alpha}_{\mathbf j}  D^{{\mathbf i}{\mathbf j}} T
- 3 i
{\bar\sigma}^{\mu\dot\alpha\beta} {\bar D}^{\mathbf i}_{\dot\alpha}
D_{\alpha\beta} T
- 12  \partial^\mu D^{\mathbf i}_\alpha T
+   a \sigma_\alpha^{\mu\nu\,\beta} \partial_\nu D_\beta^{\mathbf i}
T \right]| \label{eq307}\ ,
\end{equation}
where the term parametrized by $a$ corresponds to an
improvement term.
Note that the first term in the r.h.s.. of~(\ref{eq307})
is equal on-shell to
$192 \sigma^\mu_{\alpha\dot\alpha}
{\bar D}^{3\dot\alpha {\mathbf i}} {\bar{\cal J}}$.
As the second term in the r.h.s of (\ref{eq307}) is
traceless,  the trace of $J^{\mu {\mathbf i}}_\alpha$ is simply given by
\begin{equation}
({\bar \sigma}^\mu J_\mu)^{\dot\alpha {\mathbf i}} =
192 \left[ -4 {\bar D}^{3\dot\alpha {\mathbf i}}{\bar{\cal J}} -
\frac{3}{2} (8+a) \partial^{\dot\alpha\alpha} D_\alpha^{\mathbf i}
T \right] | . \label{eq420}
\end{equation}
Therefore, this trace vanishes when ${\mathcal J}=0$ (superconformal case)
and $a=-8$.

Next, we give the relevant part of the variation of
$J^{\mu {\mathbf i}}_\alpha$ referring to
appendix~\ref{eq608b} for details of this computation:
\begin{equation}
\delta_{\alpha {\mathbf i}} J^\mu_{\beta {\mathbf j}} =\partial_\nu \left\{
24 i \varepsilon_{\alpha\beta} \varepsilon_{{\mathbf i}{\mathbf j}}
Z^{\mu\nu}
 + 96  ( a +24) \left[ \varepsilon_{{\mathbf i}{\mathbf j}}
\sigma^{\mu\nu\,\gamma}_{(\alpha}
D_{\beta)\gamma} T |
+ \sigma^{\mu\nu}_{\alpha\beta}
D_{{\mathbf i}{\mathbf j}}T| \right] \right \} \label{eq402}
\end{equation}
where we have defined
\begin{equation}
Z^{\mu\nu} \equiv 96 \left[
{\bar\sigma}^{\mu\nu}_{\dot\alpha\dot\beta}
{\bar D}^{\dot\alpha\dot\beta} {\bar{\cal J}} - i (\frac{1}{2} -
\frac{a}{48}) \sigma^{\mu\nu}_{\alpha\beta}
D^{\alpha\beta} T\right] |. \label{eq506}
\end{equation}

\noindent One way to fix the value of $a$ corresponding to
the  canonical current is to go to components in the
particular case of classical Yang-Mills theory, where
${\cal F}({\cal A})$ is quadratic. This leads to
$a=-24$. However, it is also possible to determine this value at the
superspace level, for any ${\cal F}$,
by comparing the variation~(\ref{eq402})
of the supersymmetry current and the supersymmetry
algebra
\begin{equation}
\{ Q_{\alpha {\mathbf i}}, Q_{\beta {\mathbf j}} \} = \frac{1}{4}
\varepsilon_{{\mathbf i}{\mathbf j}}\ \varepsilon_{\alpha\beta}\ Z +
\sigma^{0k}_{\alpha\beta} \Lambda_{k({\mathbf i}{\mathbf j})}. \label
{eq615}
\end{equation}
Indeed, this Poisson algebra allows only terms having the same symmetry
properties in the $SU(2)_R$ and spinor indices. On the
other hand as $\{ Q_{\alpha {\mathbf i}},
Q_{\beta {\mathbf j}} \} = \int d^3x
\delta_{\alpha {\mathbf i}} J^0_{\beta {\mathbf j}}$,
it is clear that the r.h.s. of equation~(\ref{eq402})
has the required symmetry only when $a=-24$.
Note that this implies the vanishing of
$\Lambda_{{k}({\mathbf i}{\mathbf j})}$. According to the previous discussion,
the central charge $Z$ is then given by 
\begin{equation}
Z = \int d^3x \delta^{\alpha {\mathbf i}} J^0_{\alpha {\mathbf i}}  = 96 \int d^3x
\partial_i Z^{0i}\ .
\end{equation}
 where $Z^{0i}$  is given for $a=-24$ by
\begin{equation}
Z^{0i} = 96 i \left[   {\bar\sigma}^{0i}_{\dot\alpha\dot\beta} {\bar
  D}^{\dot\alpha\dot\beta} {\bar{ \cal J}} - i
\sigma^{0i}_{\alpha\beta} D^{\alpha\beta} T \right ]|\ . \label{eq315}
\end{equation}
Note that this result is universal in the sense
that this formula for the center is valid for any theory
whose variation of the action takes the form~(\ref{eq200}).

\medskip

Using the particular expressions~(\ref{eq304}) of $T$ and ${\bar{\cal J}}$,
  we end up with
\begin{equation}\label{cc}
Z^{0i} = \frac{i}{4\pi}  {\bar {\cal A}} \left\{ \sigma^{0i}_{\alpha\beta}
  D^{\alpha\beta} {\cal A}_D -
  {\bar\sigma}^{0i}_{\dot\alpha\dot\beta} {\bar
    D}^{\dot\alpha\dot\beta} {\bar{\cal A}}_D \right\} | -
\frac{i}{4\pi} {\bar{\cal A}}_D \left\{ \sigma^{0i}_{\alpha\beta}
  D^{\alpha\beta} {\cal A} -
  {\bar\sigma}^{0i}_{\dot\alpha\dot\beta} {\bar
    D}^{\dot\alpha\dot\beta} {\bar{\cal A}} \right\} |. \label{eq202}
\end{equation}
This result, which is the expression of the central
charge in superspace exhibits manifestly the duality
between ${\cal A}$ and ${\cal A}_D$. However, to complete the calculation
we need to identify the r.h.s. of~(\ref{cc}) with the electric and
magnetic charges.

\medskip

To identify the magnetic charge we use
that\footnote{For classical Yang-Mills theory where
${\cal F}({\cal A}) = i \frac{4\pi}{g^2} {\cal A}^2$, this definition leads to
the usual normalization $- \frac{1}{4 g^2}  F^{\mu\nu}F_{\mu\nu}$ .}
$ D_{\alpha\beta} {\cal A} | \equiv \frac{1}{2\sqrt{3}} F_{\alpha\beta}$, where
$F_{\alpha\beta} \equiv F_{\mu\nu}
\sigma^{\mu\nu}_{\alpha\beta}$ and $B^i \equiv \frac{1}{2} \varepsilon^{oijk}
F_{jk}$. It follows then that
\begin{equation}
B^i = - i \sqrt{3} \left[ \sigma^{0i}_{\alpha\beta}
  D^{\alpha\beta}{\cal A} -
  {\bar\sigma}^{0i}_{\dot\alpha\dot\beta} {\bar
    D}^{\dot\alpha\dot\beta} {\bar {\cal A}} \right] |. \label{eq204}
\end{equation}
Next we determine the conjugate momentum $\Pi^i$ of
the gauge field. In general, it is not straightforward to
determine the conjugate momentum of a field in
superspace. However, we proceed by using
that $\Pi^i$ can be extracted from the Gauss law $\partial_i
\Pi^i =0$ which is an equation of motion.
More precisely, the gauge field appears in the action
only via the field strength (in superspace
language, this is reflected by the fact that the superfield
${\cal A}$ is a gauge invariant object).
Therefore, we have the sequence of equalities:
\begin{equation}
\partial_i \Pi^i = \partial_i \frac{\partial L}{\partial
  (\partial_0 A_i)} = - \partial_i \frac{\partial L}{\partial
  (\partial_i A_0)} = \frac{\delta S}{\delta A_0}\ ,
\end{equation}
which is just expressing the fact that Gauss law is
equivalent to $\frac{\delta S}{\delta A_0}$. To obtain Gauss
law in superspace we then proceed in two steps. First we compute
the fundamental derivative
\begin{equation}
\frac{\delta}{\delta A_0(y)} (D_{\alpha\beta} {\cal A} | (x))
= -\frac{1}{\sqrt{3}} \sigma^{0i}_{\alpha\beta} \partial_{x\,i}
\delta(x-y). \label{eq117}
\end{equation}
Then we identify the terms in the action containing $D_{\alpha\beta}A$.
Now, since
\begin{equation}
D^4 {\cal F}({\cal A}) = {\cal F}' D^4 {\cal A} + 3 D^{\alpha {\mathbf i}}
{\cal
  F}' D^3_{\alpha {\mathbf i}} {\cal A} + \frac{3}{2} D^{{\mathbf i}
{\mathbf j}} {\cal F}'
D_{{\mathbf i}{\mathbf j}} {\cal A} - \frac{3}{2} D^{\alpha\beta} {\cal F}'
D_{\alpha\beta} {\cal A} - D^{3\alpha {\mathbf i}}{\cal F}'
D_{\alpha {\mathbf i}} {\cal A},
\end{equation}
the terms we are looking for are
\begin{equation}
-\frac{3}{8\pi}\mbox{Im} \int d^4x\
D^{\alpha\beta} {\cal F}' D_{\alpha\beta} {\cal A} |.
\end{equation}
Differentiating with respect to $A_0$ and using~(\ref{eq117}) we then end
up with
\begin{equation}
\frac{\delta S}{\delta A_0} = -  \frac{\sqrt{3}}{8\pi} \partial_i
\, \mbox{Im}
\left[ \sigma^{0i}_{\alpha\beta} \left(D^{\alpha\beta}
    {\cal F}' + {\cal F}'' D^{\alpha\beta} {\cal A}\right) \right]|.
\end{equation}
Thus we have
\begin{equation}
\Pi^i = - \frac{\sqrt{3}}{8\pi} \mbox{Im}
\left[ \sigma^{0i}_{\alpha\beta} \left(D^{\alpha\beta}
    {\cal F}' + {\cal F}'' D^{\alpha\beta} {\cal A}\right) \right]|.
\end{equation}
We now consider the phase space for which
$\phi \equiv {\cal A}|$ goes to a constant at infinity
and where the fermions $D^{\alpha {\mathbf i}} {\cal A}
|$ decrease sufficiently fast enough such that
\begin{equation}
 {\cal F}'' D^{\alpha\beta} {\cal A} | = D^{\alpha\beta} {\cal
   F}' | -  {\cal F}''' D^{\alpha {\mathbf i}} {\cal A} D^\beta_{\mathbf i}
 {\cal A} |
 \stackrel{|x|\to\infty}{\longrightarrow}  D^{\alpha\beta} {\cal
   F}' |.
\end{equation}
Therefore, we have, at infinity, $\Pi^i =
- \frac{\sqrt{3}}{4\pi} \mbox{Im}\
 \sigma^{0i}_{\alpha\beta} D^{\alpha\beta}
    {\cal A}_D|$, {\em i.e.}
\begin{equation}
\Pi^i  =  i \frac{\sqrt{3}}{8\pi} \left[ \sigma^{0i}_{\alpha\beta}
  D^{\alpha\beta} {\cal A}_D -
  {\bar\sigma}^{0i}_{\dot\alpha\dot\beta} {\bar
    D}^{\dot\alpha\dot\beta} {\bar {\cal A}}_D \right]
|. \label{eq203}
\end{equation}
Finally, combining  the
expressions~(\ref{eq204}) and (\ref{eq203})
respectively for the magnetic and electric fields, the
result for the central charge~(\ref{eq202}) can be rewritten as
\begin{equation}
Z^{0i} \stackrel{|x|\to\infty}{\longrightarrow}  \frac{2}{\sqrt{3}} \left\{
{\bar{\cal A}} \Pi^i +
  \frac{1}{8\pi} {\bar{\cal A}}_D B^i \right\} | \label{eq250}
\end{equation}
which corresponds to the result found in~\cite{Wolf}.

\subsubsection{Non-holomorphic Action}
\label{601}

We now consider a non-holomorphic, local action for the vector multiplet
\begin{equation}
S[{\cal A}]  \equiv \int \!d^{12}z\;\ {\cal
H}({\cal A}, {\bar{\cal A}}).
\end{equation}
To compute the corresponding $T$ and ${\cal J}$, we
start from the following form of~(\ref{eq201}):
\begin{equation}
\delta {\cal A} = \frac{1}{2} h^{\alpha\dot\alpha}
\partial_{\alpha\dot\alpha} {\cal A} + \lambda^{\alpha {\mathbf i}}
D_{\alpha {\mathbf i}} {\cal A} -  3 \sigma {\cal A} - \frac{i}{48} {\bar
D}^4 \left( (H-{\bar H}) {\bar {\cal A}} \right).
\end{equation}
The contributions from the last two terms to $T$ and
${\cal J}$ are immediately obtained. Let us so concentrate
on the first two terms.
We use the relations~(\ref{lambda}) to obtain

\begin{eqnarray}
\delta S&=&\int d^{12}z\ \left\{ h^{\alpha\dot\alpha} \left( \frac{1}{2} {\cal
H}_{\cal A} \partial_{\alpha\dot\alpha}  A + \frac{i}{8}
{\bar D}_{\dot\alpha}^{\mathbf i} D_{\alpha {\mathbf i}} {\cal H}  \right)
+ {\bar h}^{\alpha\dot\alpha} \left( \frac{1}{2}
{\cal H}_{\bar{\cal A}} \partial_{\alpha\dot\alpha}
{\bar{\cal A}} - \frac{i}{8} D^{\alpha {\mathbf i}} {\bar
D}^{\dot\alpha}_{\mathbf i} {\cal H} \right)\right\}
\nonumber\\[3mm]
&=&\int d^{12}z\ \frac{1}{2} (h^{\alpha\dot\alpha} -{\bar
h}^{\alpha\dot\alpha}) ( \frac{1}{2} {\cal
H}_{\cal A} \partial_{\alpha\dot\alpha}  A - \frac{1}{2}
{\cal H}_{\bar{\cal A}} \partial_{\alpha\dot\alpha}
{\bar{\cal A}} - \frac{i}{8} [ D_{\alpha {\mathbf i}}, {\bar
D}_{\dot\alpha}^{\mathbf i}]  {\cal H})\nonumber\\[3mm]
&=&\int d^{12}z\ \frac{1}{2} (h^{\alpha\dot\alpha} -{\bar
h}^{\alpha\dot\alpha}) (\frac{i}{4} {\cal H}_{ {\cal
A}  {\bar {\cal A}}} D_{\alpha {\mathbf i}} {\cal A} {\bar
D}^{\mathbf i}_{\dot\alpha} {\bar {\cal A}}  )\ .
\end{eqnarray}
We conclude by using the relation~(\ref{eq314})
between $h^{\alpha{\dot{\alpha}}}$ and $H$ and by integrating by parts.
Adding up all the contributions finally leads  to
\begin{eqnarray}
T &=& -\frac{1}{48} \left[ {\bar{\cal A}} {\bar D}^4 {\cal
H}_{\cal A} + {\cal A} D^4 {\cal H}_{\bar{\cal A}} + 3
[D^{\alpha {\mathbf i}} , {\bar D}^{\dot\alpha}_{\mathbf i} ] (
{\cal H}_{{\cal
A}{\bar {\cal A}}} D_{\alpha}^{\mathbf j} {\cal A} {\bar
D}_{\dot\alpha {\mathbf j}} {\bar {\cal A}}) \right],\\
{\cal J} & =& - \frac{i}{48} {\cal A} {\bar D}^4 {\cal
H}_{\cal A}.
\end{eqnarray}

\medskip

We finish this section by showing that the non-holomorphic
part of the action does not contribute to the center. We denote
generically by a subscript $h$, holomorphic terms and by
a subscript $n.h.$, non-holomorphic ones.

Among the diverse terms that appear in the
expression~(\ref{eq315}) for the center $Z^{0i}_{n.h.}$,
we concentrate on the ones which fall off slowest at infinity.
Let us take the example of ${\bar
D}^{\dot\alpha\dot\beta} {\bar {\cal J}} \propto {\bar
D}^{\dot\alpha\dot\beta} ({\bar {\cal A}} D^4 {\cal H}_{\bar
{\cal A}})$.  Remember that the dimension of
${\cal A}$ is $-1$. Furthermore, as ${\cal A} |$ goes to a
constant at infinity, all the derivatives of ${\cal H}$
 share this property. Therefore, the term that decreases
the least is the one where as less as possible
derivatives act on ${\cal H}$ namely ${\bar {\cal A}}
{\cal H}_{{\cal A}{\bar{\cal A}}} ({\bar
D}^{\dot\alpha\dot\beta} D^4 {\cal A})$. Clearly, this term
decreases sufficiently fast enough at infinity, and thus,
this holds similarly for all the other terms. As a
consequence, there is no contribution to the center
from ${\bar{\cal J}}$ and, for the same reasons, from
$T$ such that $Z^{0i}_{n.h.} \stackrel{|x|\to\infty}{\longrightarrow} 0$.
However, in the same way, by computing the Gauss
law, it is easy to prove that at infinity, the non-holomorphic terms do not
contribute to $\Pi^i$.
Thus, in a theory where both holomorphic and
non-holomorphic terms are present, we have, on one hand,
$\Pi^i = \Pi^i_{h.} + \Pi^i_{n.h.}
\stackrel{|x|\to\infty}{\longrightarrow} \Pi^i_{h.}$ and on the other
hand $Z^{0i} = Z^{0i}_{h.} + Z^{0i}_{n.h.}
\stackrel{|x|\to\infty}{\longrightarrow} Z^{0i}_{h.}$. Therefore,
\begin{equation}
Z^{0i} \stackrel{|x|\to\infty}{\longrightarrow}
\frac{2}{\sqrt{3}} \left\{
  {\bar{\cal A}} \Pi^i_{h.}
 +  \frac{1}{8\pi} {\bar{\cal A}}_D B^i \right\}\displaystyle{\biggl\vert}
= \frac{2}{\sqrt{3}} \left\{
  {\bar{\cal A}} \Pi^i
 +  \frac{1}{8\pi} {\bar{\cal A}}_D B^i \right\} \displaystyle{\biggl\vert} .
\end{equation}
This proves that the central charge is entirely determined
by the holomorphic part of the action.

\subsection{${\cal N}=2$ Tensor Multiplet}

The second example of a ${\cal{N}}=2$ theory with off-shell superfield
formulation we consider is the tensor multiplet~\cite{GS2, deWit83}.
This multiplet is described by a chiral field $\Phi$ of
dimension 1 and $R$-weight $-2$ and classical action

\begin{equation}
S[\Phi] \equiv \int \!d^4x d^8\theta\;  \Phi {\bar\Phi}
+ 24 \int \!d^4x d^4\theta \;\Phi (m^2-\hspace{0.2ex}\raisebox{.5ex}{\fbox{}}
\hspace{0.3ex}) \Phi
+ 24 \int \!d^4x d^4{\bar\theta} \; {\bar\Phi}
(m^2 - \hspace{0.2ex}\raisebox{.5ex}{\fbox{}}\hspace{0.3ex}) {\bar \Phi}.
\label{eq102}
\end{equation}
For $m=0$, this action is invariant under the gauge
transformation
$\delta \Phi \equiv {\bar D}^4 D^{{\mathbf i}{\mathbf j}}
K_{{\mathbf i}{\mathbf j}}$ where
$K_{{\mathbf i}{\mathbf j}}$ is real. The invariant ``field strength''
tensor is given by
\begin{equation}
F_{{\mathbf i}{\mathbf j}} \equiv i (D_{{\mathbf i}{\mathbf j}} \Phi -
{\bar D}_{{\mathbf i}{\mathbf j}} {\bar \Phi})
\end{equation}
which is the analogue of eq.(\ref{defphi}) for the ${\cal N}=1$ tensor.
It satisfies the following properties:
\begin{eqnarray}
 {\bar  D}_{\dot\alpha}^{({\mathbf i}} F^{{\mathbf j}{\mathbf k})} =0 &\quad&
\mbox{(completely symmetric)\ ,} \nonumber \\
{\bar D}_{\dot\alpha {\mathbf i}} F_{{\mathbf j}{\mathbf k}} = \frac{2}{3}
\varepsilon_{{\mathbf i}(
  {\mathbf j} } {\bar D}^{\mathbf l}_{\dot\alpha} F_{{\mathbf k})
{\mathbf l}}\ , &\quad&
{\bar D}_{{\mathbf i}{\mathbf j}} F_{{\mathbf k}{\mathbf l}} = \frac{1}{3}
\varepsilon_{{\mathbf i}({\mathbf k}}
\varepsilon_{{\mathbf j}{\mathbf l})} {\bar D}^{{\mathbf m}{\mathbf n}}
F_{{\mathbf m}{\mathbf n}}\ .
\label{eq111}
\end{eqnarray}
Finally, ${\bar D}_{{\mathbf i}{\mathbf j}} F^{{\mathbf i}{\mathbf j}}$ is
chiral and  so is ${\bar D}_{{\mathbf i}{\mathbf j}}
F_{{\mathbf k}{\mathbf l}}$
as a consequence of~(\ref{eq111}).

The equation of motion can be written in the two equivalent forms
\begin{equation}
{\bar D}^4 {\bar\Phi} - 48 (\hspace{0.2ex}\raisebox{.5ex}{\fbox{}}\hspace
{0.3ex} - m^2) \Phi  =0
\quad \Longleftrightarrow  \quad
i {\bar D}^{{\mathbf i}{\mathbf j}} F_{{\mathbf i}{\mathbf j}} + 48  m^2 \Phi
=0.
\end{equation}

\subsubsection{Supercurrent}

The variation of $\Phi$ under superconformal
transformations is given by eq.(\ref{rep}) with $q=\frac{3}{2}$
(and $\bar q=0$), {\em i.e.}
\begin{equation}
\delta \Phi
= - \frac{i}{48} {\bar D}^4 D^{{\mathbf i}{\mathbf j}}
(L_{{\mathbf i}{\mathbf j}} \Phi) +
\frac{i}{48} {\bar D}^4 (L_{{\mathbf i}{\mathbf j}}D^{{\mathbf i}{\mathbf j}}
\Phi)\ .
\label{eq302}
\end{equation}
As in ${\mathcal N}=1$, it is convenient to add to
the local corresponding transformation the term
$-\frac{i}{48}{\bar D}^4\left( (H-{\bar H}){\bar\Phi}\right)$ that vanishes in
the global limit.
This leads to
\begin{equation}
\delta \Phi = \frac{1}{48} {\bar D}^4 ( L_{{\mathbf i}{\mathbf j}}
F^{{\mathbf i}{\mathbf j}}) -
\frac{i}{48} {\bar D}^4 D^{{\mathbf i}{\mathbf j}} (L_{{\mathbf i}{\mathbf j}}
\Phi - {\bar
  L}_{{\mathbf i}{\mathbf j}} {\bar\Phi})\ ,
\label{eq103}
\end{equation}
which is of the same form as for the vector multiplet.
The advantage of this  form is that, for $m=0$,
the second term is in fact a gauge transformation
and can be ignored. To continue we then treat the variation of the
mass term, $S_m$, and of the gauge invariant part, $S_g$ separately.
The action $S_m$ is similar to the ${\cal N}=2$ classical
Yang-Mills action, {\em i.e.} the action~(\ref{eq303}) with
${\cal F} ({\cal A}) \equiv i 192 \pi  m^2 {\cal A}^2$.
It follows then from~(\ref{eq200})-(\ref{eq304}) that
\begin{equation}
\delta S_m =  - 2 i m^2 \int (H - {\bar H}) \Phi
{\bar\Phi}\ .\label{eq312}
\end{equation}
$S_g$ being gauge invariant, we can compute its variation
by taking only into account the first term in the
 variation given by  eq.(\ref{eq103}). This
leads to
\begin{eqnarray}
\delta S_g &=& \int \!d^4xd^4\theta \;
\delta \Phi \frac{\delta S_g}{\delta \Phi} + c.c.
 =  \int \!d^4x d^8\theta\;  (\frac{1}{48} L_{{\mathbf i}{\mathbf j}}
F^{{\mathbf i}{\mathbf j}})
 \frac{\delta S_g}{\delta \Phi} + c.c.\nonumber\\
& =& \frac{i}{48} \int L_{{\mathbf i}{\mathbf j}} F^{{\mathbf i}{\mathbf j}}
{\bar
  D}^{{\mathbf k}{\mathbf l}} F_{{\mathbf k}{\mathbf l}} + c.c.
\end{eqnarray}
Naively, one might expect that this variation
should be proportional to $H- {\bar H}$ and
$\sigma$.  However, it turns out that it is impossible
to write it in this form. Nevertheless,
it can be brought into the suggestive  form
\begin{equation}
\delta S_g = \frac{i}{48} [ \int (H-{\bar H}) F^{{\mathbf i}{\mathbf j}}
F_{{\mathbf i}{\mathbf j}} + \frac{3}{10} (H-
{\bar H})_{{\mathbf i}{\mathbf j}{\mathbf k}{\mathbf l}}
F^{{\mathbf i}{\mathbf j}} F^{{\mathbf k}{\mathbf l}}
\; ] \label{eq311}
\end{equation}
where
\begin{equation}
H_{{\mathbf i}{\mathbf j}{\mathbf k}{\mathbf l}}  \equiv
{\bar D}_{({\mathbf i}{\mathbf j}} L_{{\mathbf k}{\mathbf l})} \quad \quad
\mbox{(completely symmetric)}.
\end{equation}
This result relies on the identity
\begin{equation}
F_{{\mathbf i}{\mathbf j}} {\bar D}^{{\mathbf k}{\mathbf l}}
F_{{\mathbf k}{\mathbf l}} = {\bar D}_{{\mathbf i}{\mathbf j}}
(F^{{\mathbf k}{\mathbf l}}
F_{{\mathbf k}{\mathbf l}} ) + \frac{3}{10} {\bar D}^{{\mathbf k}{\mathbf l}}
( F_{({\mathbf i}{\mathbf j}}
F_{{\mathbf k}{\mathbf l})})\ , \label{eq106}
\end{equation}
which in turn can be obtained from the
properties~(\ref{eq111}) of the field strength.
More precisely, one shows that
\begin{eqnarray}
{\bar D}_{{\mathbf i}{\mathbf j}}(F^{{\mathbf k}{\mathbf l}}
F_{{\mathbf k}{\mathbf l}}) &=&
\frac{2}{3} F_{{\mathbf i}{\mathbf j}} {\bar D}^{{\mathbf k}{\mathbf l}}
F_{{\mathbf k}{\mathbf l}} - \frac{4}{9}
{\bar D}^{\mathbf k}_{\dot \alpha} F_{({\mathbf i}{\mathbf k}}
{\bar D}^{\dot\alpha {\mathbf l}}
F_{{\mathbf j}){\mathbf l}}\ , \label{eq105} \nonumber\\
 \\
{\bar D}^{{\mathbf k}{\mathbf l}} (F_{({\mathbf i}{\mathbf j}}
F_{{\mathbf k}{\mathbf l})}) &=&
\frac{10}{9} F_{{\mathbf i}{\mathbf j}} {\bar D}^{{\mathbf k}{\mathbf l}}
F_{{\mathbf k}{\mathbf l}} +
\frac{40}{27} {\bar D}^{\mathbf k}_{\dot \alpha} F_{({\mathbf i}{\mathbf k}}
{\bar D}^{\dot\alpha {\mathbf l}} F_{{\mathbf j}){\mathbf l}}\ .\nonumber
\label{eq104}
\end{eqnarray}
We then get the relation~(\ref{eq106}) as a consequence
of~(\ref{eq105}).

\medskip

Taking the sum of~(\ref{eq312}) and of~(\ref{eq311}), we
end up with the variation of the total action for the tensor
multiplet
\begin{equation}
\delta S =  \frac{i}{48} \int d^{12}z\ (H-{\bar H}) (F^{{\mathbf i}{\mathbf j}}
F_{{\mathbf i}{\mathbf j}} - 96 m^2 \Phi {\bar\Phi} )
+ \frac{3}{10} (H-{\bar H})_{{\mathbf i}{\mathbf j}{\mathbf k}{\mathbf l}}
F^{{\mathbf i}{\mathbf j}} F^{{\mathbf k}{\mathbf l}}
\ . \label{eq300}
\end{equation}
Therefore we define
\begin{equation}
T \equiv \frac{1}{48} \left( F^{{\mathbf i}{\mathbf j}}
F_{{\mathbf i}{\mathbf j}} - 96 m^2 \Phi {\bar\Phi} \right)
\quad \mbox{and} \quad
T^{{\mathbf i}{\mathbf j}{\mathbf k}{\mathbf l}} \equiv \frac{1}{160}
F^{({\mathbf i}{\mathbf j}}F^{{\mathbf k}{\mathbf l})}\ ,
\label{eq635}
\end{equation}
where $T^{{\mathbf i}{\mathbf j}{\mathbf k}{\mathbf l}}$ is completely
symmetric.
Indeed, it follows from the analysis of subsection~\ref{eq305},
and from  the
expression~(\ref{eq313}) of $L_{{\mathbf i}{\mathbf j}}$ for superconformal
transformations,  that for ${\cal N}=2$ super-Poincar\'e
and $U(1)_R$ transformations,  $H$ {\em and}
$H_{{\mathbf i}{\mathbf j}{\mathbf k}{\mathbf l}}$
are  both real. Thus, eq.(\ref{eq300})
makes the invariance of the action under
these transformations explicit and
the corresponding conserved
currents are expressed as components of
$T$ and $T^{{\mathbf i}{\mathbf j}{\mathbf k}{\mathbf l}}$ in analogy with the
previous examples.
For the $SU(2)_R$-transformations and the corresponding current,
the situation is more subtle. We refer the proof for
this case to subsection~\ref{nn3}.  Finally, if we define $
\tau^{{\mathbf i}{\mathbf j}}
\equiv - {\bar D}_{{\mathbf k}{\mathbf l}}
T^{{\mathbf i}{\mathbf j}{\mathbf k}{\mathbf l}}$, the conservation equation
following from~(\ref{eq300}) can be written as
\begin{equation}
{\bar D}^{{\mathbf i}{\mathbf j}} T =  \tau^{{\mathbf i}{\mathbf j}} = -
{\bar D}_{{\mathbf k}{\mathbf l}}
T^{{\mathbf i}{\mathbf j}{\mathbf k}{\mathbf l}}. \label{eq630}
\end{equation}
This form of the conservation equation, which was
proposed in~\cite{Stelle}, is similar to
that proposed in~\cite{Theisen} except from the fact that
here $\tau^{{\mathbf i}{\mathbf j}}$ is not real.

\subsubsection{Central Charge}

We end this section with the computation of the center for the ${\cal N}
=2$ tensor multiplet.  As for the vector multiplet, we
first determine the supersymmetry current. A short computation shows that 
$T^{{\mathbf i}{\mathbf j}{\mathbf k}{\mathbf l}}$ contributes to 
$J^{\mu {\mathbf i}}_{\alpha}$ a term 
\begin{equation}
768 i \sigma^{\mu}_{\alpha\dot\alpha} {\bar
D}_{\mathbf j}^{\dot\alpha} D_{{\mathbf k}{\mathbf l}} T^{{\mathbf i}
{\mathbf j}{\mathbf k}{\mathbf l}} =
-768 i\sigma^{\mu}_{\alpha\dot\alpha} {\bar
D}_{\mathbf j}^{\dot\alpha} D^{{\mathbf i}{\mathbf j}} T \label{eq632}
\end{equation}
where we have used the equation of conservation. This
means that $J^{\mu {\mathbf i}}_{\alpha}$
can be expressed in terms of $T$ only. This is due to
the fact that, for supersymmetry transformations, it is possible to
localize $L_{{\mathbf i}{\mathbf j}}$ such that
$H_{{\mathbf i}{\mathbf j}{\mathbf k}{\mathbf l}} =
{\bar H}_{{\mathbf i}{\mathbf j}{\mathbf k}{\mathbf l}}$
is still satisfied.\footnote
{This choice of $L_{{\mathbf i}{\mathbf j}}$ differs
from the one made in section $4.2.2$ 
by terms proportional to the
derivative of $\varepsilon^\alpha_{\mathbf i}$. The corresponding
currents are equal on-shell (see also remark in section~\ref{Noether}).} Taking
all the contributions we end up with
\begin{equation}
J_\alpha^{\mu {\mathbf i}} = 192 \left[
-3 i \sigma^\mu_{\alpha\dot\alpha} {\bar
  D}^{\dot\alpha}_{\mathbf j}  D^{{\mathbf i}{\mathbf j}} T
- 3 i
{\bar\sigma}^{\mu\dot\alpha\beta} {\bar D}^{\mathbf i}_{\dot\alpha}
D_{\alpha\beta} T
- 12  \partial^\mu D^{\mathbf i}_\alpha T
+   a \sigma_\alpha^{\mu\nu\,\beta} \partial_\nu D_\beta^{\mathbf i}
T \right]| \ . \label{eq631}
\end{equation}
Note that as a consequence of the
conservation equation~(\ref{eq630}), we have $D^3_{\alpha {\mathbf i}}
T =0$ on-shell. This identity
is useful for the following reason. As explained
for the vector,  to fix the value of the coefficient $a$
corresponding to the canonical case and to determine the
center, we have to compute the symmetric or
antisymmetric part of $D_\alpha^{\mathbf i} J^0_{\beta {\mathbf i}}$
respectively.
However, the corresponding variation of the first term in the
r.h.s. of~(\ref{eq631}) is proportional to $D^{\mathbf i}_\alpha
{\bar D}^{\dot\alpha {\mathbf j}} D_{{\mathbf i}{\mathbf j}} T \propto {\bar
  D}^{\dot\alpha {\mathbf j}} D^3_{\alpha {\mathbf j}} T =0$. Therefore, we
can ignore this term for our purpose.

Let us now consider the variation of the second term in
the r.h.s. of~(\ref{eq631}). Using the algebraic identity~(\ref{eq634}) and
the fact that $D^3 T$ vanishes on-shell,
we get
\begin{equation}
D^{\mathbf i}_\alpha \left[ - 3 i
{\bar\sigma}^{0\dot\alpha\gamma}  {\bar
  D}_{\dot\alpha {\mathbf i}} D_{\beta\gamma} T \right] =  -24
\left[ \sigma^{\mu 0\,\,\gamma}_\alpha \partial_\mu
  D_{\beta\gamma} T - \frac{1}{2}\partial^0 D_{\alpha\beta} T\right].
\end{equation}
Finally, the variations of the last two terms in
the r.h.s. of~(\ref{eq631}) can be
easily read from the results~(\ref{eq633}) and (\ref{eq623}).
Here we give the result.  First, as in the case of
the vector multiplet, the canonical supersymmetry current corresponds to
$a=-24$. Then, we get $Z= 96 \int d^3x
\partial_i Z^{0i}$ with
\begin{equation}
Z^{0i} = 96 \sigma^{0i}_{\alpha\beta} D^{\alpha\beta} T|\ .
\end{equation}
That is we recover the universal formula~(\ref{eq315}) with ${\cal J}=0$.
Using the expression~(\ref{eq635}) of $T$ we then end up with
\begin{equation}
Z^{0i} = 2 \sigma^{0i}_{\alpha\beta} D^{\alpha\beta}
\left( F^{{\mathbf i}{\mathbf j}} F_{{\mathbf i}{\mathbf j}} - 96 m^2 \Phi
{\bar\Phi} \right)\displaystyle{\biggl\vert}
\ . \label{eq637}
\end{equation}
However,
\begin{equation}
D_{\alpha\beta} F_{{\mathbf i}{\mathbf j}} = i (D_{\alpha\beta}
D_{{\mathbf i}{\mathbf j}} \Phi -
D_{\alpha\beta} {\bar D}_{{\mathbf i}{\mathbf j}} {\bar\Phi}) =
8\partial_{(\alpha\dot\alpha} D_{\beta)({\mathbf i}}{\bar
  D}^{\dot\alpha}_{{\mathbf j})} {\bar \Phi} =0
\end{equation}
where we have used the fact that $D_{\alpha\beta} D_{{\mathbf i}{\mathbf j}}
=0$, the result~(\ref{eq636}) of appendix~\ref{eq617}
and the anti-chirality of ${\bar\Phi}$. As a consequence,
the first term in the center~(\ref{eq637}) receives
contribution only from $(DF)^2$ {\em i.e.} from fermions.
Therefore, two cases have to be considered.

$\bullet$ $m\neq0\ $: in that case all the fields
decrease sufficiently fast
enough at infinity such that the center vanishes.

$\bullet$ $m=0\ $: as there are only contributions from the
fermions that decrease sufficiently fast enough at
infinity, the center vanishes also in that case.

\section{Linearized Supergravity}
\label{eq1000}

It is well known that the on-shell multiplet of Noether currents of
supersymmetric matter theories can be used to construct off-shell
Supergravities~\cite{OS1}, at least at the linearized level. The purpose
of this section is to discuss some elements of constructing supergravities
directly on the superfield level using the Superfield Noether Procedure
developed in this paper.

Coupling matter to gravity means gauging the
super-Poincar\'e transformations. As explained in section $2$,
super-Poincar\'e transformations are parametrized by $h^{\alpha{\dot{\alpha}}}=\bar
h^{\alpha{\dot{\alpha}}}$ and
  \begin{equation}
\sigma\ =\ {\bar\sigma}\ =\ \Lambda^{\mathbf i}{}_{\mathbf j}\ =\ 0\ .
\label{sg1}\end{equation}
On the other hand, we recall from~(\ref{16}) and (\ref{K})
that the variation of the matter
action under an arbitrary local transformation is of the general form
\begin{equation} \label{sg2}
\delta S[O^{A{\mathbf J}}] \begin{array}[t]{l}\displaystyle
=\frac{i}{16}\int d^{4+4{\mathcal N}}z\ \left\{\left( h^{\alpha{\dot{\alpha}}}
-{\bar h}^{\alpha{\dot{\alpha}}}\right)
T_{\alpha{\dot{\alpha}}} +i\left( h^{\alpha{\dot{\alpha}}} +
{\bar h}^{\alpha{\dot{\alpha}}}\right) K_{\alpha{\dot{\alpha}}}\right\}
\\[3mm]\displaystyle
\phantom{=}
-\frac{1}{2}\int  d^{4+2{\mathcal N}}z_+\ \sigma J
-\frac{1}{2}\int  d^{4+2{\mathcal N}}z_-\ {\bar\sigma}{\bar J}\ ,
\end{array}\label{h+-h2}\end{equation}
where
\begin{equation}
K_{\alpha\dot\alpha} =  -\frac{i}{4} [D^{\mathbf i}_\alpha, {\bar
D}_{{\mathbf i}\dot\alpha} ] (X - {\bar X}) - \frac{{\cal N}}{2}
\partial_{\alpha\dot\alpha} (X+{\bar X}).
\end{equation}
To construct an invariant action at the linear level, one
follows the standard procedure of coupling the currents to potentials.
Concretely we introduce the real superpotentials
$H^{\alpha\dot\alpha}$, $B$, $C$, as well as a chiral potential $\Omega$
and add the terms
\begin{equation}
\frac{i}{16} \int d^{4+4{\cal{N}}}z\
 \{
H^{\alpha\dot\alpha} T_{\alpha\dot\alpha} + B(X + {\bar
X}) + C (X - {\bar X}) \} - \frac{1}{2} \int d^{2+2{\cal{N}}}z_+\ \Omega J
- \frac{1}{2} \int d^{2+2{\cal{N}}}z_-\ {\bar\Omega} {\bar J} \label{eq300b}
\end{equation}
to the action.
The supergravity potentials must then transform like
\begin{equation}
\begin{array}{cc}
\delta H^{\alpha\dot\alpha} \equiv
-(h^{\alpha\dot\alpha} - {\bar h}^{\alpha\dot\alpha})\ ,\\[3mm]
\delta B \equiv -\frac{i {\cal N}}{2}
\partial_{\alpha\dot\alpha} (h^{\alpha\dot\alpha} + {\bar
h}^{\alpha\dot\alpha})\ ,\quad
\delta C \equiv - \frac{1}{4} [D^{\mathbf i}_\alpha, {\bar
D}_{{\mathbf i}\dot\alpha } ] (h^{\alpha\dot\alpha} + {\bar
h}^{\alpha\dot\alpha})\ ,\\[3mm]
\delta \Omega \equiv  -\sigma
\ .\end{array}
\end{equation}
By construction, the action obtained in this way has a larger
invariance than the superdiffeomorphisms. In particular, it is
invariant under linearized Weyl transformations
\begin{equation}
h^{\alpha\dot\alpha} \equiv i\frac{12}{{\cal{N}}}\,
\theta^{\alpha }_{\mathbf i}
{\bar\theta}^{\dot\alpha {\mathbf i}} \sigma(x^+), \quad {\bar
h}^{\alpha\dot\alpha} \equiv -i \frac{12}{{\cal{N}}}\,\theta^{\alpha }_{
\mathbf i}
{\bar\theta}^{\dot\alpha {\mathbf i}} {\bar \sigma}(x^-).
\end{equation}
Indeed, $h^{\alpha{\dot{\alpha}}}$ satisfies the
chirality preserving constraint
${\bar D_{\mathbf i}}^{(\dot\beta} h^{\alpha\dot\alpha)} =0$. On
the other hand,
the metric $g^{\mu\nu}$, which is proportional to
${\bar\sigma}^{(\mu\dot\alpha\alpha}
{\bar\sigma}^{\nu)\dot\beta\beta} [D^{\mathbf i}_\alpha, {\bar
D}_{\dot\alpha {\mathbf i}} ] H_{\beta\dot\beta}|$, transforms as required
for a Weyl transformation. Therefore, $\Omega$ ensures
that the gauged action is Weyl invariant.  In order to
restrict the gauge group to the superdiffeomorphisms
alone, $\Omega$ has to be set to a fixed value. Thus $\Omega$ is a
compensator.

\subsection{Improvement Terms}

From the point of view of linear supergravity, the
supersymmetry current and the stress-energy tensor are
obtained by variation of the action~(\ref{eq300b})
with respect to the gravitino and the metric
respectively.
We now explain how
to obtain the various conserved currents differing
by improvement terms.
The procedure is analogous to that used in the Weyl gauging~\cite{LRSW} to
understand the relation between non-minimal coupling and improvement terms
for the stress-energy tensor of non-supersymmetric theories.
The key point is that the divergence of the gauge potential for the
scale symmetry transforms like the Ricci scalar $R$
under diffeomorphisms and Weyl transformations. We
start by explaining how this works here at the component
level and show how the Weyl gauging is done
at the level of the multiplets
within the superfield Noether procedure.
For simplicity we first consider a generic ${\cal{N}}=1$
theory.

\medskip

{\bf Components} We can identify the component in $B$ and $C$ transforming
like the Ricci scalar $R$ by noting that
\begin{equation}
\delta (D^2{\bar D}^2 B) = -96 i \hspace{0.2ex}\raisebox{.5ex}{\fbox{}}
\hspace{0.3ex} {\bar\sigma},\qquad\qquad
\delta (D^2{\bar D}^2 C) = 96 i \hspace{0.2ex}\raisebox{.5ex}{\fbox{}}
\hspace{0.3ex} {\bar\sigma}.
\end{equation}
Therefore, the highest component, $d$, of $B$ (and
similarly for $C$) is a scalar that
transforms like $R$, which, at the
linear level, is given by
$R = \partial_\mu\partial_\nu g^{\mu\nu} - \hspace{0.2ex}\raisebox{.5ex}
{\fbox{}}\hspace{0.3ex}
g^\mu{}_\mu$.
Therefore, the component $d$ can either be
considered as independent of the metric or to be $d' + b_1 R$. It is
then clear that variation of~(\ref{eq300b})
with respect to the metric will give an improvement term  of the form
$b_1 (\eta^{\mu\nu} \hspace{0.2ex}\raisebox{.5ex}{\fbox{}}\hspace{0.3ex}
-\partial^\mu\partial^\nu) (X + {\bar X})$
for $T^{\mu\nu}$, which is what
we were looking for. We can repeat the same procedure with the second highest
component of $B$ (and $C$) to improve the supersymmetry current.

\medskip

{\bf Superfield} In order to see how this procedure lifts to superspace,
we recall the identities~(\ref{rel2}) satisfied by the parameter superfield
$h^{\alpha{\dot{\alpha}}}$. Take now a
representative triplet of currents,
$(T,X,J)$ say, and the corresponding linearized action~(\ref{eq300b}).
Now, due to~(\ref{rel2}), the variations of these potentials are
not
independent. Indeed we have
\begin{equation}
\begin{array}{lcl}
\delta B + \frac{1}{12} [D_{\alpha}, {\bar
D}_{\dot\alpha } ] \delta H^{\alpha\dot\alpha}  +
8i (\delta \Omega + \delta
{\bar{\Omega}})&=&0, \nonumber\\
&&\\
\delta C + \frac{3}{2} i \partial_{\alpha\dot\alpha}
\delta H^{\alpha\dot\alpha} + 24 i (\delta \Omega -
\delta{\bar{\Omega}} ) &=&0. \label{eq706}
\end{array}
\end{equation}
In particular, if we replace $C$ in~(\ref{eq300b})
by $-\frac{3}{2} i \partial_{\alpha\dot\alpha}
H^{\alpha\dot\alpha} - 24 i (\Omega -{\bar\Omega})$, then the total
action obtained in this way is also invariant. The
effect of this substitution is, as in the component
approach, to relate the top components of $C$
to the metric and gravitino, but this time in a supersymmetric way.

\medskip

In order to make the equivalence with adding improvement terms explicit, we
take two superfields $U$ and $V$ and rewrite~(\ref{eq300b}) as
\begin{eqnarray}\label{old}
&&  \frac{i}{16} \int d^8z\
\{
H^{\alpha\dot\alpha} T_{\alpha\dot\alpha} + B(X + {\bar
X} +U) + C (X - {\bar X} +V) -BU -CV\}\nonumber\\
&& - \frac{1}{2} \int
d^6z_+\ \Omega J
- \frac{1}{2} \int d^6z_-\ {\bar\Omega} {\bar J}.
\end{eqnarray}
Then, using~(\ref{eq706}), we replace~(\ref{old}) by
\begin{eqnarray}
&&\frac{i}{16} \int d^8z\
 \{
H^{\alpha\dot\alpha} T_{\alpha\dot\alpha} + B(X + {\bar
X} +U) + C (X - {\bar X} +V)\nonumber \\
&&\qquad\qquad- \left[ -\frac{1}{12}
[D_\alpha, {\bar D}_{\dot\alpha} ] H^{\alpha\dot\alpha} -
8i (\Omega + {\bar\Omega}) \right] U -\left[-\frac{3i}{2}
\partial_{\alpha\dot\alpha} H^{\alpha\dot\alpha} - 2 4 i
(\Omega - {\bar\Omega})  \right] V \}\nonumber \\
&&\qquad\qquad- \frac{1}{2} \int d^6z_+\ \Omega J
- \frac{1}{2} \int d^6z_-\ {\bar\Omega} {\bar J}.
\end{eqnarray}
The total action obtained in this way is, of course,
also invariant under superdiffeomorphisms and Weyl transformations.
Integrating by parts and regrouping the terms in $\Omega$ and ${\bar
\Omega}$, we end up with
\begin{eqnarray}
&& \frac{i}{16} \int d^8z\ \{
H^{\alpha\dot\alpha}  \left[ T_{\alpha\dot\alpha} +
\frac{1}{12} [D_\alpha, {\bar D}_{\dot\alpha}] U -
\frac{3i}{2} \partial_{\alpha\dot\alpha} V \right]\nonumber\\
&& \qquad\qquad\qquad + B(X + {\bar
X} +U) + C (X - {\bar X} +V) \}  \label{eq300c}\\
& &\qquad\qquad\qquad-\frac{1}{2} \int d^6z_+\
\Omega \left(J + {\bar D}^2 U + 3 {\bar D}^2V \right)\nonumber\\
& &\qquad\qquad\qquad -
\frac{1}{2} \int d^6z_-\ {\bar\Omega} \left({\bar J} + D^2U
-3 D^2V\right)\ .\nonumber
\end{eqnarray}
Therefore, starting with the coupling~(\ref{eq300b}) and the
representative $(T,X,J)$, we have constructed a different
coupling~(\ref{eq300c}) associated with another representative
$(T'$, $X'$, $J')$. This shows explicitly how the equivalence relation~(\ref
{eqv1}) relates different supergravities to each other.
In particular,
if the matter action is conformal, we can obtain the conformal ${\cal{N}}=1$
supergravity~\cite{FN1,SW2} in this way.

\medskip

{\bf Comparison with the literature:} Let us now see how we can recover the existing results in the literature~\cite
{SW1,FN1,Aku1,B1,GG1} on the various linearized ${\cal{N}}=1$
supergravities from our formalism. For this, we first recall the
conservation equation~(\ref{conG2})
\begin{equation}\label{c2}
{\bar D}^{\dot\alpha} T_{\alpha\dot\alpha} + W_\alpha +
D_\alpha \tau =0
\end{equation}
with
\begin{equation}
W_\alpha \equiv \frac{1}{4} {\bar D}^2 D_\alpha (X- 2
{\bar X}), \quad \tau \equiv \frac{1}{4} {\bar D}^2 X -
\frac{1}{6} J.
\end{equation}
Let us now furthermore impose the restriction that $X$ is purely imaginary,
$X=-\bar X$. In this case we can compare the conservation equation~(\ref{c2})
with eq. (3.4) in~\cite{GG1}. The case $n=-\frac{1}{3}$ ({\it old minimal})
in~\cite{GG1} then corresponds to $W_\alpha=0$ in~(\ref{c2}),
whereas $n=0$ ({\it new minimal}) in~\cite{GG1}
corresponds to $\tau=0$. For all other real $n$, eq.(3.4)
in~\cite{GG1},
with  $\lambda_\beta=D_\beta\Gamma$,
can be written as
\begin{equation}
\bar D^{{\dot{\alpha}}} J_{\alpha{\dot{\alpha}}}=\frac{i}{6}\bar D^2 D_\alpha (
\Gamma+\bar \Gamma)-
\frac{i}{3}\frac{1}{3n+1}D_\alpha\bar D^2\bar \Gamma\ .
\end{equation}
This then agrees with~(\ref{c2}) provided
we identify
\begin{eqnarray}
\frac{i}{6}(\bar\Gamma+\Gamma)&=&-\frac{3}{2}X{\quad\hbox{{and}}\quad}\nonumber
\\
\frac{i}{3(3n+1)}\bar D^2\bar \Gamma&=&\tau=-\frac{1}{6}J+\frac{1}{2}\bar D^2 X
\ .
\end{eqnarray}
For a general $W_\alpha$ ($X$ and $\bar X$ unrelated) and $\tau=0$,
the multiplet has $(16+16)$ components.
If furthermore $\tau\neq0$, we count $(20+20)$ components, which
agrees with the {\it non minimal}
supergravity.

Finally, if we impose no
relation between the various improvement terms,
we showed in subsection~\ref{MNC}
that we have to count all components of $X$. So, we end up with $(28+28)$
components, which agrees with the number of components of
the flexible supergravity proposed in~\cite{GOS1}.

\subsection{${\cal N}=2$ Supergravity}

To obtain a linearized ${\cal N}=2$ supergravity we can start with the
${\cal N}=2$ vector multiplet discussed is subsection~\ref{1.2}.
In analogy with
the  ${\cal N}=1$ case,  we introduce the potentials
$\Lambda$, $\Omega$ and a coupling
\begin{equation}
\int d^{12}z\ \Lambda T + \int d^8z_+\
\Omega {\cal J} + \int d^8z_-\ {\bar\Omega} {\bar{\cal J}}\ .
\end{equation}
The corresponding transformations are
\begin{equation}
\delta \Lambda \equiv - i (H - {\bar H}), \qquad \qquad
\delta \Omega \equiv   144i \sigma.
\end{equation}
Note that the Weyl multiplet $\Lambda$, which was derived in~\cite{Theisen}
within harmonic superspace, arises here in ordinary superspace as a direct
consequence of the localized transformation~(\ref{eq201})
of the vector multiplet ${\cal{A}}$.

To see how the improvement terms are traced to supergravity couplings we
again follow the same path as for ${\cal{N}}=1$. Concentrating on
 the
stress-energy tensor we notice that the improvement term is a double
derivative of lowest component in the supercurrent $T$. Thus
the highest component of $\Lambda$ transforms like the Ricci scalar and
therefore we are free to relate it to the metric.
What is different here is that there is no relation of the type~(\ref{eq706})
between the variation of the different superpotentials.
However, we can nevertheless implement the substitution
at the superspace level by expressing $\Lambda$ in terms of the
unconstrained
spinor superfield $\psi_{\alpha {\mathbf i}}$ of Poincar\'e supergravity~\cite
{GS2} as
\begin{equation}
\Lambda \equiv D^{\alpha {\mathbf i}} \psi_{\alpha {\mathbf i}} +
{\bar D}_{\dot\alpha {\mathbf i}} {\bar \psi}^{\dot\alpha {\mathbf i}}\ .
\end{equation}
This has indeed the effect of relating the highest
component of $\Lambda$ to the metric.

\section{Multiplet of Currents}
 In this section we present the algebra of the
component Noether currents and exhibit the on-shell multiplet
structure of a general, i.e. not necessarily minimal,
 supermultiplet.
\subsection{${\cal N}=1$ Multiplets}
 We start from the variation of the
energy-momentum tensor, obtained in the same manner as the supersymmetry
current (\ref{s1}) in section $3$, by choosing $L^\alpha={\bar\theta}_{\dot{\alpha}}
a^{\alpha{\dot{\alpha}}}(x_-)$. We then find
\begin{equation}
\delta S=\displaystyle
-2\int d^4 x\
a^{\alpha{\dot{\alpha}}}\displaystyle{\Biggl(}
\partial_{\alpha{\dot{\beta}}}{\bar D}^{\dot{\beta}} D^\beta
{\widetilde T}_{\beta{\dot{\alpha}}}
-\partial_{\beta{\dot{\alpha}}}D^\beta {\bar D}^{\dot{\beta}}
{\widetilde T}_{\alpha{\dot{\beta}}}
-\partial_{\beta{\dot{\beta}}}{\bar D}_{\dot{\alpha}} D^\beta
{\widetilde T}_\alpha{}^{\dot{\beta}}
+\partial_{\beta{\dot{\beta}}}D_\alpha {\bar D}^{\dot{\beta}}
{\widetilde T}^\beta{}_{\dot{\alpha}}
\displaystyle{\Biggr)}\displaystyle{\biggl\vert}\ ,
\label{translvariation}\end{equation}
from which we extract the symmetric energy-momentum tensor:
\begin{equation}
T_{\mu\nu}=\begin{array}[t]{l}\displaystyle{\Biggl(}
4\sigma_{(\mu}^{\alpha{\dot{\alpha}}} [D_\alpha,{\bar D}_{\dot{\alpha}}]
{\widetilde T}_{\nu)}
-4\eta_{\mu\nu}\sigma^{\rho\alpha{\dot{\alpha}}} [D_\alpha,
{\bar D}_{\dot{\alpha}}] {\widetilde T}_{\rho}
\\[3mm]
+(\partial_\mu\partial_\nu-\eta_{\mu\nu}\hspace{0.2ex}\raisebox{.5ex}{\fbox{}}
\hspace{0.3ex})(b_1 (X+{\bar X})+ i b_2 (X-{\bar X}))
\displaystyle{\Biggr)}\displaystyle{\biggl\vert}\ .
\end{array}
\label{tmunu}\end{equation}
The two real constants $b_1$ and $b_2$ introduced here
correspond to improvement
terms.
Its trace is given by:
\begin{equation}
T_\mu{}^\mu
=\left(
-2(D^2 {\widetilde J}+{\bar D}^2 {\kern .35em\bar{\kern -.35em{\widetilde J}}})
-3\hspace{0.2ex}\raisebox{.5ex}{\fbox{}}\hspace{0.3ex}(b_1 (X+{\bar X})+ i b_2
(X-{\bar X}))\right)\displaystyle{\biggl\vert}
\ .
\label{tracetmunu}\end{equation}
Using (\ref{s1}), the supersymmetry transformation of $T_{\mu\nu}$ can then
b written as
\begin{equation}
\begin{array}{rl}
\delta_\alpha T_{\mu\nu} =&
8\sigma_{\rho(\mu\alpha}{}^\beta\partial^\rho j_{\nu)\beta}
\\[3mm]&
+\left( (\partial_\mu\partial_\nu-\eta_{\mu\nu}\hspace{0.2ex}\raisebox{.5ex}
{\fbox{}}\hspace{0.3ex})D_\alpha
((b_1+2a_1)(X+{\bar X})+ (i b_2+2a_2)(X-{\bar X}))\right)\displaystyle
{\biggl\vert}
\ .
\end{array}
\label{multiplet1}\end{equation}
It contains not only the derivatives of the supersymmetry current,
as expected, but also the derivatives of $D_\alpha X |$,
which shows that $D_\alpha X|$ belongs to the multiplet.  The
variations of the supersymmetry current, in turn, are given by
\begin{eqnarray}
\delta_\alpha j_{\mu\beta}&=&
\sigma_{\mu\nu\alpha\beta}\partial^\nu\left(\frac{8}{3}
{\kern .35em\bar{\kern -.35em{\widetilde J}}}
+\frac{1}{2}D^2(a_1 (X+{\bar X}) + a_2 (X-{\bar X}))\right)\displaystyle
{\biggl\vert}
\ ,\label{multiplet2abis}\\[3mm]
\sigma_\mu^{\alpha{\dot{\alpha}}}{\bar\delta}_{\dot{\alpha}} j_{\nu\alpha} &=&
\frac{i}{2}T_{\mu\nu}-4\partial_\mu {\widetilde j}^{(5)}_\nu+4\eta_{\mu\nu}
\partial^\rho{\widetilde j}^{(5)}_\rho
-2i\varepsilon_{\mu\nu}{}^{\rho\lambda}\partial_\rho
{\widetilde j}^{(5)}_\lambda
\nonumber\\[3mm]&&
+\frac{3i}{8}\varepsilon_{\mu\nu}{}^{\rho\lambda}
\sigma_\lambda^{\alpha{\dot{\alpha}}}\left([D_\alpha,{\bar D}_{\dot{\alpha}}]
\partial_\rho
(a_1(X+{\bar X}) +a_2(X-{\bar X}))\right)\displaystyle{\biggl\vert}
\ ,
\label{multiplet2}\end{eqnarray}
where we have  defined
\begin{equation}
{\widetilde j}^{(5)}_\mu=\begin{array}[t]{l}
\displaystyle{\Biggl(}{\widetilde T}_\mu
+\frac{1}{16}\sigma_\mu^{\alpha{\dot{\alpha}}}[D_\alpha,
{\bar D}_{\dot{\alpha}}](a_1 (X+{\bar X}) + a_2 (X-{\bar X}))
\\[3mm]
+\frac{i}{8}\partial_\mu((b_1+2a_1)(X+{\bar X})+ (i b_2+2a_2)(X-{\bar X}))
\displaystyle{\Biggr)}\displaystyle{\biggl\vert}
\ .\end{array}
\label{tildeR}\end{equation}
 Therefore, both $X|$ and ${\widetilde J}|$
have to be included in the multiplet. In turn, this
implies that the multiplet is formed by the components of
the superfields ${\widetilde T}$, ${\widetilde J}$
and $X$.

\subsection{${\cal N}=2$ Vector Multiplet}
\label{what}
With $L_{{\mathbf i}{\mathbf j}} = -
\frac{1}{18} \theta^\alpha_{({\mathbf i}}
{\bar\theta}^{3\dot\alpha}_{{\mathbf j})}
a_{\alpha\dot\alpha}(x^+)$, one finds
\begin{equation}
\begin{array}{c}
T^{\mu\nu} = -24 \,  [
  \frac{1}{2} \eta^{\mu\nu} \{ D^{{\mathbf i}{\mathbf j}} ,
{\bar D}_{{\mathbf i}{\mathbf j}} \}
  T
- \frac{3}{2} {\bar \sigma}^{\mu \dot\alpha\alpha}
  {\bar\sigma}^{\nu\dot\beta\beta} \{ D_{\alpha\beta} ,
  {\bar D}_{\dot\alpha\dot\beta} \} T \\
\\
+48 \eta^{\mu\nu} \hspace{0.2ex}\raisebox{.5ex}{\fbox{}}\hspace{0.3ex} T
 - b ( \eta^{\mu\nu}
\hspace{0.2ex}\raisebox{.5ex}{\fbox{}}\hspace{0.3ex} -  \partial^\mu
\partial^\nu ) T \,  ] |.
\end{array} \label{eq308}
\end{equation}
Again, on-shell the first
term in the r.h.s. of~(\ref{eq308}) is equal to
$-12i \eta^{\mu\nu} (D^4{\cal J} - {\bar D}^4
{\bar{\cal J}})$.
Next we determine also the value of $b$ for the two
interesting cases.
The second term in the r.h.s. of~(\ref{eq308}) is
traceless and thus we immediately have
\begin{equation}
T_\mu^\mu = - 48 i (D^4 {\cal J} -
{\bar D}^4 {\bar{\cal J}})|  +72  (b-64) \hspace{0.2ex}\raisebox{.5ex}{\fbox{}}
\hspace{0.3ex} T|\ . \label{eq421}
\end{equation}
Therefore, this trace vanishes again when ${\mathcal J}=0$
and $b=64$.

The value of $b$ corresponding to the canonical
stress-energy tensor can only be fixed at the components level. For this
purpose, we consider classical Yang-Mills theory. All the
fields being decoupled, it is enough to determine the
contribution from the scalar field $\phi
\equiv {\cal A}|$ to the Hamiltonian density $T^{00}$.
So, suppose that ${\cal F}({\cal A}) \propto {\cal
A}^2$. It follows then from~(\ref{eq304})
that $T =  {\cal A}{\bar {\cal A}}$, up to some global
factor, and that ${\cal J} =0$. Consider then the
different terms of~(\ref{eq308}).

- As ${\cal J}=0$, the first term vanishes on-shell.

- For the second term, when all the fields except $\phi$
are set to zero, we have:
\begin{equation}
\{D_{\alpha\beta}, {\bar D}_{\dot\alpha\dot\beta} \} T
 |  =
 {\cal A} D_{\alpha\beta} {\bar D}_{\dot\alpha\dot\beta}
{\bar{\cal A}}  |   +  c.c.
= 16   {\cal A} \partial_{(\alpha(\dot\alpha}
\partial_{\beta) \dot\beta)} {\bar{\cal A}}  | + c.c.
\end{equation}
where we have used the chirality of ${\cal A}$ and the
relation~(\ref{eq230}) of appendix~\ref{eq231}.
Using the identity ${\bar \sigma}^{\mu\dot\alpha\alpha}
{\bar\sigma}^{\nu\dot\beta\beta}  \partial_{(\alpha(\dot\alpha}
\partial_{\beta) \dot\beta)} = - \eta^{\mu\nu} \hspace{0.2ex}\raisebox{.5ex}
{\fbox{}}\hspace{0.3ex} + 4
\partial^\mu\partial^\nu$,
the second term gives a contribution:
\begin{equation}
24 \phi \left(- \eta^{\mu\nu} \hspace{0.2ex}\raisebox{.5ex}{\fbox{}}\hspace
{0.3ex} {\bar\phi}  + 4
\partial^\mu\partial^\nu {\bar\phi} \right) + c.c. = 96 \phi
\partial^\mu\partial^\nu {\bar\phi} + c.c.
\end{equation}
where we have used the equation of motion
$\hspace{0.2ex}\raisebox{.5ex}{\fbox{}}\hspace{0.3ex}{\bar\phi}=0$.

- The contribution of the last terms is simply:
$(b-48) \eta^{\mu\nu}
\hspace{0.2ex}\raisebox{.5ex}{\fbox{}}\hspace{0.3ex}(\phi{\bar\phi}) - b
\partial^\mu\partial^\nu(\phi{\bar\phi}) $. Therefore, on-shell,
\begin{equation}
T^{\mu\nu} \propto 2(b-48) \eta^{\mu\nu}
\partial^\rho\phi \partial_\rho{\bar\phi} + 96 \phi
\partial^\mu \partial^\nu {\bar \phi} + 96 {\bar \phi}
\partial^\mu\partial^\nu \phi - b
\partial^\mu\partial^\nu (\phi{\bar\phi}).
\end{equation}
The Hamiltonian density is then
\begin{eqnarray}
T^{00} &\propto& -(2b-96) \left [ -
|\partial_0 \phi|^2 + |\partial_i \phi|^2 \right]
+ (96-b) \partial^0\partial^0
(\phi{\bar\phi}) - 192 \partial^0\phi \partial^0
{\bar\phi}  \nonumber\\
T^{00} &\propto&  -(2b-96) |\partial_i \phi|^2 + (2b-96-192)
|\partial_0\phi|^2 + (96-b) \partial^0\partial^0
(\phi{\bar\phi})
\end{eqnarray}
where we have used $\eta^{00} \equiv -1$. Therefore,
$T^{00}$ corresponds to the Hamiltonian  density  if
$-(2b-96) = 2b-96-192$ and $96-b=0$ {\em i.e.} when
$b=96$.

\medskip

Note that in full generality,
improvement terms proportional to
${\cal J}$ could be added to the supersymmetry
current and to the stress-energy tensor
above. However, we do not consider such a possibility here.

\paragraph{$SU(2)_R$ current} We obtain

\begin{equation}
R^\mu_{{\mathbf i}{\mathbf j}} = - 864 i \sigma^\mu_{\alpha\dot\alpha} [
D^\alpha_{({\mathbf i}}, {\bar D}^{\dot\alpha}_{{\mathbf j})} ] T |. \label
{eq309}
\end{equation}

\subsubsection{Supersymmetry Transformations of the Currents}

The next few subsections are devoted to identifying the super multiplets of
Noether currents~\cite{Sohnius79,SW1,HST1,Fisher} for a given choice of improvement terms. For this, we first give the supersymmetry transformations
of the conserved currents. The result is:
\begin{eqnarray}
\delta_{\alpha {\mathbf i}} R^\mu_{{\mathbf j}{\mathbf k}} &=& \frac{3}{2}
\varepsilon_{{\mathbf i}({\mathbf j}} \left[ J^\mu_{\alpha {\mathbf k})} - (24
+ a)
\sigma^{\mu\nu\,\,\,\beta}_\alpha \partial_\nu D_{\beta {\mathbf k})}
T|  \right], \label{eq407}\\
\nonumber \\
{\bar\delta}_{\dot\alpha}^{\mathbf i} J^{\mu {\mathbf j}}_\alpha &=&
\varepsilon^{{\mathbf i}{\mathbf j}} \sigma_{\nu\alpha\dot\alpha} \left[ -2i
T^{\mu\nu} + 48 i (b-48+2a) (\eta^{\mu\nu} \hspace{0.2ex}\raisebox{.5ex}
{\fbox{}}\hspace{0.3ex} -
\partial^\mu\partial^\nu) T| \, \right] \nonumber \\
&+& \varepsilon^{{\mathbf i}{\mathbf j}} \left\{ \frac{1}{4} \left[a+24\right]
\left[
\sigma^\mu_{\alpha\dot\alpha}
\partial^\rho {\cal R}_\rho
- \partial_{\alpha\dot\alpha} {\cal R}^\mu \right]
 - \frac{ia}{4} \varepsilon^{\mu\nu\rho\tau}
\sigma_{\tau\alpha\dot\alpha} \partial_\nu {\cal R}_\rho
 \right\}
\label{eq501}\\
&+&\frac{i}{36}(a-24) \partial_{\alpha\dot\alpha}
R^{\mu \, {\mathbf i}{\mathbf j}}
-\frac{1}{36} (a+24) \varepsilon^{\mu\nu\rho\tau}
\sigma_{\tau\alpha\dot\alpha} \partial_\nu R_\rho^{{\mathbf i}{\mathbf j}},
\nonumber  \\
\nonumber \\
\delta_{\alpha {\mathbf i}} J^\mu_{\beta {\mathbf j}} &=& \partial_\nu \left\{
24 i \varepsilon_{\alpha\beta} \varepsilon_{{\mathbf i}{\mathbf j}}
Z^{\mu\nu}
 + 96  ( a +24) \left[ \varepsilon_{{\mathbf i}{\mathbf j}}
\sigma^{\mu\nu\,\gamma}_{(\alpha}
D_{\beta)\gamma} T |
+ \sigma^{\mu\nu}_{\alpha\beta}
D_{{\mathbf i}{\mathbf j}}T| \right] \right \}, \label{eq402bis} \\
\nonumber \\
\delta_{\alpha {\mathbf i}} T^{\mu\nu} &=&
\sigma^{(\mu\rho\,\,\beta}_{\;\;\;\alpha} \partial_\rho
J^{\nu)}_{\beta{\mathbf i}}
+ 24  (b-48 + 2a) (\eta^{\mu\nu}  \hspace{0.2ex}\raisebox{.5ex}
{\fbox{}}\hspace{0.3ex} -
\partial^\mu\partial^\nu) D_{\alpha{\mathbf i}} T |\ .\label{eq619}
\end{eqnarray}
where we have introduced
\begin{equation}
{\cal R}^\mu \equiv - 48  \sigma^\mu_{\alpha\dot\alpha}
[ D^{\alpha {\mathbf i}} , {\bar D}^{\dot\alpha}_{\mathbf i} ] T |. \label
{eq310}
\end{equation}
It follows from~(\ref{eq502}) that
$\partial^\mu {\cal R}_\mu =  -8i [D^{{\mathbf i}{\mathbf j}},
{\bar D}_{{\mathbf i}{\mathbf j}}] T$ ,
or, using the equation of conservation~(\ref{eq121}), that
\begin{equation}
\partial^\mu {\cal R}_\mu =
8 (D^4 {\cal J} + {\bar D}^4 {\bar{\cal J}})\ . \label{eq503}
\end{equation}

\subsubsection{Multiplet Structure: General Discussion}

Let us now determine with which other components of $T$,
${\cal J}$, and ${\bar{\cal J}}$, the above Noether currents form a
multiplet. For this we first note that, contrary to the situation in
${\cal N}=1$, for a ${\cal N}=2$ superconformal theory,
the conserved $R$ and $SU(2)_R$ and the traceless stress-energy tensor
alone can not form a multiplet with the traceless
supersymmetry currents. This is so because the number of
bosonic components, {\em i.e.} $3+9+5=17$ differs from
the $16$ fermionic components of $J^{\mu}_{\alpha {\mathbf i}}$.
Hence the improved multiplet contains other
components than the Noether
currents above.

\medskip

The following discussion closely follows that for $N=1$.
We start with the variation~(\ref{eq619}) of the stress-energy tensor
$T^{\mu\nu}$.  Two cases have to be considered.

\medskip

$\bullet$ $b-48+2a = 0$: we consider then the variation~(\ref{eq501}) of
$J^\mu_{\alpha {\mathbf i}}$.
As $b-48+2a = 0$,  we conclude  that ${\cal R}^\mu$
belongs to the multiplet. However, we
then have
\begin{equation}
\delta_{\alpha {\mathbf i}} {\cal R}^\mu = 16 i \left[ \frac{1}{192}
J^{\mu}_{\alpha {\mathbf i}} -2 \sigma^{\mu}_{\alpha\dot\alpha}
{\bar D}^{3\dot\alpha}_{\mathbf i} {\bar{\cal J}}| -  (a+24)
\sigma^{\mu\nu\,\,\beta}_{\;\;\alpha} \partial_\nu
D_{\beta {\mathbf i}} T| \right]. \label{eq403}
\end{equation}
Using the relation~(\ref{eq420}) for the trace of
$J^\mu_{\alpha {\mathbf i}}$, we can eliminate ${\bar
  D}^{3\dot\alpha}_{\mathbf i} {\bar{\cal J}}$ to obtain
\begin{equation}
\delta_{\alpha {\mathbf i}} {\cal R}^\mu = 16i \left\{
  \frac{1}{192} \left[  \frac{3}{2} J^\mu_{\alpha {\mathbf i}} +
    \sigma_\alpha^{\mu\nu\beta} J_{\nu\beta{\mathbf i}}  \right] -
  \frac{3}{4}(8+a) \partial^\mu D_{\alpha{\mathbf i}} T| +
  \frac{1}{2} (a-24) \sigma_\alpha^{\mu\nu\beta}
  \partial_\nu D_{\beta{\mathbf i}} T| \right\}.
\end{equation}
This shows that $\chi_{\alpha {\mathbf i}} \equiv D_{\alpha {\mathbf i}} T|$
also belongs to the multiplet. We then continue with the variations
of  $\chi_{\alpha {\mathbf i}}$.  Defining $t \equiv T|$, and
$u_{\alpha\beta} \equiv  D_{\alpha\beta} T|$, we find
\begin{eqnarray}
\delta_{\alpha {\mathbf i}} \chi_{\beta {\mathbf j}} &=& \frac{1}{2}
\varepsilon_{{\mathbf i}{\mathbf j}} u_{\alpha\beta} - \frac{i}{2}
\varepsilon_{\alpha\beta} {\bar D}_{{\mathbf i}{\mathbf j}} {\bar{\cal J}}|,
\label{eq415}\\
{\bar \delta}_{\dot\alpha {\mathbf i}} \chi_{\alpha {\mathbf j}} &=&
\frac{i}{3456} R_{\alpha\dot\alpha {\mathbf i}{\mathbf j}} + \frac{1}{384
}\varepsilon_{{\mathbf i}{\mathbf j}} {\cal R}_{\alpha\dot\alpha}
 +  i \varepsilon_{{\mathbf i}{\mathbf j}} \partial_{\alpha\dot\alpha} t\ .
\label{eq416}
\end{eqnarray}
From the first variation~(\ref{eq415}) we conclude that
${\bar D}_{{\mathbf i}{\mathbf j}}{\bar{\cal J}}|$ and therefore all the
components
of ${\bar{\cal J}}$ belong to the multiplet. The second
variation~(\ref{eq416}) shows that $t= T|$ belongs to the
multiplet. As $t$ is the lowest component of $T$, this
shows that the multiplet contains $T$.
On the other hand, as $T$, ${\cal J}$ and
${\bar{\cal J}}$ are the
only superfields present, we have shown that, for any value of
$a$ and $b$ such that $b-48+2a = 0$,
the multiplet of currents corresponds to
$T$, ${\cal J}$ and ${\bar{\cal J}}$ constrained by the
equations of conservation~(\ref{eq121}).

\medskip

$\bullet$ $b-48+2a \neq 0$: in that case,  we proceed
as follows: the variation~(\ref{eq619}) of
$T^{\mu\nu}$ shows that $\chi_{\alpha {\mathbf i}}$ belongs to the
multiplet of currents. However, we consider then the variation~(\ref{eq416})
of $\chi_{\alpha {\mathbf i}}$. As ${\cal R}^\mu$ and
$t$ are real, this shows that both of them belong to the
multiplet. Therefore, the conclusion is the same as in
the preceding case.

\medskip

Let us now choose  a convenient set of independent components of
$T$, ${\cal J}$ and ${\bar{\cal J}}$.
We take $R^\mu_{{\mathbf i}{\mathbf j}}$, $J^\mu_{\alpha {\mathbf i}}$,
$T^{\mu\nu}$, ${\cal R}^\mu$, $Z^{\mu\nu}$, $t$,
$\chi_{\alpha {\mathbf i}}$, $u_{\alpha\beta}$, ${\bar{\cal J}}|$,
${\bar D}_{ \dot\alpha {\mathbf i}}{\bar{\cal  J}}|$
and ${\bar D}_{{\mathbf i}{\mathbf j}}{\bar{\cal J}}|$.
Indeed, first, ${\bar D}_{\dot\alpha\dot\beta}{\bar{\cal J}}|$ is the
anti self-dual
part of $Z^{\mu\nu}$ as given by the equation~(\ref{eq506}).
Secondly, ${\bar D}^3{\bar{\cal J}}|$ and
${\bar D}^4{\bar{\cal J}}|$ are related to other components via the 'trace'
equations~(\ref{eq420}), (\ref{eq421}) and (\ref{eq503}).
In order to count the number of components of that multiplet, we
 need to distinguish between the cases where
${\cal J}\neq 0$ and ${\cal J} =0$ respectively.

\subsubsection{Multiplet Structure: Case ${\cal J}\neq 0$}

In the general case, the theory is neither $R$-invariant nor
conformal invariant. For any value of $a$ and $b$, we have the
following number of components.
\begin{eqnarray}
{\cal R}^\mu (4), \quad R^\mu_{{\mathbf i}{\mathbf j}} (9), \quad
J^\mu_{\alpha {\mathbf i}} (-24), \quad T^{\mu\nu} (6),\quad Z^{\mu\nu} (6),
\nonumber \\
t (1), \quad \chi_{\alpha {\mathbf i}} (-8), \quad u_{\alpha\beta}
(6),\\
{\bar{{\cal J}}}| (2), \quad {\bar D}_{\dot\alpha {\mathbf i}}
{\bar {\cal J}}| (-8),
\quad
{\bar D}_{{\mathbf i}{\mathbf j}} {{\bar {\cal J}}}| (6)\ .\nonumber
\end{eqnarray}
This forms a  $(40+40)$ multiplet. The algebra
satisfied by these components is summarized
in subsection~\ref{eq609} of appendix~\ref{eq608}.

Let us examine  the conditions for $R$-invariance.
As $H ={\bar H}$ for superconformal
transformations, it follows from the variation~(\ref{eq200}) of the action,
and from the specific value~(\ref{eq313}) of $L_{{\mathbf i}{\mathbf j}}$ for
$U(1)_R$ transformations,
that the condition is that there exists $r^\mu$ such that
$-96 (D^4 {\cal J} +{\bar D}^4 {\bar{\cal J}} ) = \partial_\mu r^\mu$.
Considering local $U(1)_R$ transformations, we find that
the conserved $R$-current is in that case ${\cal R}^\mu
+ r^\mu$. However, as in ${\cal N}=1$ and as a consequence of our general
discussion above, this is ${\cal R}^\mu$ itself rather than the
$R$-current that is in the multiplet.
This is to be expected as the
supersymmetry current and stress-energy tensor are not
traceless. Therefore, the number of components
of the multiplet is again $(40+40)$.

\subsubsection{Conformal Case ${\cal J} =0$}

When ${\cal J}=0$, the theory is conformal invariant and
the equation of conservation is $D^{{\mathbf i}{\mathbf j}}T=0$. It is also
$R$-invariant and ${\cal R}^\mu$ is the conserved
$R$-current. Let us now discuss how the multiplet structure depends
on the improvement terms in the conformal case. We consider first the
improved  multiplet~\cite{SW1,HST1, Fisher} which corresponds to the values
$a=-8$ and $b=64$. It has the components
\begin{eqnarray}\label{cm}
{\cal R}^\mu (3), \quad R^\mu_{{\mathbf i}{\mathbf j}} (9), \quad
J^\mu_{\alpha {\mathbf i}} (-16), \quad T^{\mu\nu}
(5), \quad u_{\alpha\beta} (6), \nonumber\\
t (1), \quad \chi_{\alpha {\mathbf i}} (-8).
\end{eqnarray}
Its dimension is $(24+24)$. The corresponding
transformations of these components are given in~\cite{Fisher, Dahmen} and
in subsection~\ref{eq504} of appendix~\ref{eq608}.

Let us now turn to the canonical multiplet. This corresponds to
$a=-24$ and $b=96$. The difference with the improved
multiplet is that here $\chi_{\alpha {\mathbf i}}$ and
$J^{\mu {\mathbf i}}_\alpha$
are not independent.
More precisely, when ${\mathcal J}=0$ and for $a=-24$, equation~(\ref{eq420})
becomes
$({\bar \sigma}^\mu J_\mu)^{ \dot\alpha {\mathbf i}} = 4608
\partial^{\dot\alpha\alpha} \chi^{\mathbf i}_\alpha$.  Thus, the fermionic
components of the canonical multiplet are as follows: There are the $8$
components of $\chi_{\alpha {\mathbf i}}$. For $J^{\mu {\mathbf i}}_\alpha$
we have only to count its traceless part,
{\em i.e.} $16$ components, as the trace is contained in
$\chi_{\alpha {\mathbf i}}$. Hence we recover the $24$ fermionic components of
$T$.

To summarize, for a conformal theory,
the canonical multiplet contains the bosonic
components ${\cal R}^\mu \; (3)$, $R^\mu_{{\mathbf i}{\mathbf j}} \; (9)$,
 $(T^{\mu\nu}, \, t)  \; (6)$,  $Z^{\mu\nu} \; (6)$ and the
fermionic components $(J^{\mu {\mathbf i}}_\alpha, \chi_{\alpha {\mathbf i}} )
\; (-24)$. Thus the canonical multiplet is also a $(24+24)$
multiplet.

\subsection{${\cal N}=2$ Tensor Multiplet}
\label{nn3}
In this section, we  identify  the contributions from
$T^{{\mathbf i}{\mathbf j}{\mathbf k}{\mathbf l}}$ to the  various  conserved
currents.
The currents for the tensor multiplet
are then simply the sum of the terms
given  respectively in equations~(\ref{eq307}),
(\ref{eq308}), (\ref{eq309}), (\ref{eq310})
and of the terms from $T^{{\mathbf i}{\mathbf j}{\mathbf k}{\mathbf l}}$ given
explicitly below.

\paragraph{Stress-energy Tensor}

For global translations, it follows from~(\ref{eq313})
that $H_{{\mathbf i}{\mathbf j}{\mathbf k}{\mathbf l}}=
{\bar H}_{{\mathbf i}{\mathbf j}{\mathbf k}{\mathbf l}}=0$.
It is immediate to see that this also holds for the
local $L_{{\mathbf i}{\mathbf j}}$ taken in subsection~\ref{what}.
Thus, $T^{{\mathbf i}{\mathbf j}{\mathbf k}{\mathbf l}}$ does not contribute
to the stress-energy tensor.

\paragraph{$R$-current}

Again, a short computation indicates that
there is no contribution of $T^{{\mathbf i}{\mathbf j}{\mathbf k}{\mathbf l}}$
to the $R$-current.

\paragraph{$SU(2)_R$ Invariance and associated Current}

$\ $ Contrary to the ${\cal N}=2$ super-Poincar\'e and $U(1)_R$
transformations, for global $SU(2)_R$
transformations, we do have $H = {\bar H}$ but
$H_{{\mathbf i}{\mathbf j}{\mathbf k}{\mathbf l}}
\neq {\bar H}_{{\mathbf i}{\mathbf j}{\mathbf k}{\mathbf l}}$. More precisely,
it follows from~(\ref{eq313}) that
\begin{equation}
H_{{\mathbf i}{\mathbf j}{\mathbf k}{\mathbf l}} -
{\bar H}_{{\mathbf i}{\mathbf j}{\mathbf k}{\mathbf l}} = -
\frac{3i}{2} \eta_{{\mathbf m}({\mathbf i}}  [\theta^{\mathbf m}
{}_{{\mathbf j}}{\bar
\theta}_{{\mathbf k}{\mathbf l})}  +  {\bar\theta}^{\mathbf m}{}_{
{\mathbf j}}\theta_{{\mathbf k}{\mathbf l})}].
\end{equation}
It is nevertheless still possible to express the
corresponding $SU(2)_R$ current in terms of $T$
and of $T_{{\mathbf i}{\mathbf j}{\mathbf k}{\mathbf l}}$.  To see this,
we consider first global transformations and
concentrate  on the term
giving problem {\em i.e.} the one proportional to
$(H - {\bar H})_{{\mathbf i}{\mathbf j}{\mathbf k}{\mathbf l}}$. Its
contribution to the
variation of the action under $SU(2)_R$
transformations after integration on the Grassman
variables is
\begin{equation}
\frac{27i}{20} \int \!\! d^4x \,\,\,
 \eta_{{\mathbf m}{\mathbf i}} [D^{\mathbf m}{}_{{\mathbf j}}
{\bar D}_{{\mathbf k}{\mathbf l}} + {\bar
   D}^{\mathbf m}{}_{{\mathbf j}} D_{{\mathbf k}{\mathbf l}}] T^{{\mathbf i}
{\mathbf j}{\mathbf k}{\mathbf l}}|.
\end{equation}
However, we prove in appendix~\ref{appendixB} that
\begin{equation}
\left( D^{({\mathbf m}}{}_{{\mathbf j}} {\bar D}_{{\mathbf k}{\mathbf l}}
+ {\bar D}^{({\mathbf m}}{}_{{\mathbf j}}  D_{{\mathbf k}{\mathbf l}}\right)
T^{{\mathbf i}){\mathbf j}{\mathbf k}{\mathbf l}}
=
 - 8 i\partial_{\alpha\dot\alpha}
[D^\alpha_{\mathbf k}, {\bar D}^{\dot\alpha}_{\mathbf l}]
T^{{\mathbf m}{\mathbf i}{\mathbf k}{\mathbf l}}. \label{eq114}
\end{equation}
Thus, as $\eta_{{\mathbf i}{\mathbf j}}$ is symmetric, this proves the
invariance
of the action and gives the contribution of
$T^{{\mathbf i}{\mathbf j}{\mathbf k}{\mathbf l}}$ to
the $SU(2)_R$ current $R^\mu_{{\mathbf i}{\mathbf j}}$:
\begin{equation}
1728 i \sigma^\mu_{\alpha  \dot\alpha} [D^{\alpha {\mathbf k}},
{\bar D}^{\dot\alpha {\mathbf l}} ]
T_{{\mathbf i}{\mathbf j}{\mathbf k}{\mathbf l}}.
\end{equation}

\section*{Conclusions}

In this paper, we developed a method to construct the various multiplets
of Noether currents directly at the superfield level
(Superfield Noether Procedure).
This formalism is useful in view of a unified treatment of those
supersymmetric theories for which an off-shell superfield formulation
exists. In particular, it produces a manifestly supersymmetric treatment
of the various improvement terms interpolating between
canonical and improved Noether currents.

A prominent feature of this formulation
is that the various algebraic manipulations are
independent of the complexity of the action
for a given supermultiplet. This makes this approach particularly suited for
dealing with (quantum) effective actions of supersymmetric theories. As a
specific application we obtained an efficient algorithm to compute the
supersymmetry central charge for an arbitrary local action of a given
off-shell superfield. As another application we gave a
systematic derivation of the supercurrent of the ${\cal{N}}=2$
tensor multiplet as well as the multiplet of canonical Noether currents of
${\cal{N}}=2$
Yang-Mills theory. As a by-product we then also found a simple
derivation of the anomalous superconformal Ward-Identity for the effective
action of that theory.

Of course, by its very nature the application of
our procedure is limited
to those theories for which an off-shell superfield formulation exists.
This is an obvious limitation when dealing with models with
extended supersymmetry.
In view of this, it would be interesting to generalize our formalism
to harmonic superspace~\cite{Ivanov}.

\section*{Acknowledgments}
We acknowledge helpful discussions with F.~Delduc, P.~Howe, E.~Ivanov,
S.~Kuzenko and E.~Soka\-tchev.
This work has been supported by the TMR contract
FMRX-CT96-0012 of the European Union, the ACI 2078-CDR-2
program of Minist\`ere de la Recherche and by the
DFG-Stringtheorie Schwerpunktsprogramm SPP 1096.

\begin{appendix}
\renewcommand{\theequation}{\Alph{section}.\arabic{equation}}

\section{Superconformal Transformations}
Below we give the explicit form of the parameter superfields 
for ${\cal{N}}=1$ and ${\cal{N}}=2$ superconformal transformations
as well as some useful identities. 
\subsection{${\cal N}=1$}
\label{1000}
The general solution of equations (\ref{cons1b}) and
(\ref{cons1}) is~\cite{Park}:
\begin{equation}
h^{\alpha{\dot{\alpha}}}=
\begin{array}[t]{l}
a^{\alpha{\dot{\alpha}}}+4i\varepsilon^\alpha_{\mathbf i}\bar
\theta^{{\dot{\alpha}}{\mathbf i}}
+4i{\bar\varepsilon}^{{\dot{\alpha}}{\mathbf i}}\theta^\alpha_{\mathbf i}-
\omega^\alpha{}_\beta x^{\beta{\dot{\alpha}}}_-
+{\bar\omega}^{\dot{\alpha}}{}_{\dot{\beta}} x^{\alpha{\dot{\beta}}}_+-4\eta
\theta^\alpha_{\mathbf i}\bar\theta^{{\dot{\alpha}}{\mathbf i}}
-6i\eta^{\mathbf i}{}_{\mathbf j}\theta^\alpha_{\mathbf i}\bar
\theta^{{\dot{\alpha}}{\mathbf j}}
\\[3mm]
+\kappa x^{\alpha{\dot{\alpha}}}+x^{\beta{\dot{\alpha}}}_-
b_{\beta{\dot{\beta}}} x^{\alpha{\dot{\beta}}}_+
-x^{\beta{\dot{\alpha}}}_-\rho_\beta^{\mathbf i}\theta^\alpha_{\mathbf i}
+\bar\theta^{{\dot{\alpha}}{\mathbf i}}{\bar\rho}_{{\dot{\beta}}{\mathbf i}}
x^{\alpha{\dot{\beta}}}_+\ .
\end{array}\label{h}\end{equation}
The different parameters correspond to
translations $a^{\alpha{\dot{\alpha}}}$, supersymmetry transformations $
\varepsilon^\alpha_{\mathbf i}$,
Lorentz transformations $\omega^\alpha{}_\beta$ (with  $\omega_{\alpha\beta}=
\omega_{\beta\alpha}$ and
$\omega^\alpha{}_\alpha=0$), $U(1)_R$-transforma\-tions $\eta$,
$SU({\mathcal N})$-transformations $\eta^{\mathbf i}{}_{\mathbf j}$ (with
$(\eta^{\mathbf i}{}_{\mathbf j})^*=-\eta^{\mathbf j}{}_{\mathbf i}$ and
$\eta^{\mathbf i}{}_{\mathbf i}=0$),
dilations $\kappa$, special conformal
transformations $b_{\alpha{\dot{\alpha}}}$ and special
superconformal transformations $\rho_\alpha^{\mathbf i}$.
Correspondingly we have 
\begin{equation}
L^\alpha=\begin{array}[t]{l}
-\frac{1}{2}a^{\alpha{\dot{\alpha}}}\bar\theta_{\dot{\alpha}}
+i\varepsilon^\alpha\bar\theta^2-2i{\bar\varepsilon}_{\dot{\alpha}}\bar
\theta^{\dot{\alpha}}\theta^\alpha
-\frac{1}{2}\omega^\alpha{}_\beta
x^{\beta{\dot{\alpha}}}\bar\theta_{\dot{\alpha}}+\frac{1}{2}
{\bar\omega}^{\dot{\alpha}}{}_{\dot{\beta}} x^{\alpha{\dot{\beta}}}_+\bar
\theta_{\dot{\alpha}}
-\eta \theta^\alpha\bar\theta^2
\\[3mm]
+\frac{1}{2}\kappa x^{\alpha{\dot{\alpha}}}\bar\theta_{\dot{\alpha}}-\frac{i}
{2}\kappa\theta^\alpha\bar\theta^2
+\frac{1}
{2}x^{\beta{\dot{\alpha}}}b_{\beta{\dot{\beta}}}x^{\alpha{\dot{\beta}}}_+\bar
\theta_{\dot{\alpha}}
+\frac{1}{2}x^{\beta{\dot{\alpha}}}\rho_\beta\theta^\alpha\bar
\theta_{\dot{\alpha}}
-\frac{1}{4}\bar\theta^2{\bar\rho}_{\dot{\alpha}} x^{\alpha{\dot{\alpha}}}_+
\ .\end{array}
\label{L}\end{equation}

\medskip

The algebra between ${\mathcal L}$ given by (\ref{defLLgen}) and
$D_\alpha^{\mathbf i}$ is given by 
\begin{equation}
[D_\alpha^{\mathbf i},{\mathcal L}]=
\frac{3}{{\mathcal N}(4-{\mathcal N})}\left( ({\mathcal N}-2)\sigma+2
{\bar\sigma}\right)\ D_\alpha^{\mathbf i}
-i\Lambda^{\mathbf i}{}_{\mathbf j}\ D_\alpha^{\mathbf j}
-\Omega_\alpha{}^\beta\ D_\beta^{\mathbf i}
\ ,
\label{algebraLD}\end{equation}
where $\sigma$, $\Lambda^{\mathbf i}{}_{\mathbf j}$ and $\Omega_\alpha
{}^\beta$ are defined by
\begin{equation}\begin{array}{rcl}
\sigma&=&\frac{1}{6}\left( D^{\alpha{\mathbf i}}\lambda_{\alpha{\mathbf i}}
-\frac{1}{2}\partial_{\alpha{\dot{\alpha}}}h^{\alpha{\dot{\alpha}}}\right)\ ,
\\[3mm]
\Lambda^{\mathbf i}{}_{\mathbf j}&=&
-\frac{i}{4}\left( D_\alpha^{\mathbf i}\lambda^\alpha_{\mathbf j}+
{\bar D}_{{\dot{\alpha}}{\mathbf j}}{\bar\lambda}^{{\dot{\alpha}}{\mathbf i}}
-\frac{1}{{\mathcal N}}\delta^{\mathbf i}_{\mathbf j}\left(
D_\alpha^{\mathbf k}\lambda^\alpha_{\mathbf k}
+{\bar D}_{{\dot{\alpha}}{\mathbf k}}{\bar\lambda}^{{\dot{\alpha}}{\mathbf k}}
\right)\right)
\ ,\\[3mm]
\Omega_{\alpha\beta}&=&
\frac{1}{2-{\mathcal N}}\left( D_{(\alpha}^{\mathbf i}
\lambda_{\beta){\mathbf i}}
+\frac{1}{2}\partial_{(\alpha{\dot{\alpha}}}h_{\beta)}{}^{\dot{\alpha}}\right)
\ .
\end{array}\label{defsol}\end{equation}

Both $\sigma$ and $\Omega_\alpha{}^\beta$ are
chiral and $\Lambda^{\mathbf i}{}_{\mathbf j}$
is hermitian and traceless, {\em i.e.}
\begin{equation}
{\bar D}_{{\dot{\alpha}}{\mathbf i}}\sigma=0\ ,\qquad
{\bar D}_{{\dot{\gamma}}{\mathbf i}}\Omega_\alpha{}^\beta=0\ ,\qquad
(\Lambda^{\mathbf i}{}_{\mathbf j})^*=\Lambda^{\mathbf j}{}_{\mathbf i}\ ,
\qquad
\Lambda^{\mathbf i}{}_{\mathbf i}=0\ .
\label{properties}\end{equation}

From~(\ref{lambda}) and (\ref{defsol}),
we then obtain
\begin{equation}
\lambda^\alpha=-\frac{i}{4}{\bar D}^2 L^\alpha\ ,
\qquad
\Omega_{\alpha\beta}=\frac{i}{8}{\bar D}^2 D_{(\alpha}L_{\beta)}\ ,
\qquad
\sigma=-\frac{i}{24}{\bar D}^2 D^\alpha L_\alpha\ .
\end{equation}

\subsection{${\cal N}=2$}
\label{1001}

The interested reader can verify that all global
transformations are correctly parametrized by
\begin{eqnarray}
L_{{\mathbf i}{\mathbf j}} &=& - \frac{1}{18} \theta^\alpha_{({\mathbf i}}
{\bar \theta}^{3\dot\alpha}_{{\mathbf j})} \left[
  a_{\alpha\dot\alpha} - \omega_{\alpha\beta}x^\beta_{-\,\,\dot\alpha} +
{\bar\omega}_{\dot\alpha\dot\beta}
x^{\dot\beta}_{+\,\,\alpha} + \kappa x_{\alpha\dot\alpha}
\right] \nonumber \\
&&-\frac{i}{9} \varepsilon^\alpha_{({\mathbf i}} \theta_{\alpha {\mathbf j})}
{\bar\theta}^4
+ \frac{2i}{9} {\bar\varepsilon}^{\dot\alpha}_{\mathbf k}
\theta_{{\mathbf i}{\mathbf j}}
 {\bar\theta}^{3{\mathbf k}}_{\dot\alpha}
+ \frac{1}{18} \eta \theta_{{\mathbf i}{\mathbf j}} {\bar\theta}^4 +
\frac{i}{8} \eta_{({\mathbf i}{\mathbf k}} \theta^{\mathbf k}{}_{{\mathbf j})}
{\bar\theta}^4\nonumber\\
&&-\frac{1}{18} \theta^\alpha_{({\mathbf i}}
{\bar\theta}^{3\dot\alpha}_{{\mathbf j})} x_{-\dot\alpha\beta}
b^{\beta\dot\beta} x_{+\dot\beta\alpha} - \frac{i}{36}
\theta_{{\mathbf i}{\mathbf j}} {\bar\theta}^4 x_{+\alpha\dot\alpha}
b^{\alpha\dot\alpha}\label{eq313} \\
&&- \frac{5}{144} {\bar\theta}^3_{\dot\alpha ({\mathbf i}}
\theta_{{\mathbf j}){\mathbf k}} x_-^{\dot\alpha\beta} \rho_\beta^{\mathbf k} +
\frac{1}{144} {\bar\theta}^3_{\dot\alpha {\mathbf k}}
\theta_{({\mathbf i}}^{\,\,\,\,\,{\mathbf k}} x_-^{\dot\alpha\beta}
\rho_{\beta {\mathbf j})} - \frac{1}{36}
\theta^\alpha_{({\mathbf i}} {\bar\theta}^4
{\bar\rho}_{\dot\alpha {\mathbf j})} x^{\dot\alpha}_{+\alpha}.\nonumber
\end{eqnarray}
For a global superconformal transformation we then have
\begin{eqnarray}
H = {\bar H} &=& -\frac{1}{2} \theta^{\alpha {\mathbf i}}
{\bar\theta}^{\dot\alpha}_{\mathbf i} \left[ a_{\alpha\dot\alpha}
- \omega_{\alpha\beta}x^\beta_{-\,\,\dot\alpha} +
{\bar\omega}_{\dot\alpha\dot\beta}
x^{\dot\beta}_{+\,\,\alpha} + \kappa
x_{\alpha\dot\alpha}
\right]\nonumber \\
&+& \frac{4i}{3} \varepsilon^\alpha_{\mathbf i} \theta_{\alpha {\mathbf j}}
{\bar\theta}^{{\mathbf i}{\mathbf j}} + \frac{4i}{3}
{\bar\varepsilon}^{\dot\alpha}_{\mathbf i}
{\bar\theta}_{\dot\alpha {\mathbf j}}
\theta^{{\mathbf i}{\mathbf j}} -\frac{2}{3} \eta
\theta_{{\mathbf i}{\mathbf j}}
{\bar\theta}^{{\mathbf i}{\mathbf j}} -\frac{3 i}{2}
\eta_{{\mathbf i}{\mathbf k}} \theta^{\mathbf k}_{\,\,{\mathbf j}}
{\bar\theta}^{{\mathbf i}{\mathbf j}} \\
&-& \frac{1}{2} \theta^{\alpha {\mathbf i}}
{\bar\theta}^{\dot\alpha}_{\mathbf i} x_{-\dot\alpha\beta}
b^{\beta\dot\beta} x_{+\dot\beta\alpha}
+ \frac{1}{3} {\bar\theta}_{\dot\alpha {\mathbf i}}
\theta^{{\mathbf i}{\mathbf j}}
\rho_{\alpha {\mathbf j}} x_-^{\alpha\dot\alpha} + \frac{1}{3}
\theta^\alpha_{\mathbf i} {\bar\theta}^{{\mathbf i}{\mathbf j}}
{\bar\rho}_{\dot\alpha{\mathbf j}}
x^{\dot\alpha}_{+\alpha}\ .\nonumber
 \label{eq120}
\end{eqnarray}

\section{Multiplet of Currents for the ${\cal N}=2$  Vector}
\label{eq608}

In this appendix, we first give some details
how to obtain the variation~(\ref{eq402}) of the supersymmetry
current for the $N=2$ vector multiplet. Then, we summarize
the multiplet structure in the general case and in the improved case.

\subsection{Variation of the Supersymmetry Current}
\label{eq608b}

We are interested in the computation of $D_{\alpha {\mathbf i}}
J^\mu_{\beta {\mathbf j}}$
where $J^\mu_{\alpha {\mathbf i}}$ is given by eq.(\ref{eq307}) {\em i.e.}
\begin{equation}
J_\alpha^{\mu {\mathbf i}} = 192 \left[
 i \sigma^\mu_{\alpha\dot\alpha} {\bar
  D}^{\dot\alpha}_{\mathbf j}  D^{{\mathbf i}{\mathbf j}} T
- 3 i
{\bar\sigma}^{\mu\dot\alpha\beta} {\bar D}^{\mathbf i}_{\dot\alpha}
D_{\alpha\beta} T
- 12  \partial^\mu D^{\mathbf i}_\alpha T
+   a \sigma_\alpha^{\mu\nu\,\beta} \partial_\nu D_\beta^{\mathbf i}
T \right]. \label{eq707}
\end{equation}
 The method consists of course of decomposing
 all the terms into the symmetric and antisymmetric parts with respect
 to the $SU(2)_R$ and spinor indices.

For the first term in the r.h.s. of~(\ref{eq707}), we use
its equivalent form in terms of ${\bar{\cal J}}$ and the chirality of
${\bar{\cal J}}$ to get:
\begin{eqnarray}
D_{\alpha {\mathbf i}} \left[ i \sigma^\mu_{\beta\dot\alpha} {\bar
  D}^{\dot\alpha}_{\mathbf k}  D_{\mathbf j}^{\,\,\,
{\mathbf k}} T\right] &=& 3i \varepsilon_{{\mathbf i}{\mathbf j}}
\varepsilon_{\alpha\beta}
{\bar\sigma}^{\mu\nu}_{\dot\alpha\dot\beta} \partial_\nu
{\bar D}^{\dot\alpha\dot\beta} {\bar{\cal J}} - 3i
\varepsilon_{{\mathbf i}{\mathbf j}}
\sigma^\mu_{(\alpha\dot\alpha}
\sigma^\nu_{\beta)\dot\beta} \partial_\nu {\bar
  D}^{\dot\alpha\dot\beta} {\bar{\cal J}} \nonumber \\
&+& 3i \varepsilon_{\alpha\beta} \partial^\mu {\bar D}_{{\mathbf i}{\mathbf j}}
{\bar{\cal J}} - 6 i \sigma^{\mu\nu}_{\alpha\beta}
\partial_\nu {\bar D}_{{\mathbf i}{\mathbf j}} {\bar{\cal J}}. \label{eq625}
\end{eqnarray}
For the second term in the r.h.s. of~(\ref{eq707}), we
use the relation
\begin{eqnarray}
D_{\alpha {\mathbf i}} {\bar D}_{\dot\alpha {\mathbf j}} D_{\beta\gamma} T
&=& - 2i \varepsilon_{{\mathbf i}{\mathbf j}} \partial_{\alpha\dot\alpha}
D_{\beta\gamma} T + \frac{2}{3} \varepsilon_{\alpha(\beta}
{\bar D}_{\dot\alpha {\mathbf j}} D^3_{\gamma){\mathbf i}} T, \label{eq634}\\
&=&  - 2i \varepsilon_{{\mathbf i}{\mathbf j}} \partial_{\alpha\dot\alpha}
D_{\beta\gamma} T - 2
\varepsilon_{{\mathbf i}{\mathbf j}}
\varepsilon_{\alpha(\beta} \partial_{\gamma)}^{\dot\beta}
{\bar D}_{\dot\alpha\dot\beta} {\bar{\cal J}} - 2
\varepsilon_{\alpha(\beta} \partial_{\gamma)\dot\alpha}
{\bar D}_{{\mathbf i}{\mathbf j}} {\bar {\cal J}}, \label{eq620}
\end{eqnarray}
where we have used the equation of conservation~(\ref{eq121}). This enables to
get:
\begin{eqnarray}
D_{\alpha {\mathbf i}} \left[ - 3 i
{\bar\sigma}^{\mu\dot\alpha\gamma} {\bar D}_{\dot\alpha {\mathbf j}}
D_{\beta\gamma} T \right] &=& \varepsilon_{{\mathbf i}{\mathbf j}}
\varepsilon_{\alpha\beta} \left[ 6
  \sigma^{\mu\nu}_{\gamma\delta} D^{\gamma\delta} T + 9i
  {\bar\sigma}^{\mu\nu}_{\dot\alpha\dot\beta} {\bar
    D}^{\dot\alpha\dot\beta} {\bar{\cal J}} \right] \nonumber \\
&+& \varepsilon_{{\mathbf i}{\mathbf j}} \left[  6 \partial^\mu
  D_{\alpha\beta} T +12
  \sigma^{\mu\nu\,\,\,\,\gamma}_{(\alpha} \partial_\nu
  D_{\beta)\gamma} T + 3i \sigma^\mu_{(\alpha\dot\alpha}
  \sigma^\nu_{\beta)\dot\beta} \partial_\nu {\bar
    D}^{\dot\alpha\dot\beta} {\bar{\cal J}} \right]\nonumber \\
&-& 9i \varepsilon_{\alpha\beta} \partial^\mu {\bar D}_{{\mathbf i}{\mathbf j}}
{\bar{\cal J}} - 6i \sigma^{\mu\nu}_{\alpha\beta}
\partial_\nu {\bar D}_{{\mathbf i}{\mathbf j}} {\bar{\cal J}}. \label{eq621}
\end{eqnarray}
For the third and the last terms in the r.h.s. of~(\ref{eq707}) we immediately
get:
\begin{eqnarray}
D_{\alpha {\mathbf i}} \left[ - 12  \partial^\mu D_{\beta {\mathbf j}}  T
\right] &=&
-6 \varepsilon_{{\mathbf i}{\mathbf j}} \partial^\mu D_{\alpha\beta} T - 6
\varepsilon_{\alpha\beta} \partial^\mu D_{{\mathbf i}{\mathbf j}}T \label
{eq633} \\
&=& -6
\varepsilon_{{\mathbf i}{\mathbf j}} \partial^\mu D_{\alpha\beta} T + 6i
\varepsilon_{\alpha\beta} \partial^\mu {\bar
  D}_{{\mathbf i}{\mathbf j}}{\bar{\cal J}}\ , \label{eq622} \\
\nonumber \\
D_{\alpha {\mathbf i}} \left[ a \sigma_\beta^{\mu\nu\,\gamma}
  \partial_\nu D_{\gamma {\mathbf j}} T\right] &=&  \frac{a}{2} \partial_\nu
\left[
 -\frac{1}{2}\varepsilon_{{\mathbf i}{\mathbf j}} \varepsilon_{\alpha\beta}
  \sigma^{\mu\nu}_{\gamma\delta} D^{\gamma\delta} T +
  \varepsilon_{{\mathbf i}{\mathbf j}}
  \sigma^{\mu\nu\,\,\,\,\gamma}_{(\alpha}
  D_{\beta)\gamma} T +
  \sigma^{\mu\nu}_{\alpha\beta} D_{{\mathbf i}{\mathbf j}}T\right] \label
{eq623}.
\end{eqnarray}
Finally, taking the sum of~(\ref{eq625}),  (\ref{eq621}),
(\ref{eq622}) and of~(\ref{eq623}) leads to
the result~(\ref{eq402}).

\subsection{Multiplet in the General Case}
\label{eq609}

The multiplet is formed of
\begin{eqnarray*}
t &=&  T|\ ,\\
\chi_{\alpha {\mathbf i}} &=&  D_{\alpha {\mathbf i}} T|\ ,\\
u_{\alpha\beta} &=&  D_{\alpha\beta}T|\ ,\\
{\cal R}^\mu &=& - 48  \sigma^\mu_{\alpha\dot\alpha}
[ D^{\alpha {\mathbf i}} , {\bar D}^{\dot\alpha}_{\mathbf i} ] T |\ ,\\
R^\mu_{{\mathbf i}{\mathbf j}} &=& - 864 i \sigma^\mu_{\alpha\dot\alpha} [
D^\alpha_{({\mathbf i}}, {\bar D}^{\dot\alpha}_{{\mathbf j})} ] T |\ ,\\
J_\alpha^{\mu {\mathbf i}} &=& 192 \left[ i
 \sigma^\mu_{\alpha\dot\alpha} {\bar
  D}^{\dot\alpha}_{\mathbf j}  D^{{\mathbf i}{\mathbf j}} T|
- 3 i
{\bar\sigma}^{\mu\dot\alpha\beta} {\bar D}^{\mathbf i}_{\dot\alpha}
D_{\alpha\beta} T|
- 12  \partial^\mu \chi^{\mathbf i}_\alpha
+  a \sigma_\alpha^{\mu\nu\,\beta} \partial_\nu \chi_\beta^{\mathbf i}
 \right]\ ,\\
T^{\mu\nu} &=& -24 \displaystyle{\Biggl[}
  \frac{1}{2} \eta^{\mu\nu} \{ D^{{\mathbf i}{\mathbf j}} ,
{\bar D}_{{\mathbf i}{\mathbf j}} \}
  T|
- \frac{3}{2} {\bar \sigma}^{\mu \dot\alpha\alpha}
  {\bar\sigma}^{\nu\dot\beta\beta} \{ D_{\alpha\beta} ,
  {\bar D}_{\dot\alpha\dot\beta} \} T|
\\ &&
+48 \eta^{\mu\nu} \hspace{0.2ex}\raisebox{.5ex}{\fbox{}}\hspace{0.3ex} t
 - b ( \eta^{\mu\nu}
\hspace{0.2ex}\raisebox{.5ex}{\fbox{}}\hspace{0.3ex} -  \partial^\mu
\partial^\nu ) t \displaystyle{\Biggl]}\ , \nonumber \\
Z^{\mu\nu} &=& 96 \left[
{\bar\sigma}^{\mu\nu}_{\dot\alpha\dot\beta}
{\bar D}^{\dot\alpha\dot\beta} {\bar{\cal J}}| - i (\frac{1}{2} -
\frac{a}{48}) \sigma^{\mu\nu}_{\alpha\beta}
u^{\alpha\beta}\right] \ , \\
{\bar{\cal J}}|\ ,  \\
{\bar D}_{\dot \alpha {\mathbf i}} {\bar{\cal J}}|\ , \\
{\bar D}_{{\mathbf i}{\mathbf j}}{\cal J}|.
\end{eqnarray*}
The transformations of these components are
\begin{eqnarray*}
\delta_{\alpha {\mathbf i}} t &=& \chi_{\alpha {\mathbf i}}\ ,\\
\delta_{\alpha {\mathbf i}} \chi_{\beta {\mathbf j}} &=& \frac{1}{2}
\varepsilon_{{\mathbf i}{\mathbf j}} u_{\alpha\beta} - \frac{i}{2}
\varepsilon_{\alpha\beta} {\bar D}_{{\mathbf i}{\mathbf j}} {\bar{\cal J}}\ ,\\
{\bar\delta}_{\dot\alpha {\mathbf i}} \chi_{\alpha {\mathbf j}} &=&
\frac{i}{3456} R_{\alpha\dot\alpha {\mathbf i}{\mathbf j}} + \frac{1}{384}
\varepsilon_{{\mathbf i}{\mathbf j}}
{\cal R}_{\alpha\dot\alpha} +  i \varepsilon_{{\mathbf i}{\mathbf j}}
\partial_{\alpha\dot\alpha} t\ ,\\
\delta_{\alpha {\mathbf i}} u_{\beta\gamma} &=& 4
\varepsilon_{\alpha(\beta} \partial_{\gamma) \dot\alpha}
{\bar D}^{\dot\alpha}_{\mathbf i} {\bar{\cal J}}\ ,  \\
{\bar \delta}_{\dot\alpha {\mathbf i}} u_{\alpha\beta} &=&
\frac{-i}{1152} \sigma_{\mu(\alpha\dot\alpha}
J^{\mu}_{\beta){\mathbf i}} + \frac{i}{12} (a-24)
\partial_{(\alpha \dot\alpha}
\chi_{\beta){\mathbf i}}\ ,  \\
\delta_{\alpha {\mathbf i}} R^\mu_{{\mathbf j}{\mathbf k}} &=& \frac{3}{2}
\varepsilon_{{\mathbf i}({\mathbf j}} \left[ J^\mu_{\alpha {\mathbf k})} - (24
+ a)
\sigma^{\mu\nu\,\,\,\beta}_\alpha \partial_\nu \chi_{\beta
{\mathbf k})}   \right]\ ,\\
\delta_{\alpha {\mathbf i}} {\cal R}^\mu &=& 16i \left\{
  \frac{1}{192} \left[  \frac{3}{2} J^\mu_{\alpha {\mathbf i}} +
    \sigma_\alpha^{\mu\nu\beta} J_{\nu\beta{\mathbf i}}  \right] -
  \frac{3}{4}(8+a) \partial^\mu \chi_{\alpha{\mathbf i}} +
  \frac{1}{2} (a-24) \sigma_\alpha^{\mu\nu\beta}
  \partial_\nu \chi_{\beta{\mathbf i}}  \right\}\ , \\
{\bar\delta}_{\dot\alpha}^{\mathbf i} J^{\mu {\mathbf j}}_\alpha &=&
\varepsilon^{{\mathbf i}{\mathbf j}} \sigma_{\nu\alpha\dot\alpha} \left[ -2i
T^{\mu\nu} + 48 i (b-48+2a) (\eta^{\mu\nu} \hspace{0.2ex}\raisebox{.5ex}
{\fbox{}}\hspace{0.3ex} -
\partial^\mu\partial^\nu) t \, \right] \nonumber \\
&+& \varepsilon^{{\mathbf i}{\mathbf j}} \left\{ \frac{1}{4} \left[a+24\right]
\left[
\sigma^\mu_{\alpha\dot\alpha}
\partial^\rho {\cal R}_\rho
- \partial_{\alpha\dot\alpha} {\cal R}^\mu \right]
 - \frac{ia}{4} \varepsilon^{\mu\nu\rho\tau}
\sigma_{\tau\alpha\dot\alpha} \partial_\nu {\cal R}_\rho
 \right\}
{\nonumber}\\
&+&\frac{i}{36}(a-24) \partial_{\alpha\dot\alpha}
R^{\mu \, {\mathbf i}{\mathbf j}}
-\frac{1}{36} (a+24) \varepsilon^{\mu\nu\rho\tau}
\sigma_{\tau\alpha\dot\alpha} \partial_\nu R_\rho^{{\mathbf i}{\mathbf j}}
\ , \\
\delta_{\alpha {\mathbf i}} J^\mu_{\beta {\mathbf j}} &=& \partial_\nu \left\{
24 i \varepsilon_{\alpha\beta} \varepsilon_{{\mathbf i}{\mathbf j}}
Z^{\mu\nu}
 + 96  ( a +24) \left[ \varepsilon_{{\mathbf i}{\mathbf j}}
\sigma^{\mu\nu\,\gamma}_{(\alpha}
\chi_{\beta)\gamma}
- i  \sigma^{\mu\nu}_{\alpha\beta}
{\bar D}_{{\mathbf i}{\mathbf j}} {\bar{\cal J}} \right] \right \}\ , \\
\delta_{\alpha {\mathbf i}} T^{\mu\nu} &=&
\sigma^{(\mu\rho\,\,\beta}_{\;\;\;\alpha} \partial_\rho
J^{\nu)}_{\beta{\mathbf i}}
+ 24  (b-48 + 2a) (\eta^{\mu\nu}  \hspace{0.2ex}\raisebox{.5ex}
{\fbox{}}\hspace{0.3ex} -
\partial^\mu\partial^\nu) \chi_{\alpha{\mathbf i}}\ , \\
\delta_{\alpha {\mathbf i}} Z^{\mu\nu} &=& - 96 i (1+\frac{a}{24})
(\sigma^\mu_{\alpha\dot\alpha} \partial^\nu -
\sigma^\nu_{\alpha\dot\alpha} \partial^\mu )
{\bar D}^{\dot\alpha}_{\mathbf i} {\bar{\cal J}}
+ 96  (3-\frac{a}{24})\
\varepsilon^{\mu\nu\rho\tau} \sigma_{\tau \alpha\dot\alpha}
\partial_\rho {\bar D}^{\dot\alpha}_{\mathbf i} {\bar{\cal J}}
\ ,\\
{\bar\delta}_{\dot\alpha {\mathbf i}}Z^{\mu\nu} &=&
(\eta^{\rho\mu}\sigma^\nu_{\alpha\dot\alpha}
-\eta^{\rho\nu}\sigma^\mu_{\alpha\dot\alpha})
\left( \frac{1}{16} J^\alpha_{\rho {\mathbf i}}
+[\frac{1}{12}(a-24)^2+12(8+a)] \partial_\rho\chi_{\mathbf i}^\alpha\right)
\\ &&
+i\ \varepsilon^{\mu\nu\rho\tau} \sigma_{\tau\alpha\dot\alpha}
\left( -\frac{1}{1152}(a+24)\ J^\alpha_{\rho {\mathbf i}}
+[\frac{1}{12}(a-24)^2-12(8+a)]\partial_\rho\chi_{\mathbf i}^\alpha\right)
\ ,
\nonumber \\
\delta_{\alpha {\mathbf i}} { \bar{\cal J}} &=& 0\ ,  \\
{\bar\delta}_{\dot\alpha {\mathbf i}} {\bar{\cal J}}  &=& {\bar D}_{\dot
\alpha {\mathbf i}} {\bar {\cal J}}\ ,  \\
\delta_{\alpha {\mathbf i}}( {\bar D}_{\dot\alpha {\mathbf j}}{\bar{\cal J}})
&=& - 2 i
\varepsilon_{{\mathbf i}{\mathbf j}} \partial_{\alpha\dot\alpha}
{\bar {\cal J}}
\ , \\
{\bar \delta}_{\dot\alpha {\mathbf i}}  ({\bar D}_{\dot\beta {\mathbf j}}
{\bar{\cal J}})  &=&
-\frac{1}{2} \varepsilon_{\dot\alpha\dot\beta}
{\bar D}_{{\mathbf i}{\mathbf j}} {\bar{\cal J}}
+ \varepsilon_{{\mathbf i}{\mathbf j}} \frac{1}{384}
{\bar{\sigma}}^{\mu\nu}_{\dot\alpha\dot\beta} Z_{\mu\nu}\ , \\
\delta_{\alpha {\mathbf i}} ({\bar D}_{{\mathbf j}{\mathbf k}} {\bar{\cal J}}
)&=&
- 4i \varepsilon_{{\mathbf i}({\mathbf j}} \partial_{\alpha\dot\alpha}
{\bar D}^{\dot\alpha}_{{\mathbf k})}
{\bar {\cal J}}\ ,  \\
{\bar \delta}_{\dot\alpha {\mathbf i}}( {\bar D}_{{\mathbf j}{\mathbf k}}
{\bar{\cal
    J}} ) &=&
\frac{1}{6} \varepsilon_{{\mathbf i}({\mathbf j}} \left[ \frac{1}{192}
({\bar\sigma}^\mu J_{\mu\, {\mathbf k)}})_{\dot\alpha} - \frac{3}{2}
(8+a) \partial_{\alpha\dot\alpha} \chi^\alpha_{{\mathbf k})}
\right].
\end{eqnarray*}

\subsection{Improved Multiplet}
\label{eq504}

The multiplet of improved currents is composed of:
\begin{eqnarray*}
t &=& T|\ ,\\
\chi_{\alpha {\mathbf i}} &=& D_{\alpha {\mathbf i}} T|\ ,\\
u_{\alpha\beta} &=& D_{\alpha\beta} T|\ ,\\
{\cal R}^\mu &=& - 48  \sigma^\mu_{\alpha\dot\alpha}
[ D^{\alpha {\mathbf i}} , {\bar D}^{\dot\alpha}_{\mathbf i} ] T |\ ,\\
R^\mu_{{\mathbf i}{\mathbf j}} &=& - 864 i \sigma^\mu_{\alpha\dot\alpha} [
D^\alpha_{({\mathbf i}}, {\bar D}^{\dot\alpha}_{{\mathbf j})} ] T |\ ,\\
J^{\mu {\mathbf i}}_{\alpha} &=& 192 \left[
- 12  \partial^\mu \chi_\alpha^{\mathbf i}
- 3 i {\bar\sigma}^{\mu\dot\alpha\beta}
{\bar D}^{\mathbf i}_{\dot\alpha} D_{\alpha\beta} T |
-  8 \sigma_\alpha^{\mu\nu\,\beta} \partial_\nu \chi_\beta^{\mathbf i}
 \right]\ ,\\
T^{\mu\nu} &=& -24  \left[
-\frac{3 }{2} {\bar \sigma}^{\mu \dot\alpha\alpha}
  {\bar\sigma}^{\nu\dot\beta\beta} \{ D_{\alpha\beta} ,
  {\bar D}_{\dot\alpha\dot\beta} \} T|
+ 48 \eta^{\mu\nu} \hspace{0.2ex}\raisebox{.5ex}{\fbox{}}\hspace{0.3ex} t
- 64 ( \eta^{\mu\nu}
\hspace{0.2ex}\raisebox{.5ex}{\fbox{}}\hspace{0.3ex} -  \partial^\mu
\partial^\nu ) t \right].
\end{eqnarray*}
The transformation properties of these components are
given by:
\begin{eqnarray*}
\delta_{\alpha {\mathbf i}} t &=& \chi_{\alpha {\mathbf i}}\ , \\
\delta_{\alpha {\mathbf i}} \chi_{\beta {\mathbf j}} &=& \frac{1}{2}
\varepsilon_{{\mathbf i}{\mathbf j}}u_{\alpha\beta}\ ,\\
{\bar \delta}_{\dot\alpha {\mathbf i}} \chi_{\alpha {\mathbf j}} &=&
\frac{i}{3456} R_{\alpha\dot\alpha {\mathbf i}{\mathbf j}} + \frac{1}{384
}\varepsilon_{{\mathbf i}{\mathbf j}} {\cal R}_{\alpha\dot
\alpha} + i \varepsilon_{{\mathbf i}{\mathbf j}}
\partial_{\alpha\dot\alpha} t\ ,\\
\delta_{\alpha {\mathbf i}} u_{\beta\gamma} &=&0\ ,\\
{\bar \delta}_{\dot\alpha}^{\mathbf i} u_{\beta\gamma} &=&
-\frac{i}{1152} \sigma_{\mu(\beta\dot\alpha} J^{\mu {\mathbf i}}_{\gamma)}
- \frac{8i}{3} \partial_{(\beta \dot\alpha}
\chi^{\mathbf i}_{\gamma)}\ , \\
\delta_{\alpha {\mathbf i}} {\cal R}^\mu &=& 16 i \left[ \frac{1}{192}
J^{\mu}_{\alpha {\mathbf i}} - 16
\sigma^{\mu\nu\,\,\beta}_{\;\;\alpha} \partial_\nu
\chi_{\beta {\mathbf i}} \right] \ ,\\
\delta_{\alpha {\mathbf i}} R^\mu_{{\mathbf j}{\mathbf k}} &=& \frac{3}{2}
\varepsilon_{{\mathbf i}({\mathbf j}} \left[ J^\mu_{\alpha {\mathbf k})} - 16
\sigma^{\mu\nu\,\,\,\beta}_\alpha \partial_\nu
\chi_{\beta {\mathbf k})}\right]\ ,\\
{\bar\delta}_{\dot\alpha}^{\mathbf i} J^{\mu {\mathbf j}}_\alpha &=&
\varepsilon^{{\mathbf i}{\mathbf j}} \left[ -2i
\sigma_{\nu\alpha\dot\alpha} T^{\mu\nu}
- 4 \partial_{\alpha\dot\alpha} {\cal R}^\mu
 + 2i \varepsilon^{\mu\nu\rho\tau}
\sigma_{\tau\alpha\dot\alpha} \partial_\nu {\cal R}_\rho
\right] \\
&-&\frac{8i}{9} \partial_{\alpha\dot\alpha} R^{\mu {\mathbf i} {\mathbf j}}
-\frac{4i}{9} \varepsilon^{\mu\nu\rho\tau}
\sigma_{\tau\alpha\dot\alpha} \partial_\nu R_\rho^{{\mathbf i}{\mathbf j}}\ ,\\
\delta_{\alpha {\mathbf i}} J^\mu_{\beta {\mathbf j}} &=&
1536 \varepsilon_{{\mathbf i}{\mathbf j}}  \partial_\nu \left[
\varepsilon_{\alpha\beta}
\sigma^{\mu\nu}_{\gamma\delta}
u^{\gamma\delta}  +
\sigma^{\mu\nu\,\gamma}_{(\alpha}
u_{\beta)\gamma}\right]\ ,\\
\delta_{\alpha {\mathbf i}} T^{\mu\nu} &=&
\sigma^{(\mu\rho\,\,\beta}_{\;\;\;\alpha} \partial_\rho
J^{\nu)}_{\beta{\mathbf i}}\ .
\end{eqnarray*}

\section{$SU(2)_R$ Invariance for the ${\cal N}=2$ Tensor Multiplet}
\label{appendixB}

In this appendix we sketch the proof of the identity~(\ref{eq114})
\begin{equation}
\left( D^{({\mathbf m}}{}_{{\mathbf j}} {\bar D}_{{\mathbf k}{\mathbf l}}
+ {\bar D}^{({\mathbf m}}{}_{{\mathbf j}}  D_{{\mathbf k}{\mathbf l}}\right)
T^{{\mathbf i}){\mathbf j}{\mathbf k}{\mathbf l}}
=
 - 8 i\partial_{\alpha\dot\alpha}
[D^\alpha_{\mathbf k}, {\bar D}^{\dot\alpha}_{\mathbf l}]
T^{{\mathbf m}{\mathbf i}{\mathbf k}{\mathbf l}}. \label{eq2000}
\end{equation}
The starting point is to notice that
\begin{equation}
 T^{{\mathbf i}{\mathbf j}{\mathbf k}{\mathbf l}} = \frac{1}{160}
F^{({\mathbf i}{\mathbf j}} F^{{\mathbf k}{\mathbf l})} =
D^{({\mathbf i}{\mathbf j}} X^{{\mathbf k}{\mathbf l})} + c.c. \quad
\mbox{with} \quad X^{{\mathbf i}{\mathbf j}} \equiv \frac{i}{160} \Phi
F^{{\mathbf i}{\mathbf j}}.\label{eq101}
\end{equation}
As $\Phi$ is chiral,  $X^{{\mathbf i}{\mathbf j}}$ has clearly the same
properties~(\ref{eq111}) than $F^{{\mathbf i}{\mathbf j}}$.
The first step consists in writing $D^{{\mathbf m}}{}_{\mathbf j}
{\bar D}_{{\mathbf k}{\mathbf l}} = [
D^{{\mathbf m}}{}_{{\mathbf j}},  {\bar D}_{{\mathbf k}{\mathbf l}}] +
{\bar D}_{{\mathbf k}{\mathbf l}}
D^{{\mathbf m}}{}_{{\mathbf j}}$. After some algebra we find
\begin{eqnarray}
[D^{{\mathbf m}}{}_{{\mathbf j}},  {\bar D}_{{\mathbf k}{\mathbf l}}]
D^{({\mathbf i}{\mathbf j}} X^{{\mathbf k}{\mathbf l})} &=& -
4i \partial_{\alpha\dot\alpha} D^\alpha_{\mathbf k} {\bar
  D}^{\dot\alpha}_{\mathbf l} D^{({\mathbf i}{\mathbf m}} X^{{\mathbf k}
{\mathbf l})}, \label{eq2001}\\
{\bar D}_{{\mathbf k}{\mathbf l}} D^{{\mathbf m}}{}_{{\mathbf j}}
D^{({\mathbf i}{\mathbf j}}X^{{\mathbf k}{\mathbf l})} &=&
 \frac{5}{36} \varepsilon^{{\mathbf m}{\mathbf i}}
{\bar D}_{{\mathbf j}{\mathbf k}} D^4 X^{{\mathbf j}{\mathbf k}} +
\frac{5}{18} {\bar D}^{({\mathbf m}}{}_{{\mathbf j}}D^4
X^{{\mathbf i}){\mathbf j}}.\label{eq2002}
\end{eqnarray}
On the other hand,
\begin{equation}
 {\bar D}^{{\mathbf m}}{}_{{\mathbf j}}  D_{{\mathbf k}{\mathbf l}}
D^{({\mathbf i}{\mathbf j}} X^{{\mathbf k}{\mathbf l})} =
 \frac{5}{18} \varepsilon^{{\mathbf m}{\mathbf i}}
{\bar D}_{{\mathbf j}{\mathbf k}} D^4 X^{{\mathbf j}{\mathbf k}}
+    \frac{5}{9} {\bar D}^{({\mathbf m}}{}_{{\mathbf j}} D^4
X^{{\mathbf i}){\mathbf j}}.  \label{eq2003}
\end{equation}
 Adding~(\ref{eq2001}), (\ref{eq2002}) and (\ref{eq2003}), we find
\begin{equation}
 \left( D^{{\mathbf m}}{}_{{\mathbf j}} {\bar D}_{{\mathbf k}{\mathbf l}}
+ {\bar D}^{{\mathbf m}}{}_{{\mathbf j}}  D_{{\mathbf k}{\mathbf l}}\right)
D^{({\mathbf i}{\mathbf j}} X^{{\mathbf k}{\mathbf l})}
=
\begin{array}[t]{l}
 \frac{5}{12} \varepsilon^{{\mathbf m}{\mathbf i}}
{\bar D}_{{\mathbf j}{\mathbf k}}D^4X^{{\mathbf j}{\mathbf k}} +
\frac{5}{6} {\bar D}^{({\mathbf m}}{}_{{\mathbf j}} D^4
X^{{\mathbf i}){\mathbf j}}
\\[3mm]
- 4i\partial_{\alpha\dot\alpha} D^\alpha_{\mathbf k} {\bar
  D}^{\dot\alpha}_{\mathbf l} D^{({\mathbf i}{\mathbf m}} X^{{\mathbf k}
{\mathbf l})}\ .
\end{array}
\label{eq116}
\end{equation}

For the second step, we use the fact that
$$
{\bar D}^{({\mathbf m}}{}_{{\mathbf j}} D^4 X^{{\mathbf i}){\mathbf j}} = [
{\bar
  D}^{({\mathbf m}}{}_{{\mathbf j}} , D^4 ] X^{{\mathbf i}){\mathbf j}} \ ,
$$
since the other term vanishes due to the properties~(\ref{eq111}) of
$X^{{\mathbf i}{\mathbf j}}$.
Again, we find after some algebra:
$$
[ {\bar
  D}^{({\mathbf m}}{}_{{\mathbf j}} , D^4 ] X^{{\mathbf i})
{\mathbf j}}  = 48 \hspace{0.2ex}\raisebox{.5ex}{\fbox
{}}\hspace{0.3ex}
D^{({\mathbf m}}{}_{{\mathbf j}}  X^{{\mathbf i}){\mathbf j}} + \frac{32i}{5}
\partial_{\alpha\dot\alpha} D^{3\alpha({\mathbf i}} {\bar
  D}^{\dot\alpha}_{\mathbf j} X^{{\mathbf m}){\mathbf j}}.
$$
However, a direct computation shows also that
$$
\partial_{\alpha\dot\alpha} [ D^\alpha_{\mathbf k} , {\bar
  D}^{\dot\alpha}_{\mathbf l}] D^{({\mathbf m}{\mathbf i}} X^{{\mathbf k}
{\mathbf l})} = \frac{20 i}{3}
\hspace{0.2ex}\raisebox{.5ex}{\fbox{}}\hspace{0.3ex} D^{({\mathbf m}}
{}_{{\mathbf j}}  X^{{\mathbf i}){\mathbf j}}
- \frac{40}{27} \partial_{\alpha\dot\alpha}
D^{3\alpha({\mathbf i}} {\bar
  D}^{\dot\alpha}_{\mathbf j} X^{{\mathbf m}){\mathbf j}}.
$$

Thus we have proved that
\begin{equation}
{\bar D}^{({\mathbf m}}{}_{{\mathbf j}} D^4 X^{{\mathbf i}){\mathbf j}} = -
\frac{36i}{5}
\partial_{\alpha\dot\alpha} [ D^\alpha_{\mathbf k} , {\bar
  D}^{\dot\alpha}_{\mathbf l}] D^{({\mathbf m}{\mathbf i}} X^{{\mathbf k}
{\mathbf l})}\ .
\label{eq115}
\end{equation}
To conclude, we put the results~(\ref{eq116})
and (\ref{eq115}) together. This leads to the final
relation:
$$
\left( D^{{\mathbf m}}{}_{{\mathbf j}} {\bar D}_{{\mathbf k}{\mathbf l}}
+ {\bar D}^{{\mathbf m}}{}_{{\mathbf j}}  D_{{\mathbf k}{\mathbf l}}\right)
D^{({\mathbf i}{\mathbf j}} X^{{\mathbf k}{\mathbf l})}
= \frac{5}{12} \varepsilon^{{\mathbf m}{\mathbf i}}
{\bar D}^{{\mathbf j}{\mathbf k}} D^4 X_{{\mathbf j}{\mathbf k}} - 8
i \partial_{\alpha\dot\alpha} [ D^\alpha_{\mathbf k} , {\bar
  D}^{\dot\alpha}_{\mathbf l}] D^{({\mathbf m}{\mathbf i}} X^{{\mathbf k}
{\mathbf l})} \ ,
$$
from which we get~(\ref{eq2000})  as a
consequence.

\section{Conventions and Identities for the SUSY Algebra}
\label{eq231}

The conventions used in this paper are essentially
those of Wess and Bagger~\cite{WessBagger}.
The conventions about covariant
derivatives and their algebra in ${\mathcal N}=1$ as well as in extended
supersymmetry are exposed in this appendix.

The following definitions are valid for any ${\mathcal N}$.
The spinorial derivatives are defined by
\begin{equation}
\partial_\beta^{\mathbf j} \theta^\alpha_{\mathbf i} \equiv \frac{\partial
\theta^\alpha_{\mathbf i}}{\partial \theta^\beta_{\mathbf j}} =
\delta^\alpha_\beta \delta^{\mathbf j}_{\mathbf i}
\ ,\qquad\quad
{\bar\partial}_{\dot\beta {\mathbf j}}{\bar \theta}^{\dot \alpha {\mathbf i}}
\equiv \frac{\partial {\bar\theta}^{\dot\alpha
{\mathbf i}}} {\partial {\bar\theta}_{\dot\beta {\mathbf j}}} =
\delta^{\dot\alpha}_{\dot\beta}
\delta^{\mathbf i}_{\mathbf j}\ .
\end{equation}
The covariant derivative are then defined as
\begin{equation}
D_\alpha^{\mathbf i} = \partial_\alpha^{\mathbf i} + i
{\bar\theta}^{\dot\alpha {\mathbf i}} \partial_{\alpha\dot\alpha}
\ ,\qquad\quad
{\bar D}_{\dot\alpha {\mathbf i}} = -{\bar\partial}_{\dot\alpha{\mathbf i}}
- i {\theta}^\alpha_{\mathbf i} \partial_{\alpha\dot\alpha}.
\end{equation}
Finally, the algebra following from these definitions is
\begin{equation}
\{D_\alpha^{\mathbf i}, {\bar D}_{\dot\alpha {\mathbf j}} \} = -2i\
\delta_{\mathbf j}^{\mathbf i}\ \partial_{\alpha\dot\alpha}
\qquad \mbox{and}\qquad
\{D_{\alpha}^{{\mathbf i}}, D_{\beta}^{{\mathbf j}} \} =0\ .
\end{equation}
\noindent
We furthermore use the following usual definition of $x^\mu_\pm\ $:
$\ x_\pm^{\alpha\dot\alpha} \equiv x^{\alpha\dot\alpha} \pm
2 i \theta^{\alpha i} {\bar\theta}^{\dot\alpha}_i$.
The main properties of $x^\mu_\pm\ $ with respect to the covariant derivatives
are:
\[
\begin{array}{c}
{\bar D}_{\alpha{\mathbf i}} x^\mu_+ =0\ , \quad D_\alpha^{\mathbf i} x^\mu_-
=0
\ , \\[3mm]
D_{\alpha {\mathbf i}} x_{+\beta\dot\beta} = 4i
\varepsilon_{\alpha\beta}{\bar\theta}_{\dot\beta {\mathbf i}}\ , \quad
{\bar D}_{\dot\alpha {\mathbf i}} x_{-\beta\dot\beta} =  - 4 i
\varepsilon_{\dot\alpha\dot\beta} \theta_{\beta {\mathbf i}}\ .
\end{array}
\]
The different integration measures are then defined by:
\[
\int d^{4+4{\mathcal N}}z  \equiv \int d^4x\ D^{2{\mathcal N}}\
{\bar D}^{2{\mathcal N}}\ ,
\quad \int d^{4+2{\mathcal N}}z_+ \equiv \int d^4x\ D^{2{\mathcal N}}\ ,
\quad \int d^{4+2{\mathcal N}}z_- \equiv \int d^4x\ {\bar D}^{2{\mathcal N}}\ ,
\]
where $D^{2{\mathcal N}}$ is defined below for the specific values $
{\mathcal N}=1,2$.

\subsection{${\cal N}=1$ Identities}

We give there a list of useful identities for ${\mathcal N}=1$ covariant
derivatives
following from the definitions and conventions given above.

\noindent
Definition of $D^2$:
\[
D^2\ \equiv\ D^\alpha D_\alpha\ ,
\qquad\qquad
{\bar D}^2\ \equiv {\bar D}_{\dot{\alpha}} {\bar D}^{\dot{\alpha}}\ .
\]
\noindent
Products of covariant derivatives:
\[
\begin{array}{c}
D^\alpha {\bar D}^2 D_\alpha  = {\bar D}_{\dot\alpha}  D^2
{\bar D}^{\dot\alpha}
\ , \\[3mm]
D^\alpha {\bar D}_{\dot{\alpha}} D_\alpha = -\frac{1}{2}
{\bar D}_{\dot{\alpha}} D^2-\frac{1}{2}
D^2{\bar D}_{\dot{\alpha}}
\ ,\qquad
{\bar D}_{\dot\alpha} D^\alpha {\bar D}^{\dot\alpha} =
-\frac{1}{2} D^\alpha {\bar D}^2-\frac{1}{2} {\bar D}^2 D^\alpha
\ ,\\[3mm]
\left[D^\alpha,{\bar D}^{\dot{\alpha}}\right]\left[D_\alpha,
{\bar D}_{\dot{\alpha}}\right]=
2\{D^2,{\bar D}^2\}-24\hspace{0.2ex}\raisebox{.5ex}{\fbox{}}\hspace{0.3ex}\ .
\end{array}
\]
Algebra of covariant derivatives:
\[
\begin{array}{c}
\left[D_\alpha,{\bar D}^2\right] = -4i\partial_{\alpha{\dot{\alpha}}}
{\bar D}^{\dot{\alpha}}
\ ,\qquad
\left[{\bar D}_{\dot{\alpha}},D^2\right] = 4iD^\alpha
\partial_{\alpha{\dot{\alpha}}}
\ ,\\[3mm]
\left[D^2,{\bar D}^2\right] = -4i[D^\alpha,{\bar D}^{\dot{\alpha}}]
\partial_{\alpha{\dot{\alpha}}}\ .
\end{array}
\]

\subsection{${\cal N}=2$ Identities}
\label{eq617}

\noindent
Raising and lowering of {$SU(2)$}\ indices and Fierz formula:
\[
\begin{array}{c}
a^{\mathbf i}=\varepsilon^{{\mathbf i}{\mathbf j}}a_{\mathbf j}\ ,\quad
a_{\mathbf i}=\varepsilon_{{\mathbf i}{\mathbf j}}a^{\mathbf j}\ ,\\[3mm]
\mbox{with }\;\;   \varepsilon^{{\mathbf i}{\mathbf j}}=-
\varepsilon^{{\mathbf j}{\mathbf i}}\ ,\quad
\varepsilon^{{\mathbf 1}{\mathbf 2}}=1\ ,\quad
\varepsilon_{{\mathbf i}{\mathbf j}}=-\varepsilon^{{\mathbf i}{\mathbf j}}\ ,
\quad \varepsilon^{{\mathbf i}{\mathbf j}}\varepsilon_{{\mathbf j}{\mathbf k}}=
\delta^{\mathbf i}_{\mathbf k}
\ .
\end{array}
\]
\[
a_{\mathbf i} b_{\mathbf j}-a_{\mathbf j} b_{\mathbf i} =
\varepsilon_{{\mathbf i}{\mathbf j}}a^{\mathbf k} b_{\mathbf k}\ ,\qquad
a^{\mathbf i} b^{\mathbf j}-a^{\mathbf j} b^{\mathbf i} =-
\varepsilon^{{\mathbf i}{\mathbf j}}a^{\mathbf k} b_{\mathbf k}\ .
\]
\noindent
Definitions of some products of $\theta$'s:
\[
\begin{array}{lc}
\theta^2\ :&
\theta^{{\mathbf i}{\mathbf j}}\equiv\theta^{\alpha{\mathbf i}}
\theta^{\mathbf j}_\alpha
\ ,\quad
\theta^{\alpha\beta}\equiv\theta^{\alpha{\mathbf i}}
\theta^{\beta}_{{\mathbf i}}
\ ,\quad
\bar\theta^{{\mathbf i}{\mathbf j}}\equiv\bar
\theta_{{\dot{\alpha}}}^{{\mathbf i}}\bar\theta^{{\dot{\alpha}}{\mathbf j}}
\ ,\quad
\bar\theta^{{\dot{\alpha}}{\dot{\beta}}}\equiv\bar
\theta^{{\dot{\alpha}}{\mathbf i}}\bar\theta^{{\dot{\beta}}}_{{\mathbf i}}
\ .
\\[3mm]
\theta^3\ :&
\theta^3_{\alpha{\mathbf i}}\equiv \theta^{\mathbf j}_\alpha
\theta_{{\mathbf i}{\mathbf j}}
=-\theta_{\mathbf i}^\beta\theta_{\alpha\beta}
\ ,\qquad
\bar\theta^{3{\mathbf i}}_{\dot{\alpha}}\equiv
-\bar\theta_{{\dot{\alpha}}{\mathbf j}}\bar\theta^{{\mathbf i}{\mathbf j}}=\bar
\theta^{{\dot{\beta}}{\mathbf i}}\bar\theta_{{\dot{\alpha}}{\dot{\beta}}}\ .
\\[3mm]
\theta^4\ :&
\theta^4\equiv\theta^{{\mathbf i}{\mathbf j}}\theta_{{\mathbf i}{\mathbf j}}=-
\theta^{\alpha\beta}\theta_{\alpha\beta}\ ,
\quad
\bar\theta^4\equiv\bar\theta^{{\mathbf i}{\mathbf j}}\bar
\theta_{{\mathbf i}{\mathbf j}}=-\bar\theta^{{\dot{\alpha}}{\dot{\beta}}}\bar
\theta_{{\dot{\alpha}}{\dot{\beta}}}\ .
\end{array}\label{eq9b}
\]
\noindent
Properties of the products of $\theta$'s:
\[
\begin{array}{l}
\theta_{\alpha{\mathbf i}}\theta_{\beta\gamma}=
-\frac{2}{3}\varepsilon_{\alpha(\beta}\theta^3_{\gamma){\mathbf i}}\ ,\quad
\theta_{\alpha{\mathbf i}}\theta_{{\mathbf j}{\mathbf k}}=
\frac{2}{3}\varepsilon_{{\mathbf i}({\mathbf j}}\theta^3_{\alpha{\mathbf k})}\
,\quad
\bar\theta_{{\dot{\alpha}}{\mathbf i}}\bar\theta_{{\dot{\beta}}{\dot{\gamma}}}=
\frac{2}{3}\varepsilon_{{\dot{\alpha}}({\dot{\beta}}}\bar
\theta^3_{{\dot{\gamma}}){\mathbf i}}\ ,\quad
\bar\theta_{{\dot{\alpha}}{\mathbf i}}\bar\theta_{{\mathbf j}{\mathbf k}}=
\frac{2}{3}\varepsilon_{{\mathbf i}({\mathbf j}}\bar
\theta^3_{{\dot{\alpha}}{\mathbf k})}\ ,
\\[3mm]
\theta^{\beta{\mathbf j}}\theta^3_{\alpha{\mathbf i}}=\frac{1}{4}
\delta^\beta_\alpha\delta^{\mathbf j}_{\mathbf i}\theta^4\ ,\quad
\theta^{{\mathbf i}{\mathbf j}}\theta^{\alpha\beta}=0\ ,\quad
\theta^{{\mathbf i}{\mathbf j}}\theta^{{\mathbf k}{\mathbf l}}=\frac{1}{3}
\varepsilon^{({\mathbf i}|{\mathbf l}}\varepsilon^{|{\mathbf j}){\mathbf k}}
\theta^4
\ ,\quad
\theta^{\alpha\beta}\theta^{\gamma\delta}=-\frac{1}{3}
\varepsilon^{(\alpha|\gamma}\varepsilon^{|\beta)\delta}\theta^4\ ,
\\[3mm]
\bar\theta^{\dot{\beta}}_{\mathbf j}\bar\theta^{3{\mathbf i}}_{\dot{\alpha}}=-
\frac{1}{4}\delta^{\dot{\beta}}_{\dot{\alpha}}\delta^{\mathbf i}_{\mathbf j}
\bar\theta^4\ ,\quad
\bar\theta^{{\mathbf i}{\mathbf j}}\bar\theta^{{\dot{\alpha}}{\dot{\beta}}}=0\
,
\bar\theta^{{\mathbf i}{\mathbf j}}\bar\theta^{{\mathbf k}{\mathbf l}}=\frac{1}
{3}\varepsilon^{({\mathbf i}|{\mathbf l}}\varepsilon^{|{\mathbf j}){\mathbf k}}
\bar\theta^4
\ ,\quad
\bar\theta^{{\dot{\alpha}}{\dot{\beta}}}\bar
\theta^{{\dot{\gamma}}{\dot{\delta}}}=-\frac{1}{3}
\varepsilon^{({\dot{\alpha}}|{\dot{\gamma}}}
\varepsilon^{|{\dot{\beta}}){\dot{\delta}}}\bar\theta^4\ .
\end{array}\label{9e}
\]
\noindent
Definitions of products of covariant derivatives:
\[
\begin{array}{lc}
D^2\ :&
D^{{\mathbf i}{\mathbf j}}\equiv D^{\alpha{\mathbf i}}D_{\alpha}^{\mathbf j}
\ ,\quad
D^{\alpha\beta}\equiv D^{\alpha{\mathbf i}}D^\beta_{\mathbf i}\ ,\quad
{\bar D}^{{\mathbf i}{\mathbf j}}\equiv{\bar D}_{\dot{\alpha}}^{\mathbf i}
{\bar D}^{{\dot{\alpha}}{\mathbf j}}\ ,\quad
{\bar D}^{{\dot{\alpha}}{\dot{\beta}}}\equiv
{\bar D}^{{\dot{\alpha}}{\mathbf i}}{\bar D}^{\dot{\beta}}_{\mathbf i}\ ,
\\[3mm]
D^3\ :&
D^3_{\alpha{\mathbf i}}\equiv D^{\mathbf j}_\alpha
D_{{\mathbf j}{\mathbf i}}=-D^\beta_{\mathbf i} D_{\beta\alpha}\ ,\quad
{\bar D}^{3{\mathbf i}}_{\dot{\alpha}}\equiv
{\bar D}_{{\dot{\alpha}}{\mathbf j}}{\bar D}^{{\mathbf i}{\mathbf j}}
=-{\bar D}^{{\dot{\beta}}{\mathbf i}}{\bar D}_{{\dot{\beta}}{\dot{\alpha}}}\ ,
\\[3mm]
D^4\ :&
D^4\equiv D^{{\mathbf i}{\mathbf j}}D_{{\mathbf i}{\mathbf j}}
=-D^{\alpha\beta}D_{\alpha\beta}
\ ,\quad
{\bar D}^4\equiv{\bar D}^{{\mathbf i}{\mathbf j}}
{\bar D}_{{\mathbf i}{\mathbf j}}
=-{\bar D}^{{\dot{\alpha}}{\dot{\beta}}}{\bar D}_{{\dot{\alpha}}{\dot{\beta}}}
\ .
\end{array}\label{eq8}
\]
Properties of the products of covariant derivatives:
\[
\begin{array}{ll}
(12):&
D^\alpha_{\mathbf i} D_{{\mathbf j}{\mathbf k}}=
\frac{2}{3}\varepsilon_{{\mathbf i}({\mathbf j}}D^{3\alpha}_{{\mathbf k})}\ ,
\quad
{\bar D}^{\dot{\alpha}}_{\mathbf i}{\bar D}_{{\mathbf j}{\mathbf k}}=
-\frac{2}{3}\varepsilon_{{\mathbf i}({\mathbf j}}
{\bar D}^{3{\dot{\alpha}}}_{{\mathbf k})}\ ,
\\[3mm]
\,\,\,&D_{\alpha{\mathbf i}}D_{\beta\gamma}=
-\frac{2}{3}\varepsilon_{\alpha(\beta}D^3_{\gamma){\mathbf i}}\ ,\quad
{\bar D}_{{\dot{\alpha}}{\mathbf i}}{\bar D}_{{\dot{\beta}}{\dot{\gamma}}}=
-\frac{2}{3}\varepsilon_{{\dot{\alpha}}({\dot{\beta}}}
{\bar D}^3_{{\dot{\gamma}}){\mathbf i}}\ ,\quad
\\[3mm]
(13):&
D_{\alpha{\mathbf i}}D^3_{\beta{\mathbf j}}=
\frac{1}{4}\varepsilon_{\alpha\beta}\varepsilon_{{\mathbf i}{\mathbf j}}D^4\ ,
\quad
{\bar D}_{{\dot{\alpha}}{\mathbf i}}{\bar D}^3_{{\dot{\beta}}{\mathbf j}}=
\frac{1}{4}\varepsilon_{{\dot{\alpha}}{\dot{\beta}}}
\varepsilon_{{\mathbf i}{\mathbf j}}{\bar D}^4\ ,
\\[3mm]
(22):&
D_{{\mathbf i}{\mathbf j}}D_{{\mathbf k}{\mathbf l}}=
\frac{1}{3}\varepsilon_{({\mathbf i} |{\mathbf l}}
\varepsilon_{|{\mathbf j}){\mathbf k}}D^4\ ,\quad
{\bar D}_{{\mathbf i}{\mathbf j}}{\bar D}_{{\mathbf k}{\mathbf l}}=
\frac{1}{3}\varepsilon_{({\mathbf i} |{\mathbf l}}
\varepsilon_{|{\mathbf j}){\mathbf k}}{\bar D}^4\ ,
\\[3mm]&
D_{{\mathbf i}{\mathbf j}}D_{\alpha\beta}=0\ ,\quad
{\bar D}_{{\mathbf i}{\mathbf j}}{\bar D}_{{\dot{\alpha}}{\dot{\beta}}}=0\ ,
\quad
D_{{\mathbf i}{\mathbf j}}{\bar D}^4 D_{{\mathbf k}{\mathbf l}}=
{\bar D}_{{\mathbf i}{\mathbf j}}D^4 {\bar D}_{{\mathbf k}{\mathbf l}}
\ .
\end{array}\label{eq8ba}
\]

\subsubsection{Complex Conjugation}

\noindent
The general complex conjugation rule for the $SU(2)$ indices
is the following:
$\ \left({\varphi}_{\mathbf i}\right)^*={\bar{\varphi}}^{\mathbf i}\ $.
As a consequence of this, we have
$({\bar{\varphi}}_{\mathbf i})^*=
(\varepsilon_{{\mathbf i}{\mathbf j}}
{\bar{\varphi}}^{\mathbf j})^*=
\varepsilon_{{\mathbf i}{\mathbf j}}{\varphi}_{\mathbf j}
= - \varepsilon^{{\mathbf i}{\mathbf j}}{\varphi}_{\mathbf j}
=-{\varphi}^{\mathbf i}\ .
\label{eq4c}$
For the specific products of $\theta$'s defined above, this leads to:
\[
\begin{array}{l}
(\theta^{{\mathbf i}{\mathbf j}})^*=
\bar\theta_{{\mathbf i}{\mathbf j}}\ ,\quad
(\theta^{\alpha\beta})^*=
-\bar\theta^{{\dot{\alpha}}{\dot{\beta}}}\ ,\quad
(\theta^{\alpha{\dot{\alpha}}})^*=
\theta^{\alpha{\dot{\alpha}}}\ ,\quad
(\theta^3_{\alpha{\mathbf i}})^*=
\bar\theta^{3{\mathbf i}}_{\dot{\alpha}}\ ,
\quad
(\theta^4)^*=\bar\theta^4\ .
\end{array}\label{eq9bb}
\]
\noindent
For the spinorial and covariant derivatives, this gives:
\[
\begin{array}{lcllcl}
(\partial_{\alpha}^{{\mathbf i}})^*&=&
-{\bar\partial}_{\dot{\alpha}{\mathbf i}}\ ,\quad&
(D_{\alpha}^{{\mathbf i}})^* &=&
 \phantom{-}{\bar D}_{\dot\alpha{\mathbf i}}\ ,
\\[3mm]
(\partial_{\alpha{\mathbf i}})^*&=&
{\bar\partial}_{{\dot{\alpha}}}^{\mathbf i}\ ,&
(D_{\alpha {\mathbf i}})^* &=&
 -{\bar D}_{{\dot{\alpha}}}^{\mathbf i}\ ,
\end{array}
\label{eq5c}
\]
and for higher order products:
\[
(D^{{\mathbf i}{\mathbf j}})^*=
{\bar D}_{{\mathbf i}{\mathbf j}}\ ,\quad
(D^{\alpha\beta})^*=
-{\bar D}^{{\dot{\alpha}}{\dot{\beta}}}\ ,\quad
(D^{3}_{\alpha{\mathbf i}})^*=
{\bar D}^{3{\mathbf i}}_{{\dot{\alpha}}}\ ,\quad
(D^4)^*={\bar D}^4\ .
\label{eq8b}
\]

\subsubsection{Algebra of covariant Derivatives}

\begin{flushleft}
$(12):$
\end{flushleft}
\begin{eqnarray*}
\left[D_{{\mathbf i}{\mathbf j}},{\bar D}_{{\dot{\alpha}}{\mathbf k}}\right]&=&
4i\varepsilon_{{\mathbf k}({\mathbf i}}D_{{\mathbf j})}^\alpha
\partial_{\alpha{\dot{\alpha}}}\ ,\\
\left[ {\bar D}_{{\mathbf i}{\mathbf j}} , D_{\alpha{\mathbf k}} \right] &=&
4i\varepsilon_{{\mathbf k}({\mathbf i}}\partial_{\alpha{\dot{\alpha}}}
{\bar D}^{\dot{\alpha}}_{{\mathbf j})}\ , \\
\left[D_{\alpha\beta},{\bar D}_{{\dot{\alpha}}{\mathbf i}}\right] &=& 4i
\partial_{(\alpha{\dot{\alpha}}}D_{\beta){\mathbf i}}\ , \\
\left[{\bar D}_{{\dot{\alpha}}{\dot{\beta}}},D_{\alpha{\mathbf i}}\right] &=&
-4i\partial_{\alpha({\dot{\alpha}}}{\bar D}_{{\dot{\beta}}){\mathbf i}}\ , \\
\left[D_{{\mathbf i}{\mathbf j}},{\bar D}_{\dot{\alpha}}^{\mathbf j}\right]
&=& 6iD_{\mathbf i}^\alpha\partial_{\alpha{\dot{\alpha}}}\ ,\\
\left[{\bar D}_{{\mathbf i}{\mathbf j}},D_\alpha^{\mathbf j}\right] &=& 6i
\partial_{\alpha{\dot{\alpha}}}{\bar D}^{\dot{\alpha}}_{\mathbf i}\ .
\end{eqnarray*}
\begin{flushleft}
$(22):$
\end{flushleft}
\begin{eqnarray}
\left[D_{{\mathbf i}{\mathbf j}},{\bar D}_{{\mathbf k}{\mathbf l}}\right] &=&
16\varepsilon_{{\mathbf k}({\mathbf i}}\varepsilon_{{\mathbf j}){\mathbf l}}
\hspace{0.2ex}\raisebox{.5ex}{\fbox{}}\hspace{0.3ex}
-8i\varepsilon_{({\mathbf i}({\mathbf k}}D^\alpha_{{\mathbf j})}
\partial_{\alpha{\dot{\alpha}}}{\bar D}^{\dot{\alpha}}_{{\mathbf l})}\ ,
{\nonumber}\\
&=&-16\varepsilon_{{\mathbf k}({\mathbf i}}
\varepsilon_{{\mathbf j}){\mathbf l}}\hspace{0.2ex}\raisebox{.5ex}{\fbox{}}
\hspace{0.3ex}
+8i\varepsilon_{({\mathbf i}({\mathbf k}}{\bar D}^{\dot{\alpha}}_{{\mathbf l})}
\partial_{\alpha{\dot{\alpha}}}D^\alpha_{{\mathbf j})}\ ,
{\nonumber}\\
\left[D_{{\mathbf i}{\mathbf j}},{\bar D}_{{\dot{\alpha}}{\dot{\beta}}}\right]
&=&
8i\partial_{\alpha({\dot{\alpha}}}D^\alpha_{({\mathbf i}}
{\bar D}_{{\dot{\beta}}){\mathbf j})}\ , {\nonumber}\\
\left[{\bar D}_{{\mathbf i}{\mathbf j}},D_{\alpha\beta}\right]& =&
-8i\partial_{(\alpha{\dot{\alpha}}}D_{\beta)({\mathbf i}}
{\bar D}^{\dot{\alpha}}_{{\mathbf j})}\ , \label{eq636}
\\
\left[D_{\alpha\beta},{\bar D}_{{\dot{\alpha}}{\dot{\beta}}}\right] &=&
4i\partial_{(\alpha({\dot{\alpha}}}\left[D^{\mathbf i}_{\beta)},
{\bar D}_{{\dot{\beta}}){\mathbf i}}\right]\ ,
\label{eq230} \\
\left[D_{{\mathbf i}}{}^{{\mathbf j}},{\bar D}_{{\mathbf j}{\mathbf k}}\right]
&=&-4i\partial_{\alpha{\dot{\alpha}}}\left[D^\alpha_{\mathbf i}\ ,
{\bar D}^{\dot{\alpha}}_{\mathbf k}\right]
-i\varepsilon_{{\mathbf i}{\mathbf k}}\partial_{\alpha{\dot{\alpha}}}
\left[D^{\alpha{\mathbf j}},{\bar D}^{\dot{\alpha}}_{\mathbf j}\right]\ ,
{\nonumber} \\
\left[D^{{\mathbf i}{\mathbf j}},{\bar D}_{{\mathbf i}{\mathbf j}}\right] &=&
-6i\partial_{\alpha{\dot{\alpha}}}\left[D^{\alpha{\mathbf i}},
{\bar D}^{\dot{\alpha}}_{\mathbf i}\right]\ , \label{eq502}
\end{eqnarray}
\begin{flushleft}
$(13):$
\end{flushleft}
\begin{eqnarray*}
\{D_{\alpha{\mathbf i}},{\bar D}^3_{{\dot{\alpha}}{\mathbf j}}\} &=&
-3i\varepsilon_{{\mathbf i}{\mathbf j}}\partial_{\alpha{\dot{\beta}}}
{\bar D}^{\dot{\beta}}{}_{\dot{\alpha}}
-3i\partial_{\alpha{\dot{\alpha}}}{\bar D}_{{\mathbf i}{\mathbf j}}\ , \\
\{{\bar D}_{{\dot{\alpha}}{\mathbf i}},D^3_{\alpha{\mathbf j}}\} &=&
3i\varepsilon_{{\mathbf i}{\mathbf j}}D_\alpha{}^\beta
\partial_{\beta{\dot{\alpha}}}
-3i\partial_{\alpha{\dot{\alpha}}}D_{{\mathbf i}{\mathbf j}}\ .
\end{eqnarray*}
\begin{flushleft}
$(14):$
\end{flushleft}
\begin{eqnarray*}
\left[D^{\alpha{\mathbf i}},{\bar D}^4\right] &=&
8i\partial^{\alpha{\dot{\alpha}}}{\bar D}^{3{\mathbf i}}_{\dot{\alpha}} \ ,\\
\left[{\bar D}_{\dot{\alpha}}^{\mathbf i},D^4\right] &=&
8i D^{3\alpha{\mathbf i}} \partial_{\alpha{\dot{\alpha}}}\ .
\end{eqnarray*}
\begin{flushleft}
$(23):$
\end{flushleft}
\begin{eqnarray*}
\left[D^{3\alpha}_{\mathbf k},{\bar D}_{{\mathbf i}{\mathbf j}}\right] &=&
\varepsilon_{{\mathbf k}({\mathbf i}}\left(
-24D^\alpha_{{\mathbf j})}\hspace{0.2ex}\raisebox{.5ex}{\fbox{}}\hspace{0.3ex}
-6i{\bar D}^{{\dot{\beta}}}_{{\mathbf j})}D^{\alpha\beta}
\partial_{\beta{\dot{\beta}}}\right)
+6i\partial^{\alpha{\dot{\alpha}}}
{\bar D}_{{\dot{\alpha}}({\mathbf i}}D_{{\mathbf j}){\mathbf k}}\ , \\
\left[{\bar D}^{3{\dot{\alpha}}}_{\mathbf k},D_{{\mathbf i}{\mathbf j}}
\right]&=&
\varepsilon_{{\mathbf k}({\mathbf i}}\left(
24{\bar D}^{\dot{\alpha}}_{{\mathbf j})}\hspace{0.2ex}\raisebox{.5ex}{\fbox{}}
\hspace{0.3ex}
-6iD^{\beta}_{{\mathbf j})}{\bar D}^{{\dot{\alpha}}{\dot{\beta}}}
\partial_{\beta{\dot{\beta}}}\right)
-6i\partial^{\alpha{\dot{\alpha}}}D_{\alpha({\mathbf i}}
{\bar D}_{{\mathbf j}){\mathbf k}}\ , \\
\left[D^{3\alpha}_{\mathbf k},{\bar D}_{{\dot{\alpha}}{\dot{\beta}}}\right] &=&
-24\partial^\alpha{}_{({\dot{\alpha}}}
\partial_{\beta{\dot{\beta}})}D^\beta_{\mathbf k}
+6i{\bar D}_{({\dot{\alpha}}{\mathbf k}}D^{\alpha\beta}
\partial_{\beta{\dot{\beta}})}
+6i\partial^\alpha{}_{({\dot{\alpha}}}
{\bar D}^{\mathbf i}_{{\dot{\beta}})}D_{{\mathbf i}{\mathbf k}}\ , \\
\left[{\bar D}^{3{\dot{\alpha}}}_{\mathbf k},D_{\alpha\beta}\right] &=&
-24\partial_{(\alpha{\dot{\beta}}}\partial_{\beta)}{}^{{\dot{\alpha}}}
{\bar D}^{\dot{\beta}}_{\mathbf k}
-6iD_{(\alpha{\mathbf k}}{\bar D}^{{\dot{\alpha}}{\dot{\beta}}}
\partial_{\beta){\dot{\beta}}}
+6i\partial_{(\alpha}{}^{{\dot{\alpha}}}D^{\mathbf i}_{\beta)}
{\bar D}_{{\mathbf i}{\mathbf k}}\ , \\
\left[D^{3\alpha}_{\mathbf j},{\bar D}^{{\mathbf i}{\mathbf j}}\right] &=&
36\hspace{0.2ex}\raisebox{.5ex}{\fbox{}}\hspace{0.3ex} D^{\alpha{\mathbf i}}+9i
\partial_{\beta{\dot{\alpha}}}
{\bar D}^{{\dot{\alpha}}{\mathbf i}}D^{\alpha\beta}
-3i\partial^{\alpha{\dot{\alpha}}}
{\bar D}_{{\dot{\alpha}}{\mathbf j}}D^{{\mathbf i}{\mathbf j}}\ ,\\
\left[{\bar D}^{3{\dot{\alpha}}}_{\mathbf j},D^{{\mathbf i}{\mathbf j}}\right]
&=&
-36\hspace{0.2ex}\raisebox{.5ex}{\fbox{}}\hspace{0.3ex}
{\bar D}^{{\dot{\alpha}}{\mathbf i}}-9i\partial^\alpha{}_{\dot{\beta}}
D^{\mathbf i}_\alpha {\bar D}^{{\dot{\alpha}}{\dot{\beta}}}
+3i\partial^{\alpha{\dot{\alpha}}}D_{\alpha{\mathbf j}}
{\bar D}^{{\mathbf i}{\mathbf j}}\ .
\end{eqnarray*}
\begin{flushleft}
$(24):$
\end{flushleft}
\begin{eqnarray*}
\left[D_{{\mathbf i}{\mathbf j}},{\bar D}^4\right]
&=&48\hspace{0.2ex}\raisebox{.5ex}{\fbox{}}\hspace{0.3ex}
{\bar D}_{{\mathbf i}{\mathbf j}}
+16i\partial_{\alpha{\dot{\alpha}}}{\bar D}^{3{\dot{\alpha}}}_{({\mathbf i}}
D^\alpha_{{\mathbf j})}\ ,\\
&=&-48\hspace{0.2ex}\raisebox{.5ex}{\fbox{}}\hspace{0.3ex}
{\bar D}_{{\mathbf i}{\mathbf j}}
-16i\partial_{\alpha{\dot{\alpha}}} D^\alpha_{({\mathbf i}}
{\bar D}^{3{\dot{\alpha}}}_{{\mathbf j})}\ ,\\
\left[D_{\alpha\beta},{\bar D}^4\right]
&=&-48\partial_{(\alpha{\dot{\alpha}}}\partial_{\beta){\dot{\beta}}}
{\bar D}^{{\dot{\alpha}}{\dot{\beta}}}
+16i\partial_{(\alpha{\dot{\alpha}}}{\bar D}^{3{\dot{\alpha}}}_{{\mathbf i}}
D^{\mathbf i}_{\beta)}\ ,\\
&=&48\partial_{(\alpha{\dot{\alpha}}}\partial_{\beta){\dot{\beta}}}
{\bar D}^{{\dot{\alpha}}{\dot{\beta}}}
-16i\partial_{(\alpha{\dot{\alpha}}} D_{\beta)}^{\mathbf i}
{\bar D}^{3{\dot{\alpha}}}_{{\mathbf i}}\ ,\\
\left[{\bar D}_{{\mathbf i}{\mathbf j}},D^4\right]
&=&48\hspace{0.2ex}\raisebox{.5ex}{\fbox{}}\hspace{0.3ex}
D_{{\mathbf i}{\mathbf j}}
+16i\partial_{\alpha{\dot{\alpha}}}D^{3\alpha}_{({\mathbf i}}
{\bar D}^{\dot{\alpha}}_{{\mathbf j})}\ , \\
&=&-48\hspace{0.2ex}\raisebox{.5ex}{\fbox{}}\hspace{0.3ex}
D_{{\mathbf i}{\mathbf j}}
-16i\partial_{\alpha{\dot{\alpha}}} {\bar D}^{\dot{\alpha}}_{({\mathbf i}}
D^{3\alpha}_{{\mathbf j})}\ , \\
\left[{\bar D}_{{\dot{\alpha}}{\dot{\beta}}},D^4\right]
&=&-48\partial_{\alpha({\dot{\alpha}}}\partial_{\beta{\dot{\beta}})}
D^{\alpha\beta}
-16i\partial_{\alpha({\dot{\alpha}}}D^{3\alpha}_{{\mathbf i}}
{\bar D}^{\mathbf i}_{{\dot{\beta}})}\ ,\\
&=&48\partial_{\alpha({\dot{\alpha}}}\partial_{\beta{\dot{\beta}})}
D^{\alpha\beta}
+16i\partial_{\alpha({\dot{\alpha}}} {\bar D}_{{\dot{\beta}})}^{\mathbf i}
D^{3\alpha}_{{\mathbf i}}\ .
\end{eqnarray*}
\pagebreak
\begin{flushleft}
$(33):$
\end{flushleft}
\begin{eqnarray*}
\{D^{3{\mathbf i}}_{\alpha},{\bar D}^3_{{\dot{\alpha}}{\mathbf i}}\} &=&
-72i\partial_{\alpha{\dot{\alpha}}}\hspace{0.2ex}\raisebox{.5ex}{\fbox{}}
\hspace{0.3ex}
+\frac{9i}{4}
\partial_{\alpha{\dot{\alpha}}}\{D^{{\mathbf i}{\mathbf j}},
{\bar D}_{{\mathbf i}{\mathbf j}}\}
-\frac{9i}{2}\partial^{\beta{\dot{\beta}}}\{D_{\alpha\beta},
{\bar D}_{{\dot{\alpha}}{\dot{\beta}}}\} \ ,\\
\{D^3_{\alpha({\mathbf i}},{\bar D}^3_{{\dot{\alpha}}{\mathbf j})}\} &=&
-\frac{9i}{4}\left(
\partial_{\alpha{\dot{\alpha}}}\{D_{{\mathbf k}({\mathbf i}},
{\bar D}^{\mathbf k}{}_{{\mathbf j})}\}
+\partial_{\beta{\dot{\alpha}}}\{D_\alpha{}^\beta,
{\bar D}_{{\mathbf i}{\mathbf j}}\}
+\partial_{\alpha{\dot{\beta}}}\{D_{{\mathbf i}{\mathbf j}},
{\bar D}_{\dot{\alpha}}{}^{\dot{\beta}}\}
\right)\ .
\end{eqnarray*}
\begin{flushleft}
$(34):$
\end{flushleft}
\begin{eqnarray*}
\left[D^3_{\alpha{\mathbf i}},{\bar D}^4\right] &=&
-288i{\bar D}^{\dot{\alpha}}_{\mathbf i}\partial_{\alpha{\dot{\alpha}}}\hspace
{0.2ex}\raisebox{.5ex}{\fbox{}}\hspace{0.3ex}
-72D^{\mathbf j}_\alpha{\bar D}_{{\mathbf i}{\mathbf j}}\hspace{0.2ex}\raisebox
{.5ex}{\fbox{}}\hspace{0.3ex}
-72\partial_{\alpha{\dot{\alpha}}}
\partial_{\beta{\dot{\beta}}}D^\beta_{\mathbf i}
{\bar D}^{{\dot{\alpha}}{\dot{\beta}}} \nonumber\\
&&-12i\partial_{\alpha{\dot{\alpha}}}D_{{\mathbf i}{\mathbf j}}
{\bar D}^{3{\dot{\alpha}}{\mathbf j}}
-12i\partial^{\beta{\dot{\beta}}}D_{\alpha\beta}
{\bar D}^3_{{\dot{\beta}}{\mathbf i}}\ ,\\
\left[{\bar D}^3_{{\dot{\alpha}}{\mathbf i}},D^4\right] &=&
288iD^\alpha_{\mathbf i}\partial_{\alpha{\dot{\alpha}}}\hspace{0.2ex}\raisebox
{.5ex}{\fbox{}}\hspace{0.3ex}
+72{\bar D}^{\mathbf j}_{\dot{\alpha}} D_{{\mathbf i}{\mathbf j}}\hspace{0.2ex}
\raisebox{.5ex}{\fbox{}}\hspace{0.3ex}
-72\partial_{\alpha{\dot{\alpha}}}\partial_{\beta{\dot{\beta}}}
{\bar D}^{\dot{\beta}}_{\mathbf i} D^{\alpha\beta} \nonumber\\
&&-12i\partial_{\alpha{\dot{\alpha}}}
{\bar D}_{{\mathbf i}{\mathbf j}}D^{3\alpha{\mathbf j}}
+12i\partial^{\beta{\dot{\beta}}}
{\bar D}_{{\dot{\alpha}}{\dot{\beta}}}D^3_{\beta{\mathbf i}}\ .
\end{eqnarray*}
\begin{flushleft}
$(44):$
\end{flushleft}
$$
\left[D^4,{\bar D}^4\right]=
-16i\partial_{\alpha{\dot{\alpha}}}\left[D^{3\alpha{\mathbf i}},
{\bar D}^{3{\dot{\alpha}}}_{\mathbf i}\right]
+288i\hspace{0.2ex}\raisebox{.5ex}{\fbox{}}\hspace{0.3ex}
\partial_{\alpha{\dot{\alpha}}}\left[D^{\alpha{\mathbf i}},
{\bar D}^{\dot{\alpha}}_{\mathbf i}\right]\ .
$$

\subsubsection{Derivatives acting on $\theta$}

\[\begin{array}{rclrcl}
\partial_{\alpha{\mathbf i}}\theta^4&=&
-4\theta^3_{\alpha{\mathbf i}}\ ,&
{\bar\partial}_{\dot{\alpha}}^{\mathbf i}\bar\theta^4&=&
-4\bar\theta^{3{\mathbf i}}_{\dot{\alpha}}\ ,\\[3mm]
\partial_{\alpha{\mathbf i}}\theta^{3\beta{\mathbf j}}&=&
-\frac{3}{2}\left(\delta^\beta_\alpha\theta_{\mathbf i}{}^{\mathbf j}
-\delta^{\mathbf j}_{\mathbf i}\theta^\beta{}_\alpha\right)\ ,\quad&
{\bar\partial}_{\dot{\alpha}}^{\mathbf i}\bar
\theta^{3{\dot{\beta}}}_{{\mathbf j}}&=&
-\frac{3}{2}\left(
\delta^{\mathbf i}_{\mathbf j}\bar\theta^{\dot{\beta}}{}_{\dot{\alpha}}-
\delta^{\dot{\beta}}_{\dot{\alpha}}
\bar\theta^{\mathbf i}{}_{\mathbf j}\right)\ ,
\end{array}\]
\[
\begin{array}{rclrclrcl}
D^{{\mathbf i}{\mathbf j}}\theta^4&=&-12\theta^{{\mathbf i}{\mathbf j}}\ ,&
D_{{\mathbf i}{\mathbf j}}\theta^{3\alpha{\mathbf k}}&=&6
\delta^{\mathbf k}_{({\mathbf i}}\theta^\alpha_{{\mathbf j})}\ ,&
D^{{\mathbf i}{\mathbf j}}\theta^{3}_{\alpha{\mathbf j}}&=&-9
\theta^{\mathbf i}_\alpha\ ,
\\[3mm]
D^{{\mathbf i}{\mathbf j}}\theta_{{\mathbf k}{\mathbf l}}&=&-4
\delta^{\mathbf i}_{({\mathbf k}}\delta^{\mathbf j}_{{\mathbf l})}\ ,&
D^{{\mathbf i}{\mathbf j}}\theta_{{\mathbf i}{\mathbf j}}&=&-12
\ ,&&&\\[3mm]
D^{\alpha\beta}\theta_{\alpha\beta}&=&-12\ ,&
D_{\alpha\beta}\theta_{\gamma\delta}&=&-4\varepsilon_{(\alpha\gamma}
\varepsilon_{\beta)\delta}
\ ,\quad&&&\\[3mm]
D^{3\alpha{\mathbf i}}\theta^4&=&-36\theta^{\alpha{\mathbf i}}\ ,\qquad&
D^{3\alpha{\mathbf i}}\theta^3_{\beta{\mathbf j}}&=&9\delta^\alpha_\beta
\delta^{\mathbf i}_{\mathbf j}\ ,&
D^4\theta^4&=&144
\ ,\\[3mm]
{\bar D}^{{\mathbf i}{\mathbf j}}\bar\theta^4&=&-12\bar
\theta^{{\mathbf i}{\mathbf j}}\ ,&
{\bar D}_{{\mathbf i}{\mathbf j}}\bar\theta^{3{\dot{\alpha}}{\mathbf k}}&=&6
\delta^{\mathbf k}_{({\mathbf i}}\bar\theta^{\dot{\alpha}}_{{\mathbf j})}\ ,&
{\bar D}^{{\mathbf i}{\mathbf j}}\bar
\theta^{3}_{{\dot{\alpha}}{\mathbf j}}&=&-9
\bar\theta^{\mathbf i}_{\dot{\alpha}}\ ,
\\[3mm]
{\bar D}^{{\mathbf i}{\mathbf j}}\bar\theta_{{\mathbf k}{\mathbf l}}&=&-4
\delta^{\mathbf i}_{({\mathbf k}}\delta^{\mathbf j}_{{\mathbf l})}\ ,&
{\bar D}^{{\mathbf i}{\mathbf j}}\bar\theta_{{\mathbf i}{\mathbf j}}&=&-12
\ ,&&&\\[3mm]
{\bar D}^{3{\dot{\alpha}}{\mathbf i}}\bar\theta^4&=&-36\bar
\theta^{{\dot{\alpha}}{\mathbf i}}\ ,&
{\bar D}^{3{\dot{\alpha}}{\mathbf i}}\bar\theta^3_{\beta{\mathbf j}}&=&-9
\delta^{\dot{\alpha}}_{\dot{\beta}}\delta^{\mathbf i}_{\mathbf j}\ ,&
{\bar D}^4\bar\theta^4&=&144
\ ,\\[3mm]
\end{array}\label{111}\]
\begin{equation}\begin{array}{rlrl}
D^4(\theta^4 X)|&=\phantom{-} 144 X|\ ,&
{\bar D}^4(\bar\theta^4 {\bar X})|&=\phantom{-} 144 {\bar X}|\ ,\\[3mm]
D^4(\theta^3_{\alpha{\mathbf i}}
X^{\alpha{\mathbf i}})|&=36D_{\alpha{\mathbf i}} X^{\alpha{\mathbf i}}|\ ,&
{\bar D}^4({\bar\theta}^3_{{\dot{\alpha}}{\mathbf i}}
{\bar X}^{{\dot{\alpha}}{\mathbf i}})|&=\phantom{-} -36
{\bar D}_{{\dot{\alpha}}{\mathbf i}}
{\bar X}^{{\dot{\alpha}}{\mathbf i}} |\ ,\\[3mm]
D^4(\theta_{{\mathbf i}{\mathbf j}}
X^{{\mathbf i}{\mathbf j}})|&=-12D_{{\mathbf i}{\mathbf j}}
X^{{\mathbf i}{\mathbf j}}|\ ,&
{\bar D}^4({\bar\theta}_{{\mathbf i}{\mathbf j}}
{\bar X}^{{\mathbf i}{\mathbf j}})|&=-12{\bar D}_{{\mathbf i}{\mathbf j}}
{\bar X}^{{\mathbf i}{\mathbf j}} |\ ,\\[3mm]
D^4(\theta_{\alpha\beta} X^{\alpha\beta})|&=-12D_{\alpha\beta}
X^{\alpha\beta}|\ ,&
{\bar D}^4(\bar\theta_{{\dot{\alpha}}{\dot{\beta}}}
{\bar X}^{{\dot{\alpha}}{\dot{\beta}}})|&=-12
{\bar D}_{{\dot{\alpha}}{\dot{\beta}}}
{\bar X}^{{\dot{\alpha}}{\dot{\beta}}}
|\ ,\\[3mm]
D^4(\theta_{\alpha{\mathbf i}} X^{\alpha{\mathbf i}})|&=4
D^3_{\alpha{\mathbf i}}X^{\alpha{\mathbf i}}|\ ,&
{\bar D}^4(\bar\theta_{{\dot{\alpha}}{\mathbf i}}
{\bar X}^{{\dot{\alpha}}{\mathbf i}})|&=4
{\bar D}^3_{{\dot{\alpha}}{\mathbf i}}{\bar X}^{{\dot{\alpha}}{\mathbf i}}|\ .
\end{array}\label{eq10d}\end{equation}

\end{appendix}


\begin{thebibliography}{999}
\bibitem{WZ1} J. Wess and B. Zumino, Nucl. Phys. {\bf B70} (1974) 139.

\bibitem{FZ1} S. Ferrara and B. Zumino, Nucl. Phys. {\bf B87} (1975) 174.

\bibitem{Gr1} M.T.  Grisaru, {\it in} Cargese Lectures, 1979, eds. M.Levy,
D.Deser (Plenum, 1979),  p. 130.

\bibitem{NSVZ1} V. Novikov, M. Shifman, A. Vainstein and V. Zakharaov, Nucl.
Phys. {\bf B229} (1983) 381; Phys. Lett. {\bf B166} (1986) 329.

\bibitem{GW1} M.T. Grisaru and P. West, Nucl. Phys. {\bf B254} (1985) 249.

\bibitem{SW1} M. Sohnius and P. West, Phys. Lett. {\bf B100} (1981) 45.

\bibitem{HSW1} P. Howe, K.S. Stelle and P.C. West, Phys. Lett. {\bf B124}
(1983) 55.

\bibitem{PS} P.C. West, {\it in} Proceedings of the 1983 Shelter Island II
Conference on Quantum Field Theory and the Fundamental Problems of Physics;
edited by R. Jackiw, N. Khuri, S. Weinberg and E. Witten (M.I.T. Press).

\bibitem{Seiberg-Witten} N. Seiberg and E. Witten,
  Nucl. Phys. {\bf B426} (1994) 19 [Erratum {\bf B430}
  (1994) 485] [hep-th/9407087].

\bibitem{FMRSS} R. Flume, M. Magro, L. O{'}Raifeartaigh, O. Schnetz and I.
Sachs,  Nucl. Phys. {\bf B494} (1997) 331 [hep-th/9611123].

\bibitem{MRS} M. Magro, L. O{'}Raifeartaigh and I. Sachs, Nucl. Phys.
{\bf B508} (1997) 433 [hep-th/9704027].

\bibitem{OS1} V. Ogievetsky and  E. Sokatchev, Nucl. Phys. {\bf B127} (1977)
309.

\bibitem{SS1} A. Salam and J. Strathdee, Phys. Rev. {\bf D11} (1975) 1521;
Nucl. Phys. {\bf B86} (1975) 142.

\bibitem{OS2} V. Ogievetsky and  E. Sokatchev,
Sov. J. Nucl. Phys. {\bf 28}(3) (1978).

\bibitem{HST1} P. Howe, K.S. Stelle and P.K. Townsend, Nucl. Phys. {\bf B192}
(1981) 332.

\bibitem{Shizuya} K. Shizuya, Phys. Rev. {\bf D35} (1987) 1848.

\bibitem{Osborn}
H. Osborn,
Annals Phys.\ {\bf 272} (1999) 243
[hep-th/9808041].

\bibitem{Buchbinder} I.L. Buchbinder and S.M. Kuzenko,
{\em Ideas and Methods of Supersymmetry and
  Supergravity}, IOP Publ.,
Bristol and Philadelphia, 1995, Revised Edition 1998.

\bibitem{Clark-Piguet-Sibold}
T.E. Clark, O. Piguet and K. Sibold,
Nucl.  Phys.   {\bf B143} (1978) 445;\\
O. Piguet and K. Sibold,
{\it Renormalized Supersymmetry. The Perturbation Theory Of N=1
Supersymmetric Theories In Flat Space-Time},
Boston, Usa: Birkhaeuser ( 1986) 346 p. ( Progress In Physics,
12).

\bibitem{Howe96} P.S. Howe and P.C. West, Nucl. Phys.
{\bf B486} (1997) 425 [hep-th/9607239].

\bibitem{FN1} S. Ferrara and P. van Nieuwenhuizen, Phys. Lett. {\bf B74}
(1978) 333.

\bibitem{Aku1} V. Akulov, D. Volkov and V. Soroka, Theor. Math. Phys.
{\bf 31} (1977) 12; \\
M.F. Sohnious and P. West, Phys. Lett. {\bf B105} (1981) 353.

\bibitem{B1} P. Breitenlohner, Nucl. Phys. {\bf B124} (1977) 500.

\bibitem{GG1} S. J. Gates, M.T. Grisaru and W. Siegel, Nucl. Phys.
{\bf B203} (1982) 189.

\bibitem{GOS1} A. Galperin, V. Ogievetsky and  E. Sokatchev, Nucl. Phys.
{\bf B252} (1985) 435.

\bibitem{SG2} B. de Wit, J.W. van Holten and A. Van Proyen, Nucl. Phys.
{\bf B167} (1980) 186; Nucl. Phys. {\bf B148} (1981) 77;\\
B. de Wit, R. Philippe and A. Van Proyen, Nucl. Phys. {\bf B219} (1983) 143.

\bibitem{SG3} B. de Wit and A. Van Proyen,
Nucl. Phys. {\bf B245} (1984) 186;\\
B. de Wit, P.G. Lauwers and A. Van Proyen, Nucl. Phys. {\bf B255} (1985) 569.

\bibitem{SG4} L. Castellani, P. van Niewenhuizen and S.J. Gates, Phys.
Rev. {\bf D22} (1980) 2364; S.J. Gates, Nucl. Phys. {\bf B176} (1980) 397;
Phys. Lett. {\bf B96} (1980) 305.

\bibitem{SG5} P. Howe, Nucl. Phys. {\bf B199} (1982) 309.

\bibitem{SG6} M. M\"uller, {\it Consistent Classical Supergravity Theories},
LNP 336 (Springer, Berlin, 1989).

\bibitem{WessBagger} J. Wess and J. Bagger,
{\em Supersymmetry and Supergravity}, Second edition,
Princeton Series in Physics.

\bibitem{Sohnius76} M. Sohnius, Proc. 2nd Tutzing Symp. (1976) vol. 2, ed. L.Castell, M. Drieschner and C. von Weizs\"acker, (Carl Hanser Verlag, Munich).

\bibitem{Lang81} W. Lang, Nucl. Phys {\bf B179} (1981) 106.

\bibitem{Conlong-West}
B.P. Conlong and P.C. West (1993) unpublished; P.C. West, [hep-th/9805055].

\bibitem{Howe} P.S. Howe and G.G. Hartwell,
Class. Quant. Grav. {\bf 12} (1995) 1823.

\bibitem{Park}
J.H. Park,
Nucl.\ Phys.\ B {\bf 559} (1999) 455 [hep-th/9903230].

\bibitem{SW2} K.S. Stelle and P.C. West, Phys. Lett {\bf B74} (1978) 330.

\bibitem{Grisa} M.T. Grisaru, {\it in} Proceedings of
  Supersymmetry and Supergravity, Trieste 1984, p. 90.

\bibitem{Siegel1} W. Siegel, Phys. Lett. {\bf B85} (1979) 333.

\bibitem{WitRocek1} B. de Wit and M. Rocek, Phys. Lett. {\bf B109} (1982) 439.

\bibitem{shifman-chibisov}
B. Chibisov and M. Shifman,
Phys.\ Rev.\ D {\bf 56} (1997) 7990
[Erratum-ibid.\ D {\bf 58} (1997) 109901]
[hep-th/9706141].

\bibitem{GT1} G.W. Gibbons and P.K. Townsend, Phys. Rev. Lett. {\bf 83} (1999)
1727 [hep-th/9905196].

\bibitem{Brink1} L. Brink, J.H. Schwarz and J. Scherk, Nucl. Phys.
{\bf B121} (1977) 77;\\
 R. Grimm, M. Sohnius and J. Wess, Nucl. Phys. {\bf B133} (1978) 275.

\bibitem{Olive-Witten} E. Witten and D. Olive,
  Phys. Lett. {\bf B78} (1978) 97.

\bibitem{Wolf} S. Wolf, Mod. Phys. Lett. {\bf A14} (1999)
  2789 [hep/th-9905194].

\bibitem{Iorio} A. Iorio,
Phys. Lett. {\bf B487} (2000) 171 [hep-th/9905069].

\bibitem{Sohnius79} M. F. Sohnius, Phys. Lett. {\bf B81} (1979) 8.

\bibitem{Fisher} A.W. Fisher, Nucl. Phys. {\bf B229} (1983) 142.

\bibitem{Dahmen} H.D. Dahmen, S. Marculescu and L. Szymanowski,
Nucl. Phys. {\bf B383} (1992) 110.

\bibitem{GS2} S.J. Gates and W. Siegel, Nucl. Phys.
{\bf B195} (1982) 39.

\bibitem{deWit83} B. de Wit, R. Philippe and A. Van
  Proeyen, Nucl. Phys. {\bf B219} (1983) 143.

\bibitem{Stelle} K.S. Stelle,
{\it in} Proceedings of the Nuffield Workshop,
Quantum Structure of Space and Time, p. 337, August 1981,
edited by M.J. Duff and C.J. Isham.

\bibitem{Theisen} S. Kuzenko and S. Theisen,
Class. Quant. Grav. 17 (2000) 665-696  [hep-th/9907107].

\bibitem{LRSW} A. Iorio, L. O'Raifeartaigh, I. Sachs and C. Wiesendanger,
Nucl. Phys. {\bf B495} (1997) 433 [hep-th/9607110].

\bibitem{Ivanov} A.S. Galperin, E.Ivanov, S.Kalitzin, V. Ogievetsky and E. Sokatchev,
Class. Quant. Grav. {\bf 1} (1984) 469.
\end{thebibliography}
\end{document}